\documentclass[fleqn,usenatbib]{mnras}

\usepackage{hyperref}

\usepackage{epsfig}
\usepackage{myaasmacros}
\usepackage{amsmath}
\usepackage{amsfonts}
\usepackage{amssymb}
\usepackage{subfig}
\usepackage{comment}
\usepackage{color}

\newcommand{\Msun}{\hbox{M$_\sun$}}
\newcommand{\hii}{\hbox{H{\sc ii}}}
\newcommand{\ha}{\hbox{H$\alpha$}}
\newcommand{\hb}{\hbox{H$\beta$}}
\newcommand{\oiii}{\hbox{[O\,{\sc iii}]}}
\newcommand{\oiiilam}{\hbox{[O\,{\sc iii}]$\lambda5007$}}
\newcommand{\oiid}{\hbox{[O\,{\sc ii}]}}
\newcommand{\oiidlam}{\hbox{[O\,{\sc ii}]$\lambda\lambda3726,3729$}}
\newcommand{\oii}{\hbox{[O\,{\sc ii}]}}
 
\newcommand{\oi}{\hbox{[O\,{\sc i}]}}
\newcommand{\oilam}{\hbox{[O\,{\sc i}]$\lambda6300$}} 
\newcommand{\nii}{\hbox{[N\,{\sc ii}]}} 
\newcommand{\niilam}{\hbox{[N\,{\sc ii}]$\lambda6584$}}
\newcommand{\siid}{\hbox{[S\,{\sc ii}]}}
\newcommand{\siidlam}{\hbox{[S\,{\sc ii}]$\lambda\lambda6717,6731$}}

\newcommand{\oiiihb}{\hbox{[O\,{\sc iii}]/H$\beta$}}
\newcommand{\niiha}{\hbox{[N\,{\sc ii}]/H$\alpha$}}
\newcommand{\siiha}{\hbox{[S\,{\sc ii}]/H$\alpha$}}
\newcommand{\oiha}{\hbox{[O\,{\sc i}]/H$\alpha$}}

\title[Synthetic nebular emission] 
{Synthetic nebular emission from massive galaxies I: origin of the cosmic evolution 
  of optical emission-line ratios}  
\author[Hirschmann et al.]{Michaela Hirschmann$^{1}$\thanks{E-mail:
hirschma@iap.fr}, Stephane Charlot$^{1}$, Anna Feltre$^{1,2}$, Thorsten Naab$^{3}$,
\newauthor Ena Choi$^{4}$, Jeremiah P. Ostriker$^{5,6}$, Rachel S. Somerville$^{4,7}$ \\
$^{1}$Sorbonne Universit\'es, UPMC-CNRS, UMR7095, Institut d'~Astrophysique de Paris, F-75014
Paris, France\\ 
$^{2}$Univ.\,Lyon, Univ.\,Lyon1, ENS de Lyon, CNRS, Centre de Recherche Astrophysique de Lyon, UMR5574, 69230 Saint-Genis-Laval, France\\
$^{3}$Max-Planck-Institute for Astrophysics,
Karl-Schwarzschild-Strasse 1, 85741 Garching, Germany \\
$^{4}$Department of Physics and Astronomy, Rutgers, The State
University of New Jersey, NJ 08854, USA \\
$^{5}$Department of Astronomy, Columbia University, New York, NY 10027, USA \\
$^{6}$Department of Astrophysical Sciences, Princeton University,
Princeton, NJ 08544, USA\\
$^{7}$Center for Computational Astrophysics, Flatiron Institute, 162 5th Ave, New York, NY 10010 \\
}

\begin{document}

\date{Accepted ???. Received ??? in original form ???}

\pagerange{\pageref{firstpage}--\pageref{lastpage}} \pubyear{2002}

\maketitle

\label{firstpage}

\begin{abstract}
Galaxies occupy different regions of the [O\,{\sc
  iii}]$\lambda5007$/H$\beta$-versus-[N\,{\sc
  ii}]$\lambda6584$/H$\alpha$ emission-line ratio diagram in the
distant and local Universe. We investigate the origin of this
intriguing result by modelling self-consistently, for
the first time, nebular emission from young stars, accreting black
holes (BHs) and older, post-asymptotic-giant-branch (post-AGB) stellar
populations in galaxy formation simulations in a full cosmological
context. In post-processing, we couple new-generation nebular-emission
models with high-resolution,  cosmological zoom-in simulations of
massive galaxies to explore which galaxy physical properties drive the
redshift evolution of the optical-line ratios [O\,{\sc
  iii}]$\lambda5007$/H$\beta$,  [N\,{\sc ii}]$\lambda6584$/H$\alpha$,
[S\,{\sc ii}]$\lambda\lambda6717,6731$/H$\alpha$ and  [O\,{\sc
  i}]$\lambda6300$/H$\alpha$. The line ratios of simulated galaxies
agree well with observations of both star-forming and active local
SDSS galaxies. Toward higher redshifts, at fixed galaxy stellar mass,
the average \oiiihb\ is predicted to increase and \niiha, \siiha\ and
\oiha\ to decrease -- widely consistent with observations. At fixed
stellar mass, we identify star formation history, which controls
nebular emission from young stars via the ionization parameter, as the
primary driver of the cosmic evolution of \oiiihb\ and \niiha.  For
\siiha\ and \oiha, this applies only to redshifts greater than
$z=1.5$, the evolution at lower redshift being driven in roughly equal
parts by nebular emission from active galactic nuclei and post-AGB  stellar
populations. Instead, changes in the hardness of ionizing radiation,
ionized-gas density, the prevalence of BH accretion relative
to star formation and the dust-to-metal mass ratio (whose impact 
on the gas-phase N/O ratio we model at fixed O/H) play at
most a minor role in the cosmic evolution of simulated galaxy line
ratios.   
\end{abstract}

\begin{keywords}
galaxies: abundances; galaxies: formation; galaxies: evolution;
galaxies: general; methods: numerical
\end{keywords}

\section{Introduction}\label{intro}

The emission from ionized interstellar gas contains valuable
information about the nature of the ionizing radiation and the
physical conditions in the interstellar medium (ISM) in a galaxy. In
fact, prominent optical emission lines are routinely used to estimate
the density,  chemical abundances and dust content of the ISM and
whether ionization is dominated by  young massive stars (tracing the
star formation rate -- hereafter SFR), an active galactic nucleus
(hereafter AGN) or evolved, post-asymptotic giant-branch (hereafter
post-AGB) stars  \citep[e.g.,][]{Izotov99, Kobulnicky99, Kauffmann03,
  Nagao06, Kewley08, Morisset16}. These different types of ionizing
sources produce distinct, well-defined correlations between the
intensity ratios of strong lines, such as \ha, \hb, \oilam, \oiidlam,
\oiiilam, \niilam\ and  \siidlam\ (herafter simply \oi, \oiid\ ,
\oiii, \nii\ and \siid). Three of the most widely used line-ratio
diagnostic  diagrams, originally defined by \citet[][hereafter
BPT]{Baldwin81} and \citet{Veilleux87},  relate the \oiiihb\ ratio to
the \niiha, \siiha\ and \oiha\ ratios. These diagrams have proven
useful to  identify  the nature of the ionizing radiation in large
samples of galaxies in the local Universe  \citep[e.g.,][]{Kewley01,
  Kauffmann03}.  

Over the past decade, rest-frame optical spectra have become available
for increasingly large samples of more distant galaxies, at redshifts
$z \sim 0.5-3$, through near-infrared (NIR)  spectroscopy
\citep[e.g.,][]{Pettini04, Hainline09, Steidel14,  Shapley15}, in
particular with the NIR multi-object spectrographs MOSFIRE
\citep{McLean10} and FMOS \citep{Kimura10}. Interestingly, all these
observations indicate that star-forming (SF) galaxies at $z>1$ have
systematically larger \oiiihb\ ratio, at fixed \niiha\ ratio, than
their present-day counterparts from the Sloan Digital Sky Survey
\citep[SDSS; see, e.g.,][]{Shapley05, Lehnert09,
  Yabe12,  Steidel14,  Shapley15, Strom17}.  

The physical origin of this intriguing observational feature is being
heavily debated  and several explanations have been proposed:
high-redshift galaxies could have  typically higher ionization
parameters than local ones because of higher typical electron
densities, higher SFRs or a higher volume-filling factor of the
ionized gas,  originating from a higher average gas density/pressure
\citep[e.g.,][]{Brinchmann08,  Hainline09, Lehnert09, Steidel14,
  Hayashi15, Kashino17}, although a large electron  density by itself
does not seem to explain the offset in all cases
\citep[e.g.,][]{Rigby11, Hayashi15}. Other studies appeal to the
evolution of the gas-phase metallicity and an enhanced N/O abundance
ratio in high-redshift galaxies  \citep[e.g.,][]{Masters14, Shapley15,
  Masters16}. Instead, \citet[][see also
\citealt{Strom17}]{Steidel14, Steidel16} argue that the observed
offset originates primarily from a harder stellar ionizing-radiation
field in distant galaxies. Additional explanations include the
contribution by weak, unresolved AGN emission concurrent  with stellar
emission \citep[e.g.][]{Wright10} and observational selection effects
\citep{Juneau14}.  

For a large part, this diversity of explanations arises from the
intrinsic degeneracies affecting photoionization models when adopting
different prescriptions of ionizing radiation, combined with the
difficulty of distinguishing evolution from selection effects when
observing different samples of galaxies at different redshifts. In
this context, theoretical models of the nebular emission from galaxies
in a full cosmological framework could provide valuable insight into
the connection between observed emission  lines and the underlying ISM
and ionizing-source properties as a function of cosmic time. Yet,
fully self-consistent models of this kind are currently limited by the
performance of  cosmological {\it radiation}-hydrodynamic simulations
and insufficient spatial resolution on the scales of individual
ionized regions around stars and active nuclei. To circumvent these
limitations, some pioneer studies proposed the post-processing of
cosmological hydrodynamic  simulations and semi-analytic models with
photoionization models to compute the cosmic  evolution of nebular
emission \citep{Kewley13,  Orsi14,Shimizu16}. Only \citet{Kewley13}
investigate the evolution of emission-line ratios in cosmic time,
combining chemical enrichment  histories from cosmological
hydrodynamic simulations with photoionization models  of SF
galaxies. These authors explore the influence of ISM conditions on the
SF sequence  in the \oiiihb\ versus \niiha\ BPT diagram, as well as
the potential influence of an  AGN.\footnote{\citet{Kewley13} consider
  different metal enrichments and ISM conditions  in the narrow-line
  regions around AGN, but they do not rely on any simulation
  predictions for these  quantities.} \citet{Kewley13} find that the
SF sequence can be shifted to higher \oiiihb\ by  `extreme' ISM
conditions in high-redshift galaxies, such as large ionization
parameters,  high gas densities and/or hard ionizing radiation, but in
unknown relative proportions. {\em To reach more specific conclusions
  requires the self-consistent modelling of nebular emission from
  different gas components ionized by different sources in simulated
  galaxies.} 

We achieve this in the present study by modelling, for the first 
time in a largely self-consistent way, the nebular emission from
galaxies in a full cosmological context.  We account for the
integrated nebular emission from not only young stars \citep[as
in][]{Orsi14, Shimizu16}, but also AGN and post-AGB stars, based on
the star formation and chemical enrichment histories of the simulated
galaxies. Specifically, we post-process high-resolution, cosmological
zoom-in simulations  of massive galaxies with recent nebular-emission
models of galaxies and AGN.  The simulations include modern
prescriptions for star formation, chemical enrichment, stellar
feedback, black-hole (hereafter BH) growth and AGN feedback
(\citealp{Choi16, Nunez17}). The nebular-emission models of
star-forming galaxies  include improved prescriptions for the stellar
ionizing radiation and a self-consistent  treatment of metal depletion
onto dust grains (\citealt{Gutkin16}; but note that dust evolution 
is not followed explicitly in the simulations). We
extend here  these  models to include the nebular emission from
post-AGB stars. For the emission from AGN narrow-line regions, we
appeal to the models of \citet{Feltre16}. The integrated nebular 
emission of a model galaxy is then the sum of the star-forming, post-AGB 
and AGN components. This set-up provides an ideal basis to answer 
the questions we wish to address in  the present study: {\em can we 
account for the observed
  evolution of optical emission-line ratios, in particular the
  systematically larger \oiiihb\ ratio of high-redshift galaxies at
  fixed  \niiha\ ratio? If yes, what role do the different sources of
  ionizing radiation and ISM  properties play in the origin of this
  trend?}  

The paper is structured as follows. In Section~\ref{theory}, we
present the general theoretical framework of this study, including the
zoom-in simulations of massive galaxies,  the nebular-emission models
and the way in which we combine the former with the latter.
Sections~\ref{cosmicevolution} and \ref{origin} describe our main
results about the cosmic  evolution of galaxies in standard optical
line-ratio diagnostic diagrams and the potential physical origin of
this evolution. We discuss our findings in the context of previous
theoretical and  observational studies and address possible caveats of
our method in Section~\ref{discussion}.  Finally,
Section~\ref{summary} summarizes our results.

\section{Theoretical framework}\label{theory} 

\subsection{High-resolution simulations of massive haloes}\label{simulations} 

To achieve the analysis presented in this paper, we performed a set of
20 high-resolution, cosmological zoom-in simulations of massive
galaxies based on initial conditions from \citet{Oser10, Oser12}, who
computed the evolution in a full cosmological context of 39 galaxies
with present-day halo masses between $7\times10^{11}\Msun\,h^{-1}$ and
$2.7\times10^{13}\Msun\,h^{-1}$
($H_0=100\,h\,\mathrm{km\,s}^{-1}$). We performed  these simulations
with a modified version of the highly parallel, smoothed particle
hydrodynamics (SPH) code GADGET3 \citep{Springel05a},  SPHGal
\citep[][see also  \citealt{Choi16,Nunez17}]{Hu14}, as described in
the next paragraphs. We note that our simulations differ slightly 
from those presented in \citet{Choi16}, particularly
in the prescriptions for AGN  and stellar feedback. These changes hardly
affect the properties of simulated galaxies, and hence, they have a 
negligible impact on the results presented in this paper. 

\subsubsection{The hydrodynamic simulation code SPHGal}

To overcome traditional fluid-mixing problems encountered in classical
SPH  codes \citep{Agertz07}, our `modern' simulation code SPHGal
\citep{Hu14} includes a density-independent pressure-entropy SPH
formulation \citep{Ritchie01, Saitoh13, Hopkins13num}, a Wendland
C$^4$ kernel with 200 neighbouring particles \citep{Dehnen12}, an
improved artificial viscosity \citep{Cullen10} and an artificial thermal 
conductivity \citep{Read12}. Moreover, to guarantee a proper treatment 
of shock propagation and energy feedback, a limiter of the adaptive 
time-step scheme of SPH ensures that neighbouring particles have 
similar time steps \citep{Saitoh09,   Durier12}. For further details on 
these numerical schemes and their performance in test runs, we refer 
the reader to \citet{Hu14}. 

SPHGal also follows baryonic
processes, such as star formation, chemical enrichment, metal-line
cooling, stellar and AGN feedback and ultraviolet photo-ionization
background. Specifically, star formation and chemical evolution is
modelled as described in \citet{Aumer13}, assuming chemical
enrichment via type-Ia and type-II supernovae (SNe) and AGB stars, with
 chemical yields from \citet{Woosley95}, \citet{Iwamoto99}  and
\citet{Karakas10}, respectively. We trace 11 elements (H, He, 
C, N, O, Ne, Mg, Si, S, Ca and Fe) in both gas and
star particles. For gas particles, we include metal diffusion
to allow a more realistic mixing of metals released into the 
ambient (possibly more metal-poor) gas. The net cooling rates are 
calculated from the individual element abundances, gas temperatures 
and densities, accounting for a redshift-dependent 
ultraviolet background \citep{Haardt01}. 

Stars are assumed to form stochastically out of gas particles if the gas
density exceeds a threshold value $n_{\mathrm{th}} = n_0
(T_{\mathrm{gas}}/T_0)^3 (M_0/M_{\mathrm{gas}})^2$, where 
$T_{\mathrm{gas}}$ and $M_{\mathrm{gas}}$ are the temperature
and mass of the gas particle, and $n_0 = 2\,\mathrm{cm}^{-3}$ and 
$T_0 = 30000\,$K (see Section~\ref{setup}). Gas particles with densities
above $n_{\mathrm{th}}$ are Jeans unstable. Their star formation rate 
is calculated as $d\rho_*/dt = \eta\rho_{\mathrm{gas}}/t_{\mathrm{dyn}}$, 
where $\rho_*$, $\rho_{\mathrm{gas}}$ and $t_{\mathrm{dyn}}$ are the
stellar and gas densities and gas dynamical time-scale. The star formation 
efficiency, $\eta$, is set to a value of $0.025$ reproducing the observed 
Schmidt-Kennicutt relation.

Star formation is regulated by both stellar and AGN feedback. We adopt
the approach outlined in \citet{Nunez17} for early stellar and SN feedback. 
Early feedback from young, massive stars includes ultraviolet radiative 
heating (within a Stroemgren sphere) and mass, energy,  momentum and 
metal injection by stellar winds. SN feedback includes mass and metal release 
into the ambient gaseous medium, together with energy and momentum
input during the momentum-conserving free-expansion phase of
type-I and type-II SN blast waves ($v_{\mathrm{out,SN}} =
4500\,$km\,s$^{-1}$).\footnote{This is a simplified version of the
  full, 3-phase  blast-wave model adopted in \citet{Nunez17} and
  \citet{Choi16}.} Mass, metals,  momentum and energy from low- and
intermediate-mass stars are also transferred  to surrounding gas
particles in the form of slow winds ($v_{\mathrm{out,AGB}} =
10\,$km\,s$^{-1}$), mimicking an AGB phase with mass loss. Finally,
AGN Feedback is tied to the prescription for BH growth.  BHs are
represented by collisionless sink particles, a BH seed of $10^5\Msun$
being placed at the density minimum of any dark-matter halo whose mass
exceeds $10^{11}\Msun$.\footnote{These halo-threshold and  BH-seed
  masses were chosen to roughly reproduce the \citet{Magorrian98}
  relation and  follow theoretical calculations of BH formation by
  \citet{Stone16}.} BHs can further grow via two channels: gas
accretion and merger events with other BHs. Gas accretion is assumed
to follow a statistical Bondi-Hoyle approach  \citep{Bondi52}, whereby
a gas particle is accreted onto a BH with a probability given by the
volume fraction of the gas particle lying within the (unresolved)
Bondi radius of the BH \citep[e.g.,][]{Choi12}.  

To compute AGN feedback from this prescription, we 
do not make the widely used assumption of considering {\em only}
(spherical) thermal energy release into the ambient medium (as is the
case in, e.g., the  Illustris, Magneticum and EAGLE simulations; see
\citealt{Genel14, Hirschmann14,  Schaye15} and the recent reviews by
\citealt{Naab16, Somerville15}). Instead,  we rely on a more
physically motivated approach including both mechanical and radiative
feedback \citep{Ostriker10, Choi16}. Specifically, we incorporate 
the effect of  AGN-driven winds (motivated by observed 
broad-absorption-line winds) by randomly selecting gas particles in the
vicinity of the BH (with a probability given by  the  feedback
efficiency), which are given a velocity kick of 10,000~km\,s$^{-1}$
perpendicular to the gaseous disk. Kicked particles share momentum
with their two nearest neighbours, the residual energy being deposited
into the gas particles as thermal energy. This allows us to roughly
capture the Sedov-Taylor expansion phase of a blast wave (roughly 70
per  cent in thermal energy, 30 per cent in kinetic energy).  
Radiative feedback from Compton and photoionization heating due to 
X-ray radiation from the accreting BH, radiation pressure associated
with X-ray heating and the Eddington force are also included. 
Coupling between X-ray radiation and the surroundings follows 
detailed small-scale simulations by \citet{Sazonov05}. 
Accretion is not limited by the Eddington rate, but the Eddington
force acting on electrons is self-consistently included. We refer the
reader to \citet{Choi15, Choi16} for more details about AGN feedback
modelling. We note that our set of 20 zoom-in simulations do
not include any metallicity-dependent heating prescription. This is 
justified by the fact that, as shown by \citet{Choi16}, such
refinements are not found to have any significant impact on basic
properties of massive galaxies.

\citet{Choi16} show how the hydrodynamic simulation described 
above, and in particular the sophisticated prescription for AGN feedback,
can generate many realistic properties of massive galaxies, such as 
star formation histories, baryon conversion efficiencies, sizes, gas 
fractions and hot-gas X-ray luminosities. It is worth mentioning that 
these last two quantities are often over- or underestimated when 
adopting `traditional' prescriptions for AGN feedback.

\subsubsection{The simulation set-up}\label{setup}

The dark matter haloes chosen for zoom-in re-simulations were selected
from  a dark matter-only N-body simulation with a co-moving periodic
box length $L=100\ \mathrm{Mpc}$ and $512^3$ particles
\citep{Moster10}. The  cosmological parameters, based on WMAP3
measurements, are taken to be $\sigma_8 = 0.77$,
$\Omega_{\mathrm{m}}=0.26$, $\Omega_{\Lambda}=0.74$ and $h= 0.72$
\citep[see, e.g.,][]{Spergel03}.
The simulation was started at $z=43$ and run to $z=0$, with a
dark-matter  particle mass $M_{\mathrm{DM}} = 2 \times 10^8
\Msun\,h^{-1}$ and a fixed co-moving gravitational softening length of
$2.52\ h^{-1} \mathrm{kpc}$. We refer the reader to the original
papers of \citet{Oser10, Oser12} for more details about the simulation
setup.  

From this simulation, \citet{Oser10} selected 39 haloes with masses in
the range  $7\times10^{11}$--$2.7\times10^{13}\Msun\,h^{-1}$ at $z=0$
for re-simulation. To construct the initial conditions for the
high-resolution re-simulations, individual haloes are traced back in
time, and all particles closer to the halo centre than twice  the
radius where the mean density drops below 200 times the critical
density of the universe at any given snapshot are identified. These
dark matter particles  are replaced with dark matter as well as gas
particles at higher resolution ($\Omega_{\mathrm{b}}=0.044,
\Omega_{\mathrm{dm}}=0.216$). The new dark  matter particles have a
mass $m_{\mathrm{dm}} = 2.5\times 10^7 \Msun\,h^{-1}$, i.e., 8 times
smaller than the original ones, while the gas particles have a mass
$m_{\mathrm{gas}} = 4.2\times 10^6\Msun\,h^{-1}$, equal to that  of
star particles. The co-moving gravitational softening length of the
dark matter particles is $890\ h^{-1}\mathrm{pc}$, and that of the gas
and star  particles $400\ h^{-1} \mathrm{pc}$.  Here, we select for
re-simulation 20 of the  most massive haloes identified by
\citet{Oser10}, with $z=0$ virial masses between  $3\times 10^{12}
\Msun$ and $3 \times 10^{13} \Msun$, and associated central  galaxy
masses (computed as the stellar mass within a tenth of the virial
radius)  between $3 \times 10^{10} \Msun$ and $3 \times 10^{11} \Msun$
($h=0.72$). 

\begin{table*}
\centering
\begin{tabular}{ | p{3.5cm} || p{4cm} p{4cm} p{4cm} p{0.001cm}|} 
\centering{{\bf Parameter space}} & \centering{{\bf SF models}\\ (Gutkin et al. 2016)}
 & \centering{{\bf AGN models}\\
  (Feltre et al. 2016)}
  & \centering{{\bf PAGB models}\\ (this work)} & \\ \hline \hline

\centering{\bf Ionizing spectrum}\\ ({\it matched/fixed})  &
\centering{{\bf 10~Myr-old} stellar population with {\bf const SFR}
                                                                                 ({\it fixed}),\\ 
stellar metallicity same as that of gas ({\it matched})} 
& \centering{UV slope $\alpha$ =
                                       $-1.2$, $-1.4$, $\boldsymbol{-1.7}$,  $-2.0$\\
  ({\it fixed})} & \centering{
                                                           3, 5, 7, 9 Gyr-old stellar
                                                     populations\\
  ({\it matched})\\
$Z_{\diamond, \mathrm{stars}}$ = 0.008, 0.014, 0.017, 0.02 ({\it matched})} &
  \\ \hline 

\centering{\bf Interstellar metallicity} \\ $Z$ ({\it matched})  &  
\centering{$Z_{\star} =$\\ 0.0001, 0.0002, 0.0005, 0.001, 0.002, 0.004,
0.006, 0.008, 0.014, 0.017, 0.02, 0.03}
& 
\centering{$Z_{\bullet} = $\\ 0.0001, 0.0002, 0.0005, 0.001, 0.002, 0.004,
0.006, 0.008, 0.014, 0.017, 0.02, 0.03, 0.04, 0.05, 0.06, 0.07}   
&  \centering{$Z_{\diamond} = $\\ 0.0001, 0.0002, 0.0005, 0.001, 0.002, 0.004,
0.006, 0.008, 0.014, 0.017, 0.02, 0.03, 0.04, 0.05, 0.06, 0.07} &
  \\ \hline

\centering{\bf Ionization parameter}  $\log U$, function of the average gas
  density\\ ({\it matched})  &  
\centering{\bf $\log U_{\star} = -0.65, -1.15, -1.65, -2.15, $\\
                                                                              $-2.65,
  -3.15,  -3.65$}
& \centering{\bf $\log U_{\bullet} = -0.65,
                              -1.15, -1.65, -2.15, $\\ $-2.65, -3.15, -3.65,
  -4.65$} 
  &  \centering{\bf $\log U_{\diamond} = -2.15, -2.65, -3.15, -3.65,
    $\\ $-4.15, -4.65$} &
  \\ \hline
    
\centering{\bf Dust/metal mass ratio}\\ $\xi_d$ ({\it fixed})  &
                                                              \centering{0.1,
                                                              {\bf 0.3}, 0.5} &
 \centering{0.1, {\bf 0.3}, 0.5} &  \centering{0.1, {\bf 0.3}, 0.5} &  \\ \hline 

\centering{\bf Ionized-gas density
  $\log (n_{\mathrm{H}}/\mathrm{cm}^3)$}\\ ({\it fixed})  &
 \centering{$\log (n_{\mathrm{H}, \star})$={\bf 2.0}, 3.0, 4.0} &
                                                            \centering{
                                                            $\log (n_{\mathrm{H},\bullet})$= 2.0,
                                                              {\bf 3.0}, 4.0} &
                                                                \centering{$\log
                                                                (n_{\mathrm{H},\diamond})$=
                                                                {\bf 1.0}, 2.0, 3.0} &
  \\ \hline 

\centering{\bf C/O abundance ratio} in solar units \\ ({\it matched})  &  
\centering{(C/O)$_{\star}$/(C/O)$_\odot =0.1, 0.2, 0.27, 0.38,0.52,0.72,1.0$}& 
\centering{(C/O)$_{\bullet}$/(C/O)$_\odot =0.1, 0.2, 0.27, 0.38,0.52,0.72,1.0$} 
  &  \centering{(C/O)$_{\bf \diamond}$/(C/O)$_\odot = 1.0$} &
  \\ \hline

\centering{\bf Model normalization} \\ ({\it matched})  &  
\centering{Star formation rate}\\{SFR}&
\centering{AGN luminosity} \\{$L_{\mathrm{AGN}}$} 
  &  \centering{Mass of evolved stars} \\
    {$M_{\diamond, \mathrm{stellar}}$} &  \\ \hline

\end{tabular}
\caption{Overview of the parameter space of the nebular-emission
 models for young stars, AGN and post-AGB stars. To select the SF, AGN and
 PAGB models appropriate for each galaxy at each simulation time step, we 
adopt a fixed spectral slope of AGN ionizing radiation ($\alpha=-1.7$), fixed 
dust-to-metal mass ratio ($\xi_\mathrm{d}=0.3$), and fixed ionized-gas density
($n_{\mathrm{H}, \star} =10^2\,$cm$^{-3}$, $n_{\mathrm{H},\bullet} = 10^3\,$cm$^{-3}$
and $n_{\mathrm{H}, \diamond} = 10\,$cm$^{-3}$). We further match the AGN, SF and
 PAGB ionization parameters, interstellar (i.e. gas+dust-phase) metallicity, C/O 
 abundance ratio and age and metallicity of post-AGB stars to those of the simulated
galaxy. The emission-line luminosities are scaled to the SFR, AGN luminosity 
and mass of post-AGB stellar population of the galaxy.}
\label{Table_1}
\end{table*}

\subsubsection{Mass assembly histories}

To investigate the redshift evolution of different galaxy properties 
(including emission-line ratios), we construct stellar merger trees
for  the sample of 20 model galaxies described in the previous
section. As in  \citet{Oser12}, we start by using a friends-of-friends
algorithm to identify, at  any simulation snapshot, a central galaxy
-- the host (i.e., the most massive  galaxy sitting at the minimum of
the halo potential well) -- and its surrounding  satellite (less
massive) galaxies. We require a minimum of 20 stellar particles
(i.e., a minimum mass of about $1.2 \times 10^8 \Msun$) to identify a
galaxy.  At $z = 2$, all galaxies in our sample are more massive than
about $10^{10}\Msun$, implying that, at $z<2$, we resolve mergers down
to a mass ratio of at least $1:100$. In the analysis presented in the
remainder of this paper, we trace back at every time step only the
most massive progenitor of a present-day galaxy, i.e., we focus on
central galaxies.

\subsection{Modeling of nebular emission}\label{nelms} 

We post-process the re-simulations of 20 galaxies presented in
Section~\ref{simulations} to include nebular emission. To achieve
this, we adopt the recent prescriptions of \citet{Gutkin16} and
\citet{Feltre16} to compute the nebular emission arising from young
massive stars  (Section~\ref{emlines_sf}) and narrow-line regions of
AGN (Section~\ref{emlines_agn}). We also design a prescription to
account for the contribution by post-AGB stars to nebular  emission
(Section~\ref{emlines_pagb}). All emission-line models  presented in
this paper  were computed using version c13.03 of the photoionization
code \textsc{Cloudy} (\citealp{Ferland13}), always accounting
for  the interaction of photons, electrons and atomic ions with dust
  grains in HII regions. 

\subsubsection{Nebular emission from star-forming galaxies}\label{emlines_sf}

We adopt the grid of nebular-emission models of
star-forming galaxies computed by \citet{Gutkin16}. These calculations
combine the latest version of the \citet{Bruzual03} stellar
population synthesis model (Charlot \& Bruzual, in preparation) with
\textsc{Cloudy}, following the method outlined by
\citet{Charlot01}. In brief, the approach consists in convolving the spectral
evolution of a typical, ionization-bounded \hii\ region powered by a
new-born star cluster with a star formation history, to compute the nebular
emission of a whole galaxy. In this context, the parameters of the
photoionization model  of this typical \hii\ region should be interpreted as effective
(i.e. galaxy-wide) ones, describing the ensemble of \hii\ regions and
the diffuse gas ionized by stars throughout the galaxy (see 
section 2.3 of \citealt{Charlot01} and section 2.2 of
\citealt{Gutkin16} for details). A `closed geometry' is used in
\textsc{Cloudy} to perform these calculations, as appropriate for
spherical \hii\ regions.  

The grid computed by \citet[][see their table~3]{Gutkin16} encompasses
models in  wide ranges of interstellar (i.e. gas+dust-phase)
metallicities, $Z_{\star}$,  ionization parameters, $U_{\star}$,
dust-to-metal mass ratios, $\xi_d$, \hii-region densities,
$n_{\mathrm{H}, \star}$ and carbon-to-oxygen abundance ratios,
(C/O)$_{\star}$. A main feature of these models (which incorporate
secondary nitrogen production) is the  self-consistent treatment of
metal depletion onto dust grains, which allows one to relate the
gas-phase metallicity measured from nebular emission to the total 
interstellar metallicity $Z_{\star}$. 
The ionization parameter is defined as the dimensionless ratio of 
the number density  of H-ionizing photons to that of
hydrogen. \citet{Gutkin16} adopt the same metallicity for the ionizing
stars as for the ISM. This is consistent with our simulations, in
which the metallicity of newly born stars is (by construction) very
similar to that of the warm/cold gas. The parameters of this model
grid are summarized in Table~\ref{Table_1}.\footnote{ The values of
the ionization parameter reported in Table~\ref{Table_1} are a factor
of 9/4 larger than those in table~3 of \citet{Gutkin16}. This is because 
we label models here in terms of the volume-averaged ionization parameter
(see equation~\ref{logU} of Section~\ref{sfmatch}).} It is worth noting
that, as pointed out by \citet[][see also \citealt{Gutkin16}]{Charlot01}, 
the actual mass of the effective star cluster has no influence on the 
predicted nebular emission at fixed $U_{\star}$ and $n_{\mathrm{H}, \star}$,
due to a degeneracy between this mass and the gas filling factor 
entering the definition of the ionization parameter (see Section~\ref{sfmatch} 
below). We adopt here the
emission-line predictions of \citet{Gutkin16} for 10\,Myr-old stellar
populations with constant SFR (sufficient to reach a steady population 
of \hii\ regions) and a standard \citet{Chabrier03} IMF
(consistent with the IMF adopted in the simulations), truncated at 0.1
and 100 $\Msun$. As we shall see in Section~\ref{discussion},
increasing the upper mass cutoff of the IMF to $300 \Msun$ hardly
affects our results.     

\subsubsection{Nebular emission from AGN}\label{emlines_agn}

For the narrow-line regions of AGN, we adopt the grid of
nebular-emission  models of \citet{Feltre16}. In this prescription,
the spectrum of  an AGN is approximated by a broken power law 
of adjustable index $\alpha$ in the frequency range of ionizing
photons \citep[equation 5 of][]{Feltre16}. In the narrow-line region,
gas is assumed  to be distributed in clouds of a single type. These
models include dust and radiation pressure. An `open geometry' is used
in  \textsc{Cloudy}, as appropriate for gas with a small covering 
factor. 

The grid of AGN nebular-emission models is parametrized
in terms of the interstellar metallicity in the narrow-line region,
$Z_{\bullet}$, the ionization parameter of this gas, $U_{\bullet}$,
the dust-to-metal  mass ratio, $\xi_d$, the density of gas clouds,
$n_{\mathrm{H}, \bullet}$,  and the carbon-to-oxygen  abundance ratio,
(C/O)$_\bullet$ (see  Table~\ref{Table_1} for details). We note that
the introduction of non-solar (C/O)$_\bullet$ ratios is an improvement
over \citet{Feltre16}, who considered only models with the solar
value. This refinement hardly affects the predictions for optical
emission-line  ratios presented in this paper, but it has a major
influence on the predicted ultraviolet  emission-line ratios
(Hirschmann et al. in prep.).

\subsubsection{Nebular emission from post-AGB stars}\label{emlines_pagb}

To describe the nebular emission from quiescent, passively evolving
galaxies, we must also account for the emission from diffuse gas
heated by evolved,  post-AGB stars. We
build a grid of such `PAGB' models using spectra of single-age,
evolved stellar populations (computed with the same  version  of the
\citealt{Bruzual03} stellar population synthesis code as used by
\citealt{Gutkin16}) as input to the photoionization code {\scshape
  CLOUDY}.  We compute models for a range of stellar populations ages
(between 3 and 9 Gyr) and metallicities, $Z_{\diamond, {\rm stars}}$ (see
Table~\ref{Table_1}). In contrast to the SF models
(Section~\ref{emlines_sf}), we do not impose the  interstellar
metallicity to be the same as that of the ionizing stars in the PAGB
models.  Instead, we treat it as an independent parameter,
Z$_{\diamond}$. This is motivated by the fact that old stars do not
necessarily have the same metallicity as  the diffuse ISM in which
they evolve (we are interested here in the emission from diffuse 
gas ionized by all hot post-AGB stars, including  beyond the 
short-lived planetary-nebula phase). In
fact, our simulations show that, particularly at redshifts  $z<2$, old
stellar populations can be significantly more metal-rich than
cold/warm gas,  whose metallicity is often diluted by late infall of
metal-poor gas. The other model  parameters controlling the nebular
emission from post-AGB stars, summarised in  Table~\ref{Table_1}, are
consistent with those of the SF models: the gas ionization  parameter,
$U_{\diamond}$, dust-to-metal mass ratio, $\xi_{\rm d}$, and hydrogen
density, $n_{\mathrm{H}, \diamond}$.  

\subsection{Coupling nebular-emission models with zoom-in simulations}\label{model} 

We couple the extensive grid of nebular-emission models
described in Section~\ref{nelms}, with the simulations of massive
galaxies described in Section~\ref{simulations}, by selecting an SF, an AGN 
and a PAGB emission-line model for each simulated galaxy at each redshift
step.\footnote{Note that in this study, we select only 
one  emission-line model per galaxy since we are focusing on integrated
  spectral properties. Future studies will, instead, focus on
  spatially resolved emission properties requiring a more refined
  coupling procedure.} The sum of these three components makes
  up the integrated nebular emission of a model galaxy. 
In practice, we select the SF/AGN/PAGB models 
appropriate for each galaxy by self-consistently matching all model 
parameters possibly available from the simulations (e.g., metallicity
of the star-forming gas).  The quantities derived from the
  simulated galaxies are calculated within 1/10 of the halo virial
  radius for the SF and PAGB models, and within 1~kpc of the black hole
  for the AGN models. Those parameters that cannot be retrieved from the
simulation  are set to standard values (indicated in bold characters
in Table~\ref{Table_1}). This is the case for, e.g., the dust-to-metal
mass ratio, $\xi_{\rm d}$, and the hydrogen gas density in individual
ionized regions,  $n_{\mathrm{H}}$, since we do not
model dust physics in the simulation and also cannot resolve
individual \hii\ regions, AGN narrow-line regions and PAGB-star
environments. We adopt $\xi_{\rm d}=0.3$ in all SF, AGN and PAGB
models of nebular emission in this paper. This value, close to that of
$\xi_{\rm d,\odot}=0.36$ in the Solar  neighbourhood \citep{Gutkin16},
lies in the middle of the parameter  range. Moreover, we adopt
$n_{\mathrm{H}, \star} =10^2\,$cm$^{-3}$ and $n_{\mathrm{H},\bullet} =
10^3\,$cm$^{-3}$, as typical gas densities estimated from  
optical line-doublet analyses of \hii\ regions and AGN \citep[see,
e.g., sections 5.6 and 13.4 of][]{Osterbrock06}.\footnote{Densities
  measured from emission-line doublets are those of the emitting gas
  clumps.} For the PAGB models, we adopt $n_{\mathrm{H}, \diamond}  =
10\,$cm$^{-3}$, which should be more appropriate for the diffuse
environment of old, post-AGB stars in the ISM of mature galaxies
 (adopting $n_{\mathrm{H}, \diamond}  = 1\,$cm$^{-3}$ would not
change significantly our results). The
potential impact of adopting different values for these fixed
parameters is discussed in Section~\ref{discussion}.   

In the next paragraphs, we describe in more detail the way in which 
we couple the SF, AGN and PAGB nebular models with our zoom-in galaxy
simulations.    

\subsubsection{Matching SF models to simulated galaxies}\label{sfmatch}

With each galaxy at each simulation time step, we associate the SF
emission-line  model from the \citet{Gutkin16} grid described in
Section~\ref{emlines_sf} with closest star and gas parameters. 
 As mentioned earlier, the parameters of the \citet{Gutkin16} models
are effective (i.e. galaxy-wide) ones, describing the ensemble of \hii\ 
regions and the diffuse gas ionized by young stars throughout the galaxy. 
We select the grid metallicity $Z_{\star}$ and carbon-to-oxygen ratio
(C/O)$_{\star}$ closest to the simulated global (i.e. galaxy-wide)
metallicity $Z_{\mathrm{gas, glob}}$ and abundance ratio 
(C/O)$_{\mathrm{gas, glob}}$  of the warm-gas phase. Here, we consider gas
particles with temperatures  between 10,000\,K, typical of giant \hii\ 
regions and narrow-line regions of AGN, and 100\,K, the
lowest  temperature achievable via atomic gas cooling in the simulations (the
remainder of the gas in the simulations is in the diffuse, hot ionized
phase). We compute the volume-averaged ionization parameter of the
simulated galaxy as \citep[see, e.g., equation~B.6
of][]{Panuzzo03}\footnote{ The definition of the volume-averaged
  ionization parameter in equation~\eqref{logU} 
differs from that of the quantity $\langle U\rangle$ in equation~(7)
of \citet{Charlot01} 
by a factor of 3/4, and from that of $U_{\mathrm S}$ in equation~(7) of 
\citet{Gutkin16} by a factor of 9/4. These different model-labelling choices
have no influence on the actual CLOUDY calculations, for
which the input parameter is the rate of ionizing photons
\citep[equation~8 of][]{Charlot01}.}
\begin{equation}\label{logU}
U_{\mathrm{sim}, \star} = \frac{3\alpha_B^{2/3}}{4c} \left(
    \frac{3Q_{\mathrm{sim},\star}\epsilon^2n_{\mathrm{H},\star}}{4 \pi} \right)^{1/3}.
\end{equation}
Here $Q_{\mathrm{sim},\star}$ is the rate of ionizing photons
(obtained by multiplying the  SFR of the simulated galaxy by the rate
output by a 10\,Myr-old stellar population with unit SFR and
metallicity $Z_{\star}$), $n_{\mathrm{H}, \star} =10^2\,$cm$^{-3}$
(see above), $\alpha_B$ is the case-B hydrogen recombination
coefficient and $\epsilon$ is the volume-filling factor of the gas,
defined by 
\begin{equation}\label{fillfac}
\epsilon = \frac{n_{\mathrm{gas, glob}}}{n_{\mathrm{H},\star}},
\end{equation}
where  $n_{\mathrm{gas, glob}}$  is the volume-averaged,
  global  (hydrogen) gas density, again considering gas particles
  with temperatures between 10,000\,K and 100\,K.  In the rare cases where
the volume-averaged gas density exceeds the adopted hydrogen density in
the ionized regions, i.e  $\epsilon > 1$, we set the filling factor to unity, 
i.e. $\epsilon = 1$. Note that a larger  $n_{\mathrm{gas, glob}}$ at fixed
$n_{\mathrm{H},\star}$ implies a larger filling factor $\epsilon$,
i.e., a more  compact arrangement of individual gas clumps around the
ionizing source, and hence, a larger incident flux of ionizing photons
per unit gas area. In our approach, therefore, $U_{\mathrm{sim}, \star}$ 
depends on the simulated SFR via $Q_{\mathrm{sim},\star}$. This is justified  
by the observed relation between ionization parameter and metallicity 
\citep[fig. 2 of][]{Carton17}, together with that between metallicity and specific 
SFR \citep[fig. 7 of][]{Mannucci10}, for SDSS star-forming galaxies, 
which imply a positive correlation between $\log U_{\star}$ and SFR at fixed 
galaxy stellar mass. Combined with the filling factor derived from the simulation 
(equation~\ref{fillfac}), this uniquely defines the quantity $Q_{\mathrm{sim},\star}\epsilon^2$ 
entering the definition of $U_{\mathrm{sim}, \star}$ at fixed $n_{\mathrm{H},\star}$ 
(equation~\ref{logU}). We select the SF model with ionization parameter
$\log U_{\star}$ closest to $\log U_{\mathrm{sim}, \star}$ (computed by
\citealt{Gutkin16} for an arbitrary combination of effective star-cluster mass and gas 
filling factor yielding the same $Q_{\star}\epsilon^2$; see Section~\ref{emlines_sf}). 
This uniquely defines the \citet{Gutkin16} model associated to each
simulated galaxy at each time step. 
\bigskip

\subsubsection{Matching AGN models to simulated galaxies}\label{agnmatch}

We adopt a procedure similar to that outlined in the previous section
to associate nuclear activity of a galaxy at any simulation time step
with an AGN emission-line model from the \citet{Feltre16} grid
described in Section~\ref{emlines_agn}. The ISM  conditions for the
AGN model are taken to be the {\it central} (and not global) ones  of
the simulated galaxy, i.e., in the vicinity of the black
hole. Specifically, we select gas particles  in a co-moving sphere of
1-kpc radius around the black hole to compute the central warm-gas
metallicity, $Z_{\mathrm{gas, 1kpc}}$, central carbon-to-oxygen ratio,
(C/O)$_{\mathrm{gas, 1kpc}}$, and central volume-averaged gas density,
$n_{\mathrm{gas, 1kpc}}$ (we have checked that  adopting radii in
the range 0.4--3\,kpc instead of 1\,kpc hardly affects our
results). This should be roughly appropriate to probe the
narrow-line regions around AGN with luminosities in the range found
in our simulations (see fig.~3 of \citealt{Hainline14} and
the model AGN luminosities in Fig.~\ref{Propevol} below). 
As before (equation~\ref{fillfac}), a larger $n_{\mathrm{gas,
    1kpc}}$ at fixed $n_{\mathrm{H},  \bullet}$ implies a larger
volume-filling factor of the narrow-line region. We compute the AGN
luminosity from the simulated black-hole accretion rate (BHAR) as in
\citet{Hirschmann14}. For  the AGN ionizing spectrum, we adopt a
  fixed ultraviolet slope $\alpha =-1.7$ of the flux per unit
  frequency, ($\propto \nu^\alpha$; indicated in bold face in
  Table~\ref{Table_1}) and an amplitude scaled to the simulated
  AGN luminosity. This allows us to compute the  rate of ionizing 
photons produced by the AGN, noted $Q_{\mathrm{sim},\bullet}$,  and
the corresponding ionization parameter, noted
$U_{\mathrm{sim},\bullet}$, via equation~\eqref{logU}.   Then, we
select the \citet{Feltre16} model with closest $Z_{\bullet}$, $\log
U_{\bullet}$ and (C/O)$_\bullet$.

\subsubsection{Matching PAGB models to simulated galaxies}\label{pagbmatch}

To select a PAGB emission-line model from the grid presented in
Section~\ref{emlines_pagb} for each galaxy at each simulation time
step, we start by computing the average age and metallicity of all
star particles older than 3\,Gyr (when a significant  population of
post-AGB stars starts to build up in a simple stellar population). We
identify the available grid age and metallicity ($Z_{\diamond,
  \mathrm{stars}}$) closest to these  values in Table~\ref{Table_1}
and compute the rate of ionizing photons from PAGB stars, noted
$Q_{\mathrm{sim},\diamond}$, based on the mass in stars older than
3\,Gyr in the  simulation. Since our simulation does not allow us to
distinguish gas in  star-forming clouds from that in the diffuse ISM
(which would require to resolve scales of a few tens of parsecs), for
the PAGB model we adopt the same global interstellar  metallicity
$Z_{\mathrm{gas,glob}}$, abundance ratio (C/O)$_{\mathrm{gas, glob}}$
and  volume-averaged gas density  $n_{\mathrm{gas, glob}}$ as for
the SF model above.  Assuming $n_{\mathrm{H}, \diamond} =
10\,$cm$^{-3}$ accounts for the fact that the gas seen by old stars is
more diffuse (equation~\ref{fillfac}) than that seen by young stars
($n_{\mathrm{H}, \star} =10^2\,$cm$^{-3}$). Then, using
equation~\eqref{logU}, we compute the ionization parameter of the gas
ionized by post-AGB stars, noted  $U_{\mathrm{sim},\diamond}$, and
select the model with closest  $Z_{\diamond}$, $\log U_{\diamond}$ and
(C/O)$_\diamond$ in Table~\ref{Table_1}. 

\subsubsection{Total emission-line luminosities and line ratios}

The procedure described in the previous paragraphs allows us to
compute the contributions of young stars, an AGN and post-AGB stars to
the luminosities of various emission lines (such as
$L_{\mathrm{H}\alpha}$, $L_{\mathrm{H}\beta}$,  $L_{\mathrm{OIII}}$,
etc.) in a simulated galaxy. The {\it total} emission-line
luminosities of the galaxy can then be calculated by summing over
these three contributions. For line luminosity ratios, we adopt for
simplicity the notation
$L_{\mathrm{OIII}}/L_{\mathrm{H}\beta}=\oiiihb$. In this study, we
focus on exploring four line ratios in the optical regime: 
[O\,{\sc iii}]$\lambda5007$/H$\beta$, [N\,{\sc
  ii}]$\lambda6584$/H$\alpha$,  [S\,{\sc
  ii}]$\lambda\lambda6717,6731$/H$\alpha$ and [O\,{\sc
  i}]$\lambda6300$/H$\alpha$. The strengths of metal lines reflect  
a combination of photo-ionization processes of the corresponding
elements  and excitation processes of the resulting ions via
collisions with photo-electrons.  We note that, in this paper, we 
do not consider attenuation by dust outside \hii\ regions
and compare our predictions with observed emission-line ratios
corrected for this effect (if not provided by the original studies, we 
applied a correction based on the \ha/\hb\ ratio and the \citealt{Calzetti00} 
attenuation curve). By design, 
the above optical-line ratios are anyway little sensitive to dust, as
they are defined by lines close in wavelength (for reference, the 
corrections are less than $\sim0.015$~dex for a $V$-band attenuation 
of $A_V\sim1$\,mag).     
 
\section{Cosmic evolution of optical emission-line ratios}\label{cosmicevolution} 

Knowing the most massive (i.e. main) progenitor of every (central)
galaxy at any simulation time step allows us to explore the cosmic
evolution of emission-line ratios. We start by investigating the
dependence of optical emission-line ratios on the evolution of stars,
black holes and the ISM in three  galaxies illustrating different
physical properties (Section~\ref{casestudies}).  Then, we explore the
predicted evolution of optical emission-line ratios for  the full set
of 20 simulated massive galaxies described in
Section~\ref{simulations}  and compare this with observations at
various redshifts (Section~\ref{popstudy}). Our sample of zoom-in
simulations of massive, mostly quiescent, present-day  galaxies is
well-suited for this analysis, since it allows us to probe a mass
range of star-forming galaxies at high redshift (corresponding to the
main progenitors) similar to that sampled by available observations
(Section~\ref{massbias}). A natural consequence of following the
evolution of the main progenitors of present-day  massive galaxies is
that the average galaxy mass increases from high to low redshift in
our  simulated sample. To understand how this feature can affect our
results, we also investigate  emission-line evolutionary trends for
galaxies in a fixed stellar mass range at all redshift. 

\subsection{Three case studies}\label{casestudies} 

\begin{figure*}
\centering
\vspace{-0.5cm}
\epsfig{file=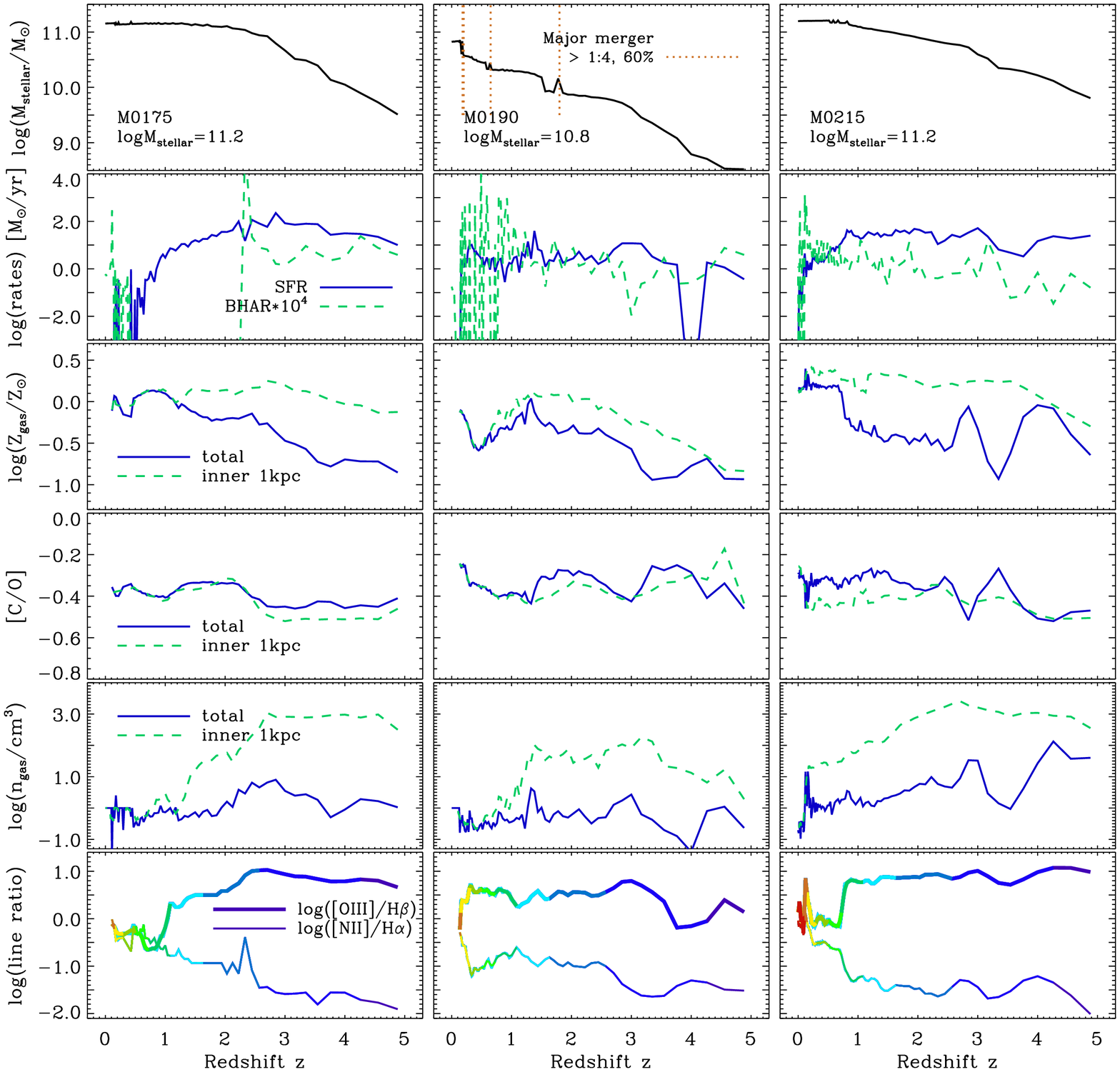,
 width=0.95\textwidth}\vspace{-0.3cm}
\epsfig{file=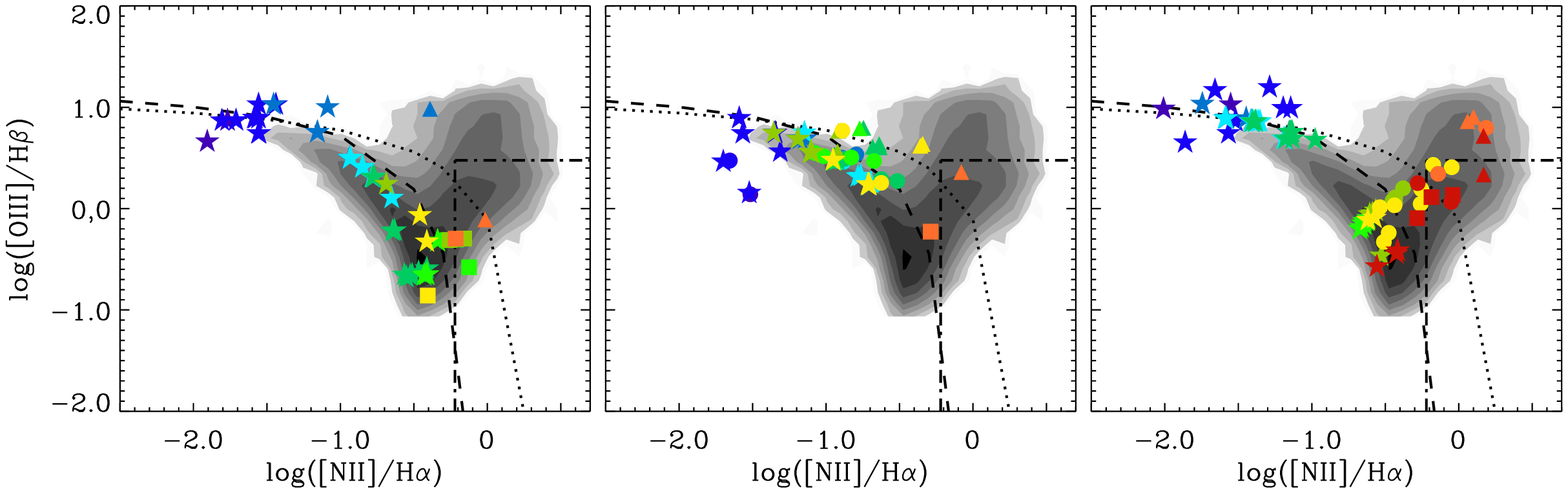,
 width=0.95\textwidth}\vspace{-0.4cm}
\caption{Redshift evolution of different properties of three re-simulated 
galaxies with (middle column) and without  (left and right
  columns) major-merger events: stellar mass (first row);
  BH accretion rate (green dashed lines, second row); SFR 
  (blue solid lines, second row);  global (blue solid) and
  central (green dashed) interstellar metallicity (third row);
  global (blue solid) and central (green dashed) C/O ratio
  (fourth row);  and global and central gas density
  (fifth row). The sixth and seventh rows show the implied redshift 
  evolution of the integrated \oiiihb\ and \niiha\ ratios and
  the corresponding \oiiihb\ versus \niiha\ diagram  (where
    symbols are colour-coded according to redshift as along the 
    x-axis of the sixth
    row). Different symbols refer to  different galaxy types (stars:
  SF; circles: composite; triangles: 
  AGN-dominated; and squares: post-AGB-dominated galaxies; see 
  Section~\ref{casestudies} for details). Also shown for reference are
  observations of local SDSS galaxies (grey shaded areas and contours), 
  together with standard observational criteria to distinguish SF galaxies 
  (below the dashed line) from composites (between the dashed and dotted lines), 
  AGN (above the dotted line) and LI(N)ER (in the bottom-right quadrant defined 
  by dot-dashed lines), according to \citet[][dotted line]{Kewley01} and 
  \citet[][dashed and dot-dashed lines]{Kauffmann03}.}\label{Evolution}    
\end{figure*}

Fig.~\ref{Evolution} illustrates the evolution of various quantities
pertaining  to stars, black holes and the ISM, for three galaxies of
our simulation set, M0175, M0190 and M0215 (from left to right). These
quantities are: mass assembly history (first row); SFR and BH
accretion rate (second row); global and central warm-gas metallicities
(third row); global and central C/O ratios (fourth row); and global
and central volume-averaged gas densities (fifth row). The sixth row
shows, as an example, the redshift evolution of two emission-line
ratios, \oiiihb\ (thick line) and \niiha\ (thin line). In the last
row, we show the evolution of the same models in the \oiiihb\  versus
\niiha\ diagram (colour-coded by redshift as in the sixth
row). Different symbols  refer to three ranges of BAHR-to-SFR ratio,
which we associate to: SF galaxies, with $\mathrm{BHAR/SFR <10^{-4}}$
(stars); composite galaxies, with $10^{-4}\lid\mathrm{BHAR/SFR}\lid
10^{-2}$ (circles); and active galaxies, with $\mathrm{BHAR/SFR} >
10^{-2}$ (triangles). In addition, galaxies in which H$\beta$ emission
from post-AGB stars exceeds that from both AGN and young stars are
indicated by squares. Such galaxies are thought to constitute a
sub-group of the classical population of low-ionization (nuclear)
emission-line regions \citep[LINER/LIER; e.g.][]{Singh13,
  Belfiore16}. For reference, we also indicate the location of local
($z\sim0.1$) SDSS galaxies in these line-ratio diagrams (grey shaded
areas and contours), together with standard observational criteria to
distinguish SF galaxies (below the dashed line) from composites
(between the dashed and dotted lines), AGN (above the dotted line) and
LI(N)ER (in the bottom-right quadrant defined by dot-dashed lines),
according  to \citet[][dotted line]{Kewley01} and \citet[][dashed and
dot-dashed lines]{Kauffmann03}.  We now describe in more detail our
findings for this illustrative sample of three galaxies.  

\subsubsection{Galaxy M0175}\label{galm0175}

Galaxy M0175 (left-most column in Fig.~\ref{Evolution}) has a fairly  
quiet mass assembly history dominated by `in situ' star formation
without any major merger (top panel). Yet, around $z=2$, an infall of
gas clouds drives a peak in BH accretion (second panel). AGN feedback
from this event stops further BH accretion and also makes star
formation drop on a longer time scale. Only at low redshift does new
gas infall (not merger-driven) trigger a second peak in BH
accretion. These peaks in BH accretion induce peaks in the otherwise
gradually rising \niiha\ ratio. This is because the harder ionizing
radiation of an AGN relative to stars makes the electronic temperature
higher, and hence, the collisionally-excited [N\,{\sc ii}] line
stronger  \citep[see figure~1 of ][]{Feltre16}. During these events,
the galaxy makes excursions in the regions of the \oiiihb\ versus
\niiha\ diagram populated by  AGN and composites (blue and orange
triangles in the bottom panel). We note that, unlike the \niiha\
ratio, the \oiiihb\ ratio starts to decline below $z=3$, and drops
further by almost an order of magnitude around $z \sim 1$. This
behaviour results from  the combination of two factors: the drop in
star formation rate at additionally slightly  decreasing global
average gas density, which makes the ionization parameter also  drop
(equations~\ref{logU}--\ref{fillfac}); and the rise in metallicity at
nearly constant C/O ratio. This rise makes cooling more efficient in
\hii\ regions, reducing the electronic temperature and in turn the
luminosity of [O{\sc iii}]$\lambda5007$, whose  excitation potential
is higher than that of [N{\sc ii}]$\lambda6584$
\citep{Stasinska80,Gutkin16}.  We note that the rise in \niiha\ toward
low redshift is boosted by the inclusion of secondary nitrogen
production in the nebular models (Section~\ref{emlines_sf}). The drop
in average central gas density toward low redshift further contributes
to the  above trends in \oiiihb\ and \niiha\ by making the AGN
ionizing parameter smaller.  

As a result of these trends, galaxy M1075 `moves' from the top-left to
the bottom part  of the SF branch in the \oiiihb\ versus \niiha\
diagram in Fig.~\ref{Evolution} (area below  the dashed line, along
the observed sequence of SDSS star-forming galaxies). At later  times,
when the radiation from young stars and the AGN is very weak, post-AGB
stars become the dominant source of ionizing photons, causing the
galaxy to appear as a LI(N)ER (yellow and green squares). At $z=0$,
M1075 disappears from  line-ratio diagram because no warm gas is left.

\subsubsection{Galaxy M0190}\label{galm0190}

The mass assembly history of galaxy M0190  (top panel, middle column
in Fig.~\ref{Evolution}) is very different from that of M0175. Galaxy
M0190 experiences three major mergers with mass ratio above 1:4 -- one
at $z =1.8$, one at $z = 0.7$ and one at $z = 0.2$ (indicated by orange
dashed lines), through which roughly 60 per cent of the final galaxy
stellar mass is accreted. Interestingly, the BH accretion history
(second panel) reveals peaks of AGN activity associated with the two
mergers at $z<1$, but not the high-redshift one. This is presumably
because a turbulent environment with torques and radial gas flows at
high redshift maintains the BHAR at a high fraction of the Eddington
rate. At $z<1$, instead, spikes of AGN activity (resulting in AGN
luminosities of $10^{45}$--$10^{46}$\,erg\,s$^{-1}$) can be triggered
even by small events during a `smooth'  accretion phase. The impact of
AGN feedback on star formation for M0190 is also moderated by the
repeated inflow of gas, which maintains the SFR above
$1\,\Msun\,\mathrm{yr}^{-1}$ over most of the galaxy's history. Only
after the last major merger at $z = 0.2$ do both star formation and
black-hole accretion shut down, as a consequence of feedback-driven
gas heating and outflows.  High levels of (pristine) gas accretion
also affect the central and global interstellar  metallicities, as can
be seen from the dip in metallicity evolution between $z\ga1$ and
$z=0.2$  (third panel). 

The above histories of star formation, BH accretion and chemical
enrichment have  consequences for emission-line properties: despite
several pronounced nuclear-activity peaks at $0.2<z<1$, M0190 remains
in the star-forming  and composite regions of the \oiiihb\ versus
\niiha\ diagram without strong  excursions in the AGN region during
this evolutionary phase (green and turquoise triangles in  the bottom
panel). This is primarily because sustained star formation (and hence,
high  ionization parameter) and low metallicity contribute to
maintaining high \oiiihb\ and  low \niiha\ at redshifts down to
$z\sim0.2$ (Section~\ref{galm0175}; see also fig.~2 of
\citealt{Feltre16}). Only below $z=0.2$, when star formation drops and
metallicity rises,  can the galaxy move to the AGN region of the
diagram (yellow symbols). 

\subsubsection{Galaxy M0215}\label{galm0215}

As a third example, galaxy M0215 (top panel, right-most column in
Fig.~\ref{Evolution})  exhibits a smooth mass-assembly history with
only three minor mergers at redshifts  $z<2$. Star formation remains
elevated, at a rate of 10--100\,\Msun\,yr$^{-1}$, down $z=0.7$.
Black-hole accretion is also smooth down to $z=0.3$, when spikes of
AGN activity set in  (second panel). Below $z=0.7$, the SFR starts to
decline as new gas supply does not compensate consumption through star
formation and BH accretion, until AGN-driven  winds shut down entirely
star formation at $z<0.2$. The lack of new supply of metal-poor gas
together with the cooling of a hot enriched halo also make the average
interstellar metallicity  rise sharply from sub-solar to super-solar
values at $z<0.7$ (third panel). This sudden change  in metallicity
triggers a sharp drop by nearly an order of magnitude in \oiiihb\
ratio and a rise in \niiha\ ratio, which translate into a `bi-modal'
feature in the corresponding line-ratio diagram (bottom panel). During
the SF phase at $z\ga1$, the galaxy remains on the top left of the SF
branch, with $\log(\oiiihb)> 0.8$  and $\log(\niiha)<-1$, while around
$z=0.7$, it jumps from the SF-galaxy to composite  regions in the 
bottom right area, with $\log(\oiiihb)< 0.5$ and $\log(\niiha)>-1$.
Later on,  enhancement in \niiha\ from nuclear activity
(Section~\ref{galm0175}) triggers excursions  of the galaxy in the AGN
region of the diagram. This illustrates how BH accretion causing
low-redshift galaxies to populate the AGN region of the \oiiihb\
versus \niiha\ diagram does not necessarily require merger events
\citep{Li08}. At $z<0.2$, after star formation  and BH accretion have
been suppressed, post-AGB stars take over the production of ionizing
photons, making the galaxy appear as a LI(N)ER (red squares). 

In summary, therefore, a generic feature of all three examples of
massive-galaxy  evolution shown above is the drop in \oiiihb\ and the
rise in \niiha\ from high to low redshift.  These general trends arise
from a combination of overall increasing metallicity and decreasing
SFR. Peaks in the BHAR typically trigger peaks in \niiha, provided
that the (central) metallicity  is large enough and not diluted by
infall of (metal-poor) gas. Around these trends, the exact mass
assembly and merger history of a galaxy can strongly affect nebular
emission on a case  by case basis, which leads to substantial scatter
in the predicted line ratios at given redshift and  mass, complicating
the interpretation. The C/O abundance ratio and average gas density,
instead,  appear to have a  minor influence on the evolution of
\oiiihb\ and \niiha. In Section~\ref{origin} below, we explore in more
detail the potential role of various physical parameters in driving
the observed evolution of optical-line ratios in galaxies at different
redshifts.

\subsection{Population study}\label{popstudy}

\begin{figure*}
\centering\vspace{-0.5cm}
\epsfig{file=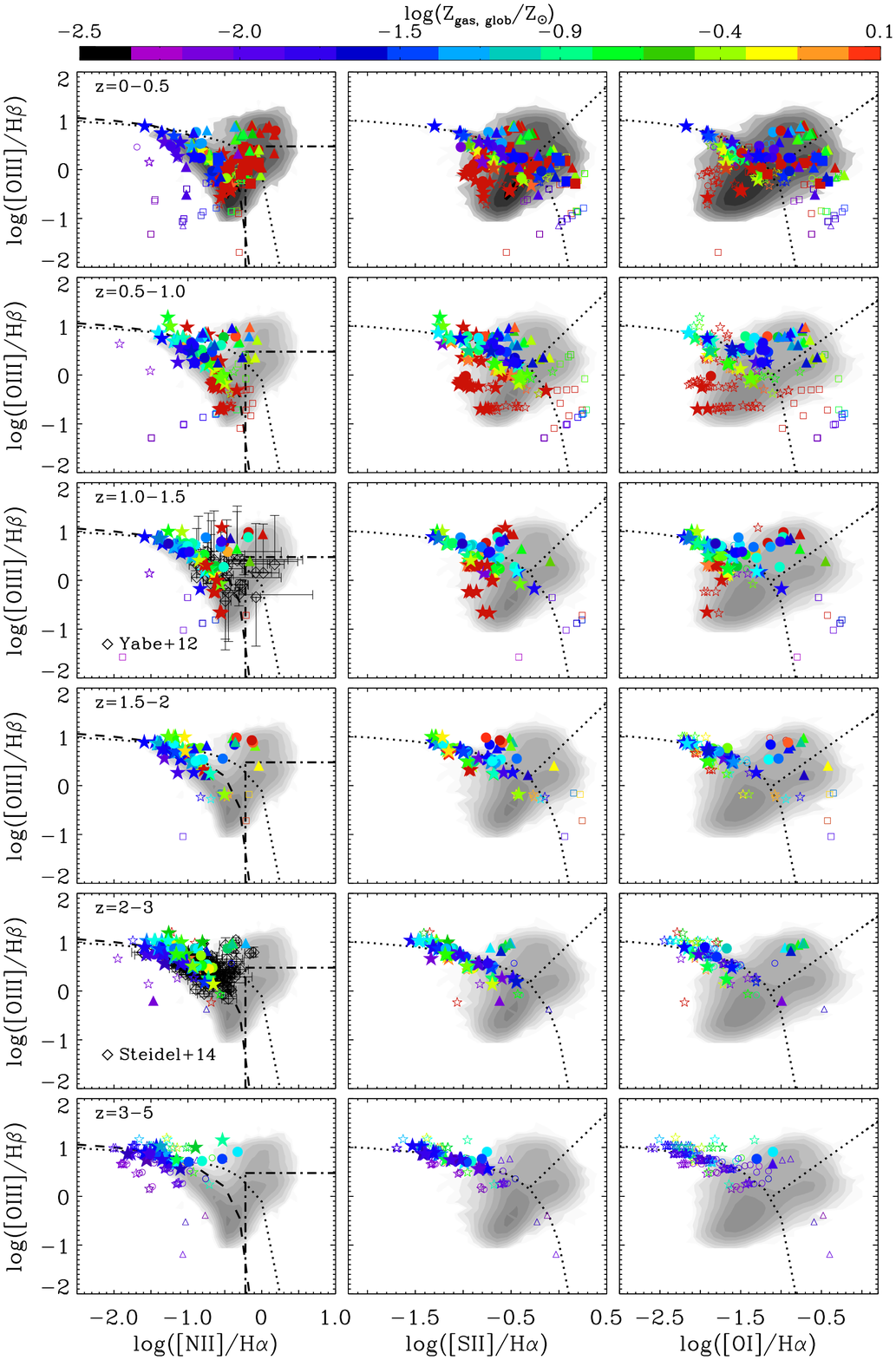, width=0.8\textwidth}
 \caption{Optical emission-line ratio diagrams, \oiiihb\ versus \niiha\
   (left column), \siiha\ (middle column) and \oiha\ (right column), for the 20
   simulated massive galaxies of Section~\ref{theory} and their main high-redshift
   progenitors (symbols, colour-coded according to global interstellar metallicity, 
    as indicated on the top scale), 
   extracted from all simulation snapshots in different redshift intervals
   (different rows), as described in Section~\ref{popstudy}. The symbols and
   different lines have the same meaning as in Fig.~\ref{Evolution}. 
   The simulations are compared to observations of
   local SDSS galaxies (grey shaded area, first row; also reported in light grey in
   higher-redshift bins for reference) as well as of distant galaxies
  (black diamonds) by \citet[$z\sim1.4$]{Yabe12}  and
  \citet[$z\sim2.3$]{Steidel14}.  The small open symbols show the synthetic line ratios
   of {\it all} galaxies regardless of luminosity, while the large filled symbols show
   galaxies above a  flux limit of $\mathrm{5\times 10^{-17}\,erg\,s^{-1}cm^{-2}}$ in all lines. }\label{BPT}    
\end{figure*}
We now investigate the evolution of the {\it full} set of 20 zoom-in
simulations of massive galaxies and their main progenitors in various
optical emission-line ratio diagrams at different redshifts. We
consider redshift bins including several simulation snapshots, and
hence, potentially several tens of emission-line galaxies. As noted
earlier in this section, such a sample is naturally characterised by
an increase in average galaxy stellar mass from high to low redshift,
accompanied by an increase in metallicity. In what follows, we start
by investigating the properties of the full set of simulated galaxies
in optical line-ratio diagrams at different  redshifts, including the
evolution of the SF sequence and of average emission-line ratios
(Sections~\ref{lowzBPT}--\ref{avgratios}). Then, we  assess the
potential influence of observational selection effects on these
properties (Sections~\ref{selection}) and, by considering only
galaxies and progenitors in a fixed  stellar-mass bin of 0.3--$1.0
\times 10^{11}\Msun$ at all redshifts, that of the mass  distribution
of our sample (Section~\ref{massbias}). 

\subsubsection{Line-ratio diagrams at redshifts $z<0.5$}\label{lowzBPT}

\begin{figure*}
\centering
\epsfig{file=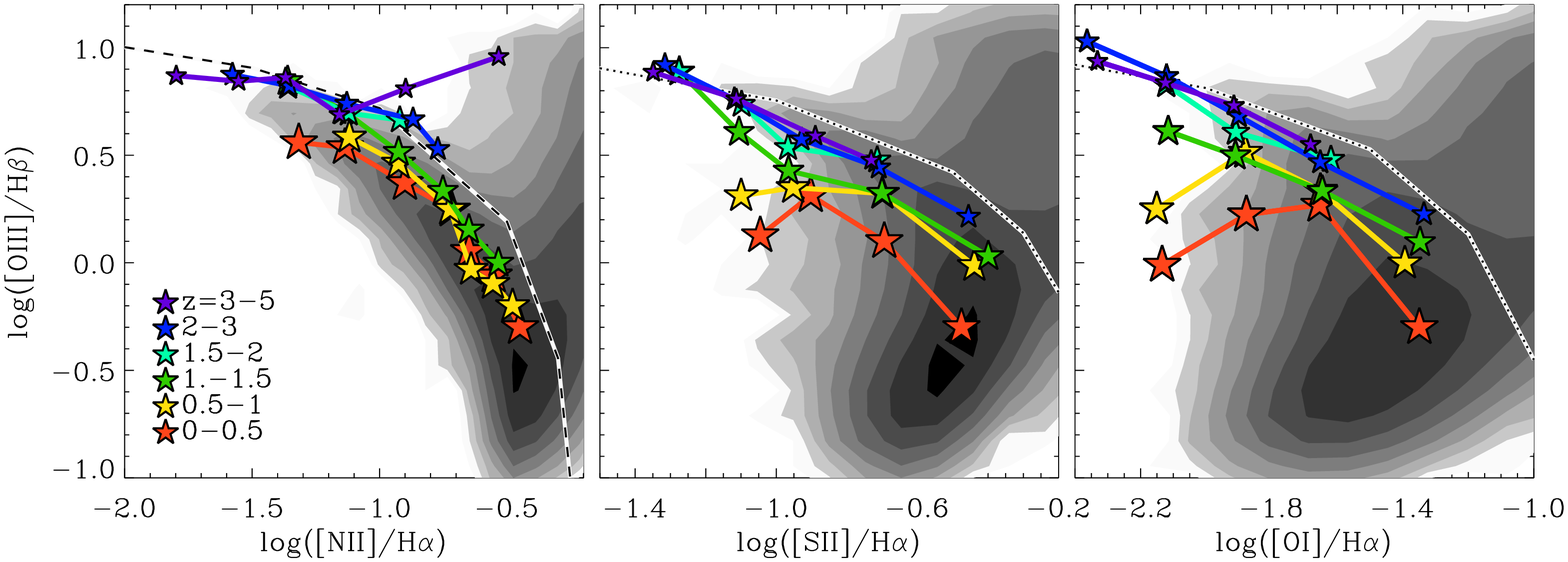, width=1.0\textwidth}
 \caption{Average \oiiihb\ emission-line ratio in bins of \niiha\
   (left panel), \siiha\ (middle panel) and \oiha\   (right panel) for the {\it star-forming subset} [i.e., with
   $\log(\mathrm{BHAR/SFR})< -4$] of the 20 simulated massive galaxies of Section~\ref{theory} 
   and their main high-redshift progenitors, in different redshift ranges (connected stars
   of different colours). The SDSS data (in grey) and dashed and dotted lines are the
   same as in Fig.~\ref{BPT}.}\label{BPT_SFoffset}     
\end{figure*}
In the top row of Fig.~\ref{BPT}, we show the locations of galaxies and their main
progenitors extracted from all simulation snapshots at redshifts $z<0.5$, in three line-ratio
diagrams defined by \oiiihb, \niiha, \siiha\ and \oiha. The symbol types have the same
meaning as in the bottom panels of Fig.~\ref{Evolution} (stars: SF; circles: composite;
triangles: AGN; squares: PAGB), but are now colour-coded according to global interstellar metallicity 
(as indicated in the middle panel). The grey shaded areas and contours indicate the location
of SDSS galaxies in each diagram. To perform a meaningful comparison between models and
observations, we show the effect of requiring a typical  flux limit
of $\mathrm{5\times 10^{-17}\,erg\,s^{-1}cm^{-2}}$ for all simulated emission
lines \citep[e.g., table~1 of][]{Juneau14}. The filled symbols correspond to galaxies 
satisfying this criterion, and the open symbols to those too faint to be detected. Also
shown in these diagrams are standard observational criteria to distinguish SF galaxies from 
composites, AGN and LI(N)ER. The dashed, dotted and dot-dashed lines in the \oiiihb\ versus \niiha\
diagram have the same meaning as in Fig.~\ref{Evolution}, while the dotted lines in the other
two diagrams distinguish SF galaxies (bottom left) from AGN (top) and LI(N)ER (bottom right;
\citealt{Kewley01}).

Fig.~\ref{BPT} shows that, at $z<0.5$, simulated galaxies satisfying our conservative flux 
detection limit occupy the same areas as SDSS galaxies in all three line-ratio diagrams (galaxies
with line luminosities below the flux limit are typically very quiescent). Moreover, in general, 
simulated galaxies of SF, composite, AGN and PAGB types appear to fall in regions of the diagrams 
corresponding to the observationally defined SF, composite, AGN and LI(N)ER categories. This is remarkable in 
that, in our approach, the different types are connected to physical parameters, such as the BHAR/SFR
ratio (for SF galaxies, composites and AGN) and the contribution to total H$\beta$ luminosity 
(for PAGB-dominated galaxies). An in-depth investigation of the usefulness of this connection 
for the interpretation of nebular emission from distant galaxies will be the subject of a future 
study. For the remainder of the present paper, we conclude
on the basis of this comparison with SDSS galaxies in the nearby Universe that our sample of 
zoom-in simulations of massive galaxies is well anchored at $z<0.5$ to investigate the evolution 
of nebular emission with cosmic time.

\subsubsection{Line-ratio diagrams at redshifts $z>0.5$}\label{highzBPT}

\begin{figure*}
\epsfig{file=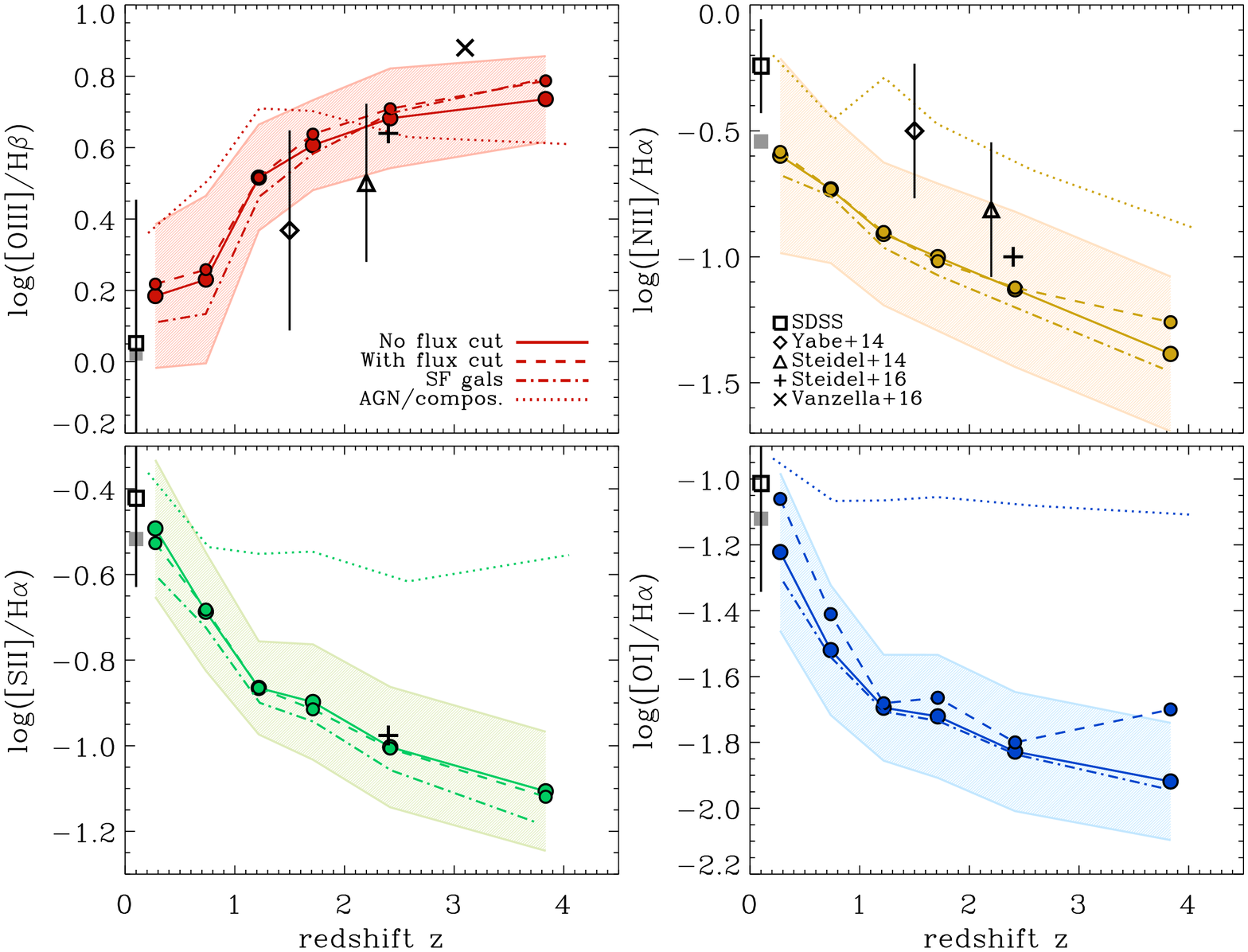, width=0.9\textwidth}
\caption{Redshift evolution of the average \oiiihb\ (red, top left), \niiha\ 
(beige, top right), \siiha\ (green, bottom left) and \oiha\ (blue, bottom right) 
ratios of the 20 simulated massive galaxies of Section~\ref{theory} and their 
main high-redshift progenitors (large filled circles and solid lines), together 
with the $\pm1\sigma$ scatter around the mean relations (shaded areas). The
small filled circles and dashed lines show the results obtained when considering 
only galaxies above a   flux limit of $\mathrm{5\times 10^{-17}\,erg\,s^{-1}cm^{-2}}$ 
in all lines. Dot-dashed and dotted lines show the mean relations for SF- and 
composite/AGN-dominated galaxies, respectively. Also shown for references are
observed mean ratios of SDSS emission-line galaxies in the local Universe 
(black open square: all galaxies; grey filled square: SF galaxies only) and of
different samples of galaxies at various redshifts, as indicated \citep{Yabe14,
Steidel14, Steidel16,Vanzella16}.}\label{Evol_lineratios}      
\end{figure*}

Rows 2 to 6 of Fig.~\ref{BPT} show the analog of the the first row for 
the redshift bins $z=0.5$--1.0, 1.0--1.5, 1.5--2.0, 2.0--3.0 and 3.0--5.0
(from top to bottom). As expected from the evolution of the three example
galaxies in Fig.~\ref{Evolution}, Fig.~\ref{BPT} confirms that, for the whole
sample of simulated galaxies, \oiiihb\ tends to increase and \niiha\ to 
decrease from low to high redshift, and that part of this trends at least
is attributable to the lower global interstellar metallicity of
high-redshift galaxies. The \siiha\ and \oiha\ ratios also tend to
decrease toward high redshift. A most notable result from
Fig.~\ref{BPT} is the consistency between the emission-line properties
of simulated galaxies brighter than the flux-detection limit (filled
symbols) and the observed properties of star-forming galaxies from the
sample of \citet{Yabe12} at $z\sim1.4$ (black diamonds with error bars
in the third row) and that of \citet{Steidel14} at $z\sim2.3$ (black
diamonds with error bars in the fifth  row). In Section~\ref{origin},
we will exploit our fully self-consistent simulations  to gain insight
into the physical parameters that, along with global interstellar
metallicity,  are likely to contribute to the observed evolution of
galaxies in these optical line-ratio  diagrams. 

\subsubsection{Evolution of the SF-galaxy sequence in optical line-ratio diagrams}\label{SFsequence}\label{SFseq}

Motivated by a number of observational studies \citep[e.g.,][]{Steidel14,
Shapley15,Kashino17,Strom17}, we quantify in Fig.~\ref{BPT_SFoffset} the 
redshift evolution of the average `star-forming-galaxy sequence' in optical line-ratio
diagrams. We define this sequence as the average \oiiihb\ of SF galaxies
(i.e., those with $\mathrm{BHAR/SFR <10^{-4}}$; see Section~\ref{casestudies}) 
in bins of \niiha\ (left panel), \siiha\ (middle panel) and \oiha\ (right panel), as shown by 
stars and solid lines in the three diagrams of Fig.~\ref{BPT_SFoffset}. The different 
colours correspond to different redshift bins, as indicated. As in Fig.~\ref{BPT}, the 
shaded areas show the distributions of local ($z\sim0.1$) SDSS galaxies in these diagrams, 
while the dashed and dotted curves mark the upper boundary of the SF-galaxy region 
according to \citet{Kauffmann03} and \citet{Kewley01}.
  
Fig.~\ref{BPT_SFoffset} shows that the rise in \oiiihb\ as a function of redshift
predicted by our zoom-in simulations amounts to roughly 
0.4--0.5\,dex at fixed $\niiha=0.1$, $\siiha=0.1$ and $\oiha=0.01$. These values
are widely consistent with observed trends for the relation between \oiiihb\ 
and \niiha\ \citep[e.g.,][]{Steidel14,Kashino17,Strom17}, while the existence of
similar observational evidence for the relation between \oiiihb\ and \siiha\ is 
still debated \citep[e.g.,][see also Section~\ref{discussion}]{Shapley15,Kashino17}. 
We note that the predicted evolutionary trend of \oiiihb\ is a specific outcome 
from our zoom-in simulations including AGN feedback. In the runs without 
AGN feedback, instead, \oiiihb\ almost does not evolve with redshift. We discuss 
this in more detail in Appendix~\ref{AGNfeedback}.

\begin{figure*}
\epsfig{file=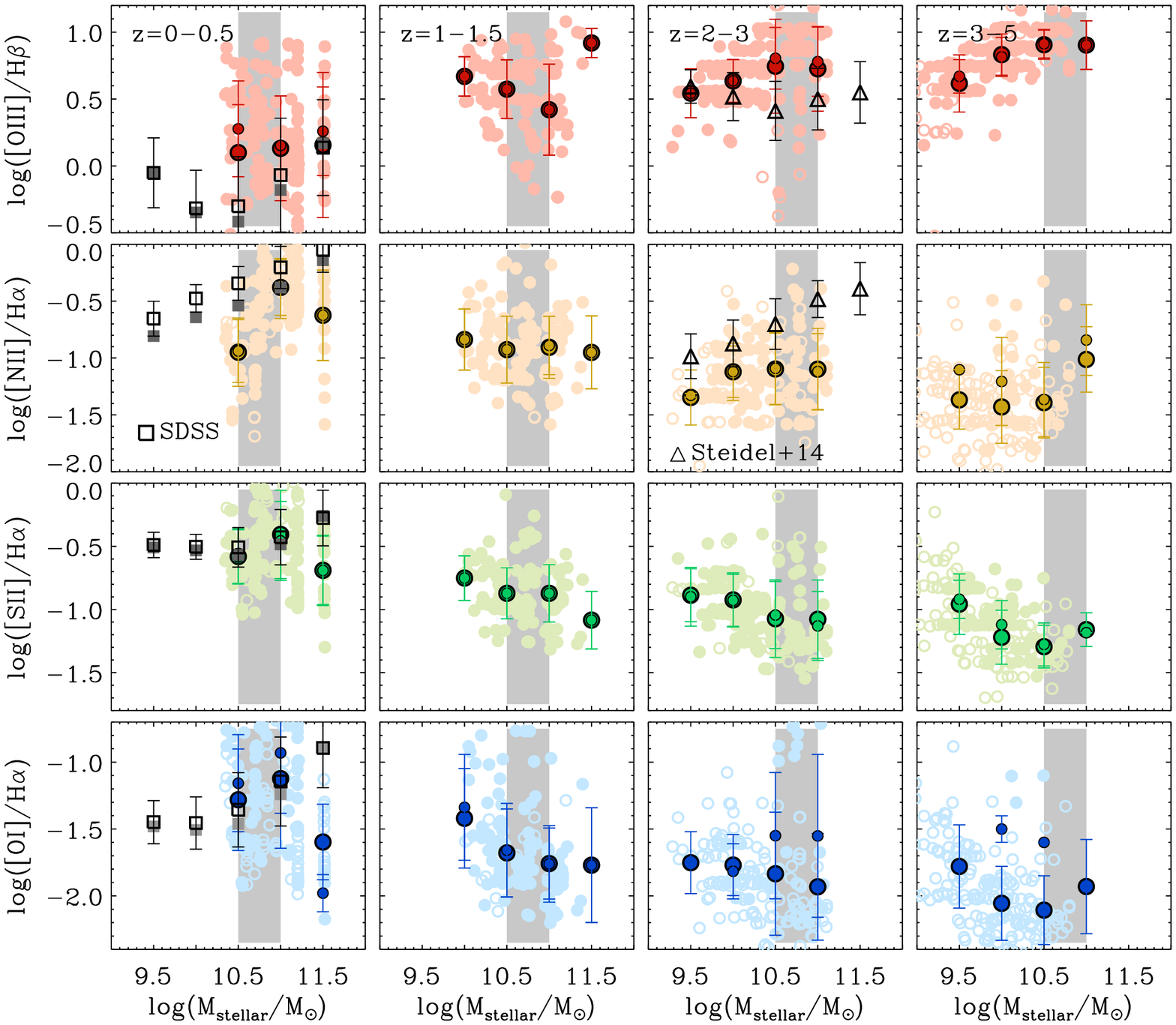, width=0.85\textwidth}
\caption{\oiiihb\ (red, top row), \niiha\ (beige, second row), \siiha\ (green, third row) and 
\oiha\ (blue, bottom row) as a function of stellar mass, in different redshift bins 
(different columns), for the 20 simulated massive galaxies of Section~\ref{theory} and their 
main high-redshift progenitors. In each panel, individual (pastel circles) and 
average (bright circles with error bars) line ratios are shown with and without
including flux limit of $\mathrm{5\times 10^{-17}\,erg\,s^{-1}cm^{-2}}$ 
in all lines (filled/big and open/small circles, respectively). The grey
shaded area highlights the stellar-mass range $10.5<\log (M_{\rm stellar}/\Msun)<11.0$.
Also shown are observations of local SDSS (black open square: all galaxies; grey filled
square: SF galaxies only) and of galaxies at $z\sim2.3$ from 
\citet[black open triangles]{Steidel14}.} \label{Lineratios_mass}        
\end{figure*}

\subsubsection{Evolution of the global galaxy population in optical line-ratio diagrams}\label{avgratios}

It is also of interest to examine the evolution of the emission-line
properties of the global  galaxy population, beyond that of purely SF
galaxies. In Fig.~\ref{Evol_lineratios}, we show  the redshift
evolution of the average \oiiihb, \niiha, \siiha\ and \oiha\ for our
full  sample of simulated galaxies (large filled circles and solid
lines), together with the $\pm1\sigma$  scatter around this mean
evolution (shaded areas). The small filled circles and dashed lines
show the (very similar) evolution obtained when considering only
galaxies above the flux-detection limit of $\mathrm{5\times
  10^{-17}\,erg\,s^{-1}cm^{-2}}$ (as in Fig.~\ref{BPT}). As expected
from our previous findings, Fig.~\ref{Evol_lineratios} shows that the
trends of increasing \oiiihb\ and decreasing \niiha, \siiha\ and
\oiha\ from low to high redshift predicted by our simulations are in
good  agreement with various observational results at $0\la z\la3$
\citep[from SDSS;][black symbols, as indicated in the top right panel]
{Yabe14,Steidel14,Steidel16,Vanzella16}, despite the sparse statistics
of the simulations.  

In the lowest redshift bin, the predicted average \oiiihb,
\siiha\ and \oiha\ agree reasonably well with the observed ones of
SDSS galaxies at $z\sim0.1$,  but the predicted \niiha\ ratio appears
significantly lower than observed out to $z\sim2.5$. A  mismatch in
gas-phase metallicity is not likely to account for this discrepancy,
given the good  general agreement between models and observations for
the other three ratios.  At $z\sim0.1$,  part of the discrepancy could
arise from the larger fraction of AGN-dominated galaxies  in the SDSS
sample relative to the simulations (which, because of low statistics, 
do not represent the same galaxy populations as in the SDSS and 
other observational samples), as the average \niiha\ of  AGN-dominated
galaxies (dotted line in the top-right panel of
Fig.~\ref{Evol_lineratios}) is  significantly larger than that  of  SF
galaxies (dot-dashed line). In fact, selecting only SF galaxies --
i.e., those below the \citet{Kauffmann03} criterion in the \oiiihb\
versus \niiha\ diagram -- in the SDSS sample (grey filled square in
each panel of Fig.~\ref{Evol_lineratios}) brings down \niiha\ far more
significantly than \oiiihb, \siiha\ and \oiha. The same argument does
not hold for the mismatch with the \citet{Yabe12} measurement of
\niiha\ at $z\sim1.4$ and the \citet{Steidel14} one at $z\sim2.5$,
since these samples include, respectively, zero and 4 per cent of
AGN. The higher-than-predicted \niiha\ (and marginally
lower-than-predicted \oiiihb)  of these samples could arise from a
difference in global interstellar metallicity. For reference, we
checked  that the evolution of the galaxy mass-metallicity relation
out to $z\sim3$ in our simulations is broadly  consistent with that in
fig.~8 of \citet{Maiolino08}.

\subsubsection{Influence of flux-selection effects}\label{selection}

The small differences between the dashed and solid lines in
Fig.~\ref{Evol_lineratios} indicate that, for our sample of simulated
massive galaxies and their main high-redshift  progenitors, evolution
effects dominate over flux-selection effects in determining the
redshift trends of \oiiihb, \niiha, \siiha\ and \oiha. It is important
to stress the significance of this finding in the context of the
recent study by \citet{Juneau14}, who showed the difficulty in
disentangling evolution from selection effects in analyses of
emission-line  properties of distant galaxy samples. While a more
robust conclusion would require a  larger, statistically complete
sample of simulated galaxies (not achievable with our current  limited
set of zoom-in simulations; see Section~\ref{discussion}), the results
of Fig.~\ref{Evol_lineratios} already demonstrate the usefulness of
our fully consistent modelling  to interpret observations of the
nebular emission from distant galaxy populations. 

\subsubsection{Influence of the stellar-mass evolution of simulated galaxies}\label{massbias}

Despite the generally encouraging consistency between model and
observed trends in Fig.~\ref{Evol_lineratios}, we must remember that
our predictions rely on a  statistically small set of simulated
massive galaxies, whose progenitor masses at high redshift are
systematically smaller than today. In this regard, it is important to
check the influence of the global evolution of stellar mass (and
metallicity)  in the simulations on the predicted redshift trends of
emission-line ratios. To this goal, we show in
Fig.~\ref{Lineratios_mass} the individual (pastel circles) and
average (bright circles) line ratios as a function of stellar mass for
galaxies and their main progenitors in different redshift bins
(different columns), with and without including flux selection
(filled/big and open/small circles, respectively). The grey shaded
area highlights the stellar-mass range  $10.5<\log (M_{\rm
  stellar}/\Msun)<11.0$ in all panels. Fig.~\ref{Lineratios_mass}
shows that, even in a fixed stellar-mass range, \oiiihb, \niiha,
\siiha\ and \oiha\ for  galaxies in our simulations exhibit strong
redshift evolution, in good agreement with that inferred from the
comparison between SDSS and the  \citet{Steidel14} data for \oiiihb\
and \niiha\ (both samples being available over a range of stellar
masses; we note in passing the good overlap between the observed and
model stellar-mass ranges at $z=2$--3). Moreover, at a given redshift,
the  average line ratios show only a modest dependence on stellar
mass, suggesting  that any stellar-mass bias would have a limited
influence on population-wide line-ratio evolution. Such evolution may
instead depend more sensitively on the evolution of other ISM and
radiation properties, even at fixed stellar mass. This will  be the
topic of the next section.    

\begin{figure*}
\centering{\large Physical quantities for SF models}
\epsfig{file=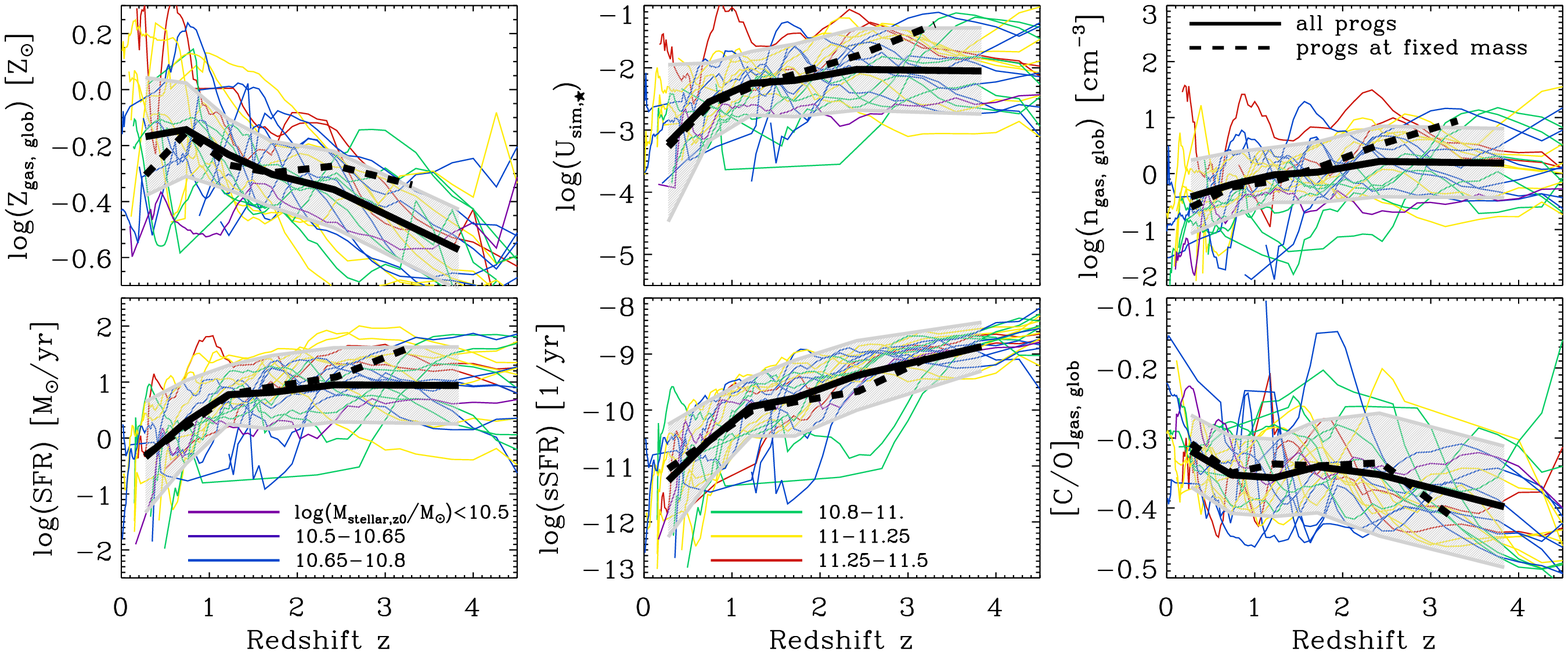, width=0.98\textwidth}
\centering{\large Physical quantities for AGN models}
\epsfig{file=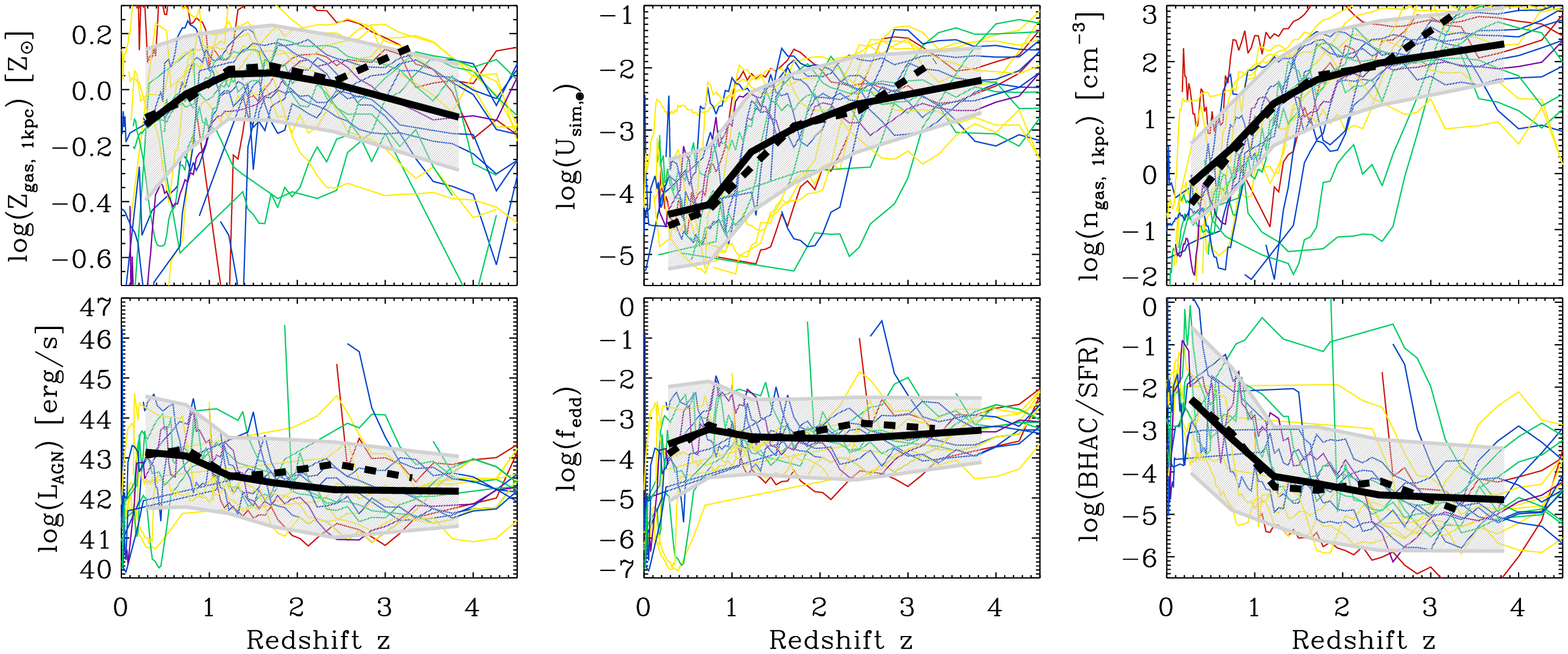,
  width=0.98\textwidth}
\centering{\large Physical quantities for PAGB models}
\epsfig{file=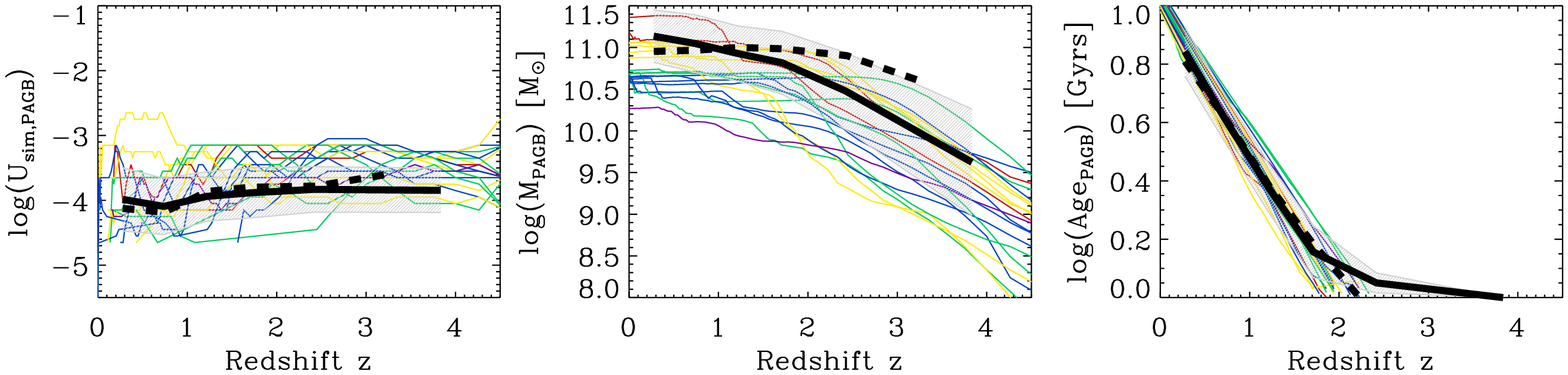,
  width=0.98\textwidth}
\caption{Redshift evolution of the different physical
  quantities used to select SF (first and second rows), AGN (third and fourth
  rows) and PAGB (bottom row) nebular-emission models, for the 20 simulated massive 
  galaxies of Section~\ref{theory} and their main high-redshift progenitors
  (thin lines, colour-coded according to final stellar mass, as indicated):
global interstellar metallicity ($Z_{\mathrm{gas, glob}}$); SF ionization
  parameter ($U_{\mathrm{sim,}\star}$); global gas density ( $n_{\mathrm{gas, glob}}$); 
  SFR; specific SFR; global C/O ratio; central interstellar metallicity 
  ($Z_{\mathrm{gas, 1kpc}}$); AGN ionization parameter ($U_{\mathrm{sim,}\bullet}$);
central gas density ($n_{\mathrm{gas, 1kpc}}$); AGN luminosity ($L_\mathrm{AGN}$);
AGN Eddington ratio $\log(f_{\mathrm{edd}})$; BHAR/SFR ratio; PAGB ionization 
parameter ($U_{\mathrm{sim,\diamond}}$); mass ($M_\mathrm{PAGB}$) and average
age (${\rm Age}_\mathrm{PAGB}$) of stars older than 3\,Gyr. In each panel, the thick 
black line and shaded area show the average
  evolution and $\pm1\sigma$ scatter around it, including all galaxies and progenitors, 
  while the thick dashed line shows the average evolution of galaxies and progenitors in the 
  fixed stellar-mass range $10.5<\log (M_{\rm stellar}/\Msun)<11.0$.}\label{Propevol}       
\end{figure*}

\section{Physical origin of the evolution of optical emission-line ratios}\label{origin} 

The physical origin of the observed evolution of optical emission-line
ratios in galaxies is a heavily debated issue (Section~\ref{intro}). 
In this section, we use our models to investigate the origin of
the predicted redshift evolution of \oiiihb, \niiha, \siiha\ and
\oiha\ in terms of ISM and radiation properties of simulated
galaxies. We focus on those physical quantities that control the
evolution of the  SF, AGN and PAGB models (Section~\ref{nelms}). Our
approach offers  the unique advantage of exploring separately the {\it
  relative} influence  of different physical parameters on the
evolution of emission-line properties.  In the next subsections, we
carry out such an analysis for the whole set of simulated galaxies and
their main progenitors, as well as for galaxies and progenitors in a
fixed stellar-mass range.  

\subsection{Redshift evolution of the parameters controlling nebular emission}\label{zevolparam}

To identify the physical processes responsible for the drop in
\oiiihb\ and the rise in \niiha, \siiha\ and \oiha\ from high to low
redshift, we start by exploring how the parameters controlling the SF,
AGN and PAGB nebular-emission models  change with redshift. These
parameters are potential drivers of the cosmic evolution  of optical
emission-line ratios. We show the redshift evolution of these
parameters  in the different panels of Fig.~\ref{Propevol}, for the SF
(first and second rows), AGN  (third and fourth rows) and PAGB (bottom
row) models, and for the 20 simulated galaxies  and their main
progenitors in our sample (thin lines, colour-coded according to final
stellar  mass, as indicated). In each panel, we also show the
corresponding average evolution  (thick solid line) and associated
$\pm1\sigma$ scatter (grey shaded area). 

Some parameters exhibit little variation with redshift, such as the
average global  carbon-to-oxygen abundance ratio, (C/O)$_{\mathrm{gas,
    glob}}$, ISM density, $n_{\mathrm{gas, glob}}$, AGN
Eddington ratio,  $f_\mathrm{fedd}$, central
interstellar metallicity  $Z_\mathrm{gas, 1kpc}$, and PAGB ionization
parameter, $U_{\rm sim,\diamond}$.  These parameters are not  likely
to drive the evolutionary trends in emission-line ratios  seen in the
simulations. Other parameters rise significantly from high to low
redshift,  such as the global interstellar metallicity, 
$Z_\mathrm{gas,glob}$, AGN luminosity,  $L_\mathrm{AGN}$,
and mass and age of PAGB stellar
populations. Instead, the SF  ionization parameter, $U_{\rm
sim,\star}$ (driven by the SFR), and the AGN ionization
parameter, $U_{\rm sim,\bullet}$ (driven by the average  central
ISM density, $n_\mathrm{gas,1kpc}$)  show a marked decline.  We
note that, since
$U_{\rm sim,\bullet}$ scales as $L_\mathrm{AGN}^{1/3}$ and 
$n_\mathrm{gas,1kpc}^{2/3}$
(equations~\ref{logU} and \ref{fillfac}), the drop in $U_{\rm sim,\bullet}$ 
by itself indicates that $L_\mathrm{AGN}$ has a negligible 
influence on this parameter (and hence on optical-line ratios; 
see \citealt{Feltre16}). The rise in BHAR/SFR ratio further suggests a  growing
relevance of AGN  nebular emission toward low redshift, and the drop
in specific SFR an increasing importance  of post-AGB stellar
populations.  These different evolving quantities, aside from $L_\mathrm{AGN}$, 
can potentially drive the
trends  seen in emission-line ratios.

\subsection{Relation between line-ratio and parameter evolution}\label{levolparam}

To reveal which of the potential driving parameters identified in
Section~\ref{zevolparam}   effectively contribute to the redshift
evolution of emission-line ratios predicted by the  models (and seen
in the observations), we plot in Fig.~\ref{Lineratios_prop} the
average  \oiiihb\ (red lines), \niiha\ (beige lines), \siiha\ (green
lines) and \oiha\ (blue lines) ratios  against the physical parameters
controlling the SF (top row), AGN (middle and bottom rows) and PAGB
(top and bottom rows) models, together with the $\pm1\sigma$ scatter
about these mean  relations (pastel shaded areas). We note that, while
the average trends in Fig.~\ref{Lineratios_prop} were derived by
considering all simulated galaxies and their main progenitors at
redshift $z<5$, we have checked that the trends hardly depend on the
exact redshift range  chosen within this interval. We now examine the
dependence of each optical-line ratio in Fig.~\ref{Lineratios_prop} on
those model parameters identified above as potential  drivers of the
cosmic evolution of galaxy nebular emission, separating between the
influence  of the SF, AGN and PAGB components. 

\subsubsection{\oiiihb\ ratio}\label{OIII}

\begin{figure*}
\epsfig{file=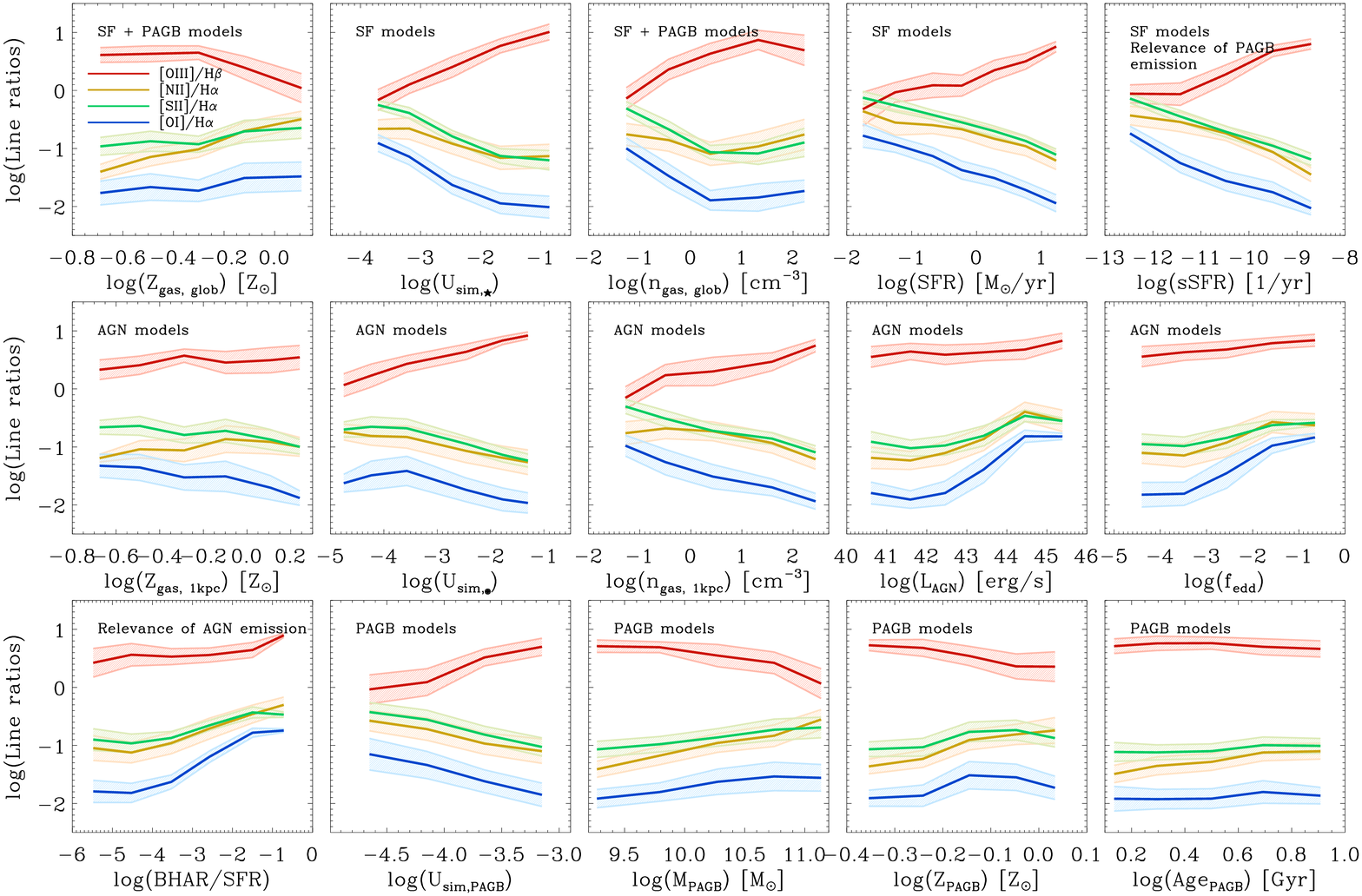, width=1.0\textwidth}
\caption{Average \oiiihb\ (red), \niiha\ (beige), \siiha\ (green) and \oiha\ (blue) ratios of 
the 20 simulated massive galaxies of Section~\ref{theory} and their main high-redshift 
progenitors (thick lines), plotted against the same galaxy physical parameters as in 
Fig.~\ref{Propevol}, plus the mean metallicity of stars older than 3\,Gyr ($Z_\mathrm{PAGB}$).
Shaded areas show the $\pm1\sigma$ scatter around the mean relations.}\label{Lineratios_prop}        
\end{figure*}

\begin{itemize}

\item{\it Influence of SF-related parameters:}
The top left panel of Fig.~\ref{Lineratios_prop} shows that \oiiihb\
(in red)  decreases with increasing global interstellar
metallicity. This is primarily because  a rise in metallicity makes
cooling more efficient (and oxygen is a major coolant),  which reduces
the electronic temperature in the ionized gas, and thus,  the
probability for collisional excitation of [O\,{\sc iii}] optical
transitions  \citep[e.g.,][]{Stasinska80,Gutkin16}. A secondary effect
is the softer ionizing radiation of metal-rich stars relative to
metal-poor ones, which  lowers the number density of O$^{++}$ ions,
new-born  stars having the same metallicity as star-forming gas in our
simulations  \citep[Section~\ref{emlines_sf}; see,
e.g.,][]{Gutkin16}. The other panels in the top row of
Fig.~\ref{Lineratios_prop} further show that \oiiihb\ rises with
increasing (specific) SFR and volume-averaged gas density, and hence,
by virtue of equations~\eqref{logU} and \eqref{fillfac}, with
increasing SF  ionization parameter. The rise in \oiiihb\ with
 $n_{\mathrm{gas, glob}}$ follows from the implied larger
volume-filling factor at fixed $n_{\mathrm{H},\star}$
(equation~\ref{fillfac}), which increases the probability for oxygen
to be photoionized twice. Double ionization is also favoured when a
rise in the rate of ionizing photons is induced by a larger SFR at
fixed $n_{\mathrm{gas, glob}}$. 

Combining these findings with the results of Fig.~\ref{Propevol}
suggests that  a rising global interstellar metallicity and declining
strength of the stellar radiation field  (i.e., SFR) can both
contribute to the drop in \oiiihb\ from high to low redshift in
Fig.~\ref{Evol_lineratios}.  Instead, the roughly constant
 $n_{\mathrm{gas, glob}}$  with redshift argues against a primary
influence of this parameter on cosmic \oiiihb\  evolution.

\item{\it Influence of AGN-related parameters:}
The middle row of Fig.~\ref{Lineratios_prop} shows that \oiiihb\
strongly increases with increasing central gas density,
$n_{\mathrm{gas, 1kpc}}$ (third panel),  and hence, increasing AGN
ionization parameter, $U_{\mathrm{sim,\bullet}}$ (second panel), for
reasons analogous to those described above for the dependence of
\oiiihb\  on  $n_{\mathrm{gas, glob}}$ and
$U_{\mathrm{sim,\star}}$. Based on the results  from
Fig. \ref{Propevol}, we conclude that the drop in $n_{\mathrm{gas,
    1kpc}}$  from high to low redshift can also contribute to the
cosmic evolution of \oiiihb.   

\item{\it Influence of PAGB-related parameters:}
We already noted above that global interstellar metallicity, which
enters PAGB as well  as SF nebular-emission models, could contribute
to the drop in \oiiihb\ from high to low redshift. The bottom row of
Fig. \ref{Lineratios_prop} further shows that,  among the other
potential driving parameters identified in Section~\ref{zevolparam},
\oiiihb\ appears to vary significantly with only the mass -- and not
the age -- of post-AGB stellar populations. We note that the drop in
\oiiihb\ at increasing $M_\mathrm{PAGB}$  follows from the greater
average metallicity of the most massive PAGB stellar populations,
hosted by the most massive galaxies. These stellar populations
produce softer ionizing spectra, and hence fewer O$^{++}$ ions,
than their more metal-poor  counterparts in less massive galaxies, an
effect which dominates over the rise in  total number of PAGB stars
from low- to high-mass galaxies. The dependence of  $M_\mathrm{PAGB}$
on redshift in Fig.~\ref{Propevol} confirms that this parameter can
also contribute to the cosmic evolution of \oiiihb. 

\end{itemize}

\subsubsection{\niiha\ ratio}\label{NII}

\begin{itemize}

\item{\it Influence of SF-related parameters:}
The top row of Fig.~\ref{Lineratios_prop} shows that \niiha\ (in
beige)  rises significantly with increasing global interstellar
metallicity and decreasing (specific) SFR, and in turn, decreasing SF
ionization parameter. The rise of \niiha\ with $Z_{\mathrm{gas,
    glob}}$ follows primarily from the implied  increased abundance
of nitrogen, for which secondary production is included in the models
(Section~\ref{emlines_sf}). Additionally, the softer ionizing
radiation of metal-rich stars relative to metal-poor ones lowers the
production probability of multiply-ionized nitrogen relative to
N$^+$. A smaller probability of  multiply-ionized nitrogen is also the
reason for the rise of  \niiha\  with decreasing (specific) SFR and
$U_{\mathrm{sim,\star}}$ (by analogy with  our discussion of \oiiihb\
above). In light of the dependence of  $Z_{\mathrm{gas, glob}}$ and
(specific) SFR on redshift in Fig.~\ref{Propevol},  we conclude that
both parameters can contribute to the rise in \niiha\ from  high to
low redshift in Fig.~\ref{Evol_lineratios}. 

\item{\it Influence of AGN-related parameters:}
Fig.~\ref{Lineratios_prop} show that \niiha\  rises significantly with
decreasing  central gas density, and hence, decreasing AGN ionization
parameter (middle row). This is because a lower $n_{\mathrm{gas,
    1kpc}}$ at fixed $n_{\mathrm{H},\bullet}$  (lower
$U_{\mathrm{sim,\bullet}}$) leads to a lower probability of multiply
ionizing nitrogen at the expense of N$^+$. Also, \niiha\ rises with
increasing BHAR/SFR ratio  (bottom row), because the harder ionizing
spectrum of an AGN relative to young stars makes the electronic
temperature -- and collisional excitation of [N{\sc ii}] --
larger. The redshift dependence of $n_{\mathrm{gas, 1kpc}}$ and the
BHAR/SFR ratio in Fig.~\ref{Propevol} further suggest that these
parameters are likely to contribute to  the cosmic evolution of \niiha. 

\item{\it Influence of PAGB-related parameters:}
As noted above, global interstellar metallicity, which enters PAGB as
well  as SF nebular-emission models, can contribute to the rise in
\niiha\ from high to low redshift. The bottom row of
Fig.~\ref{Lineratios_prop} further  shows that \niiha\ rises with
increasing mass and increasing  metallicity of post-AGB stellar
population. This is because metal-rich stellar  populations in massive
galaxies produce softer ionizing spectra (and hence,  less
multiply-ionized nitrogen) than their more metal-poor counterparts in
lower-mass galaxies. The dependence of $M_\mathrm{PAGB}$ on redshift
in Fig.~\ref{Propevol} indicates that this parameter (linked to
$Z_\mathrm{PAGB}$)  can also contribute to the cosmic evolution of
\niiha.

\end{itemize}

\begin{figure*}
\center{\large All progenitor galaxies}
\epsfig{file=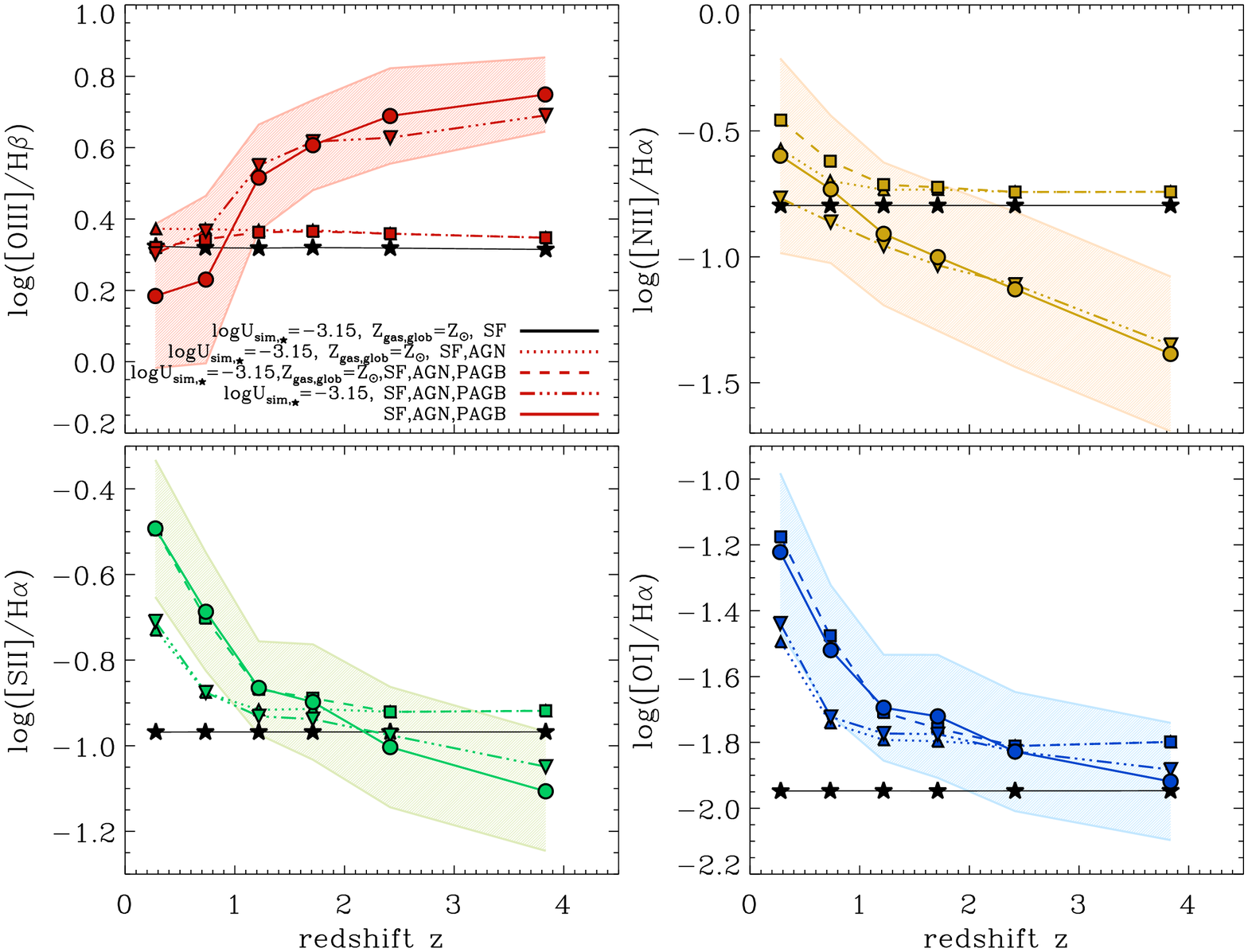, width=0.9\textwidth}
\caption{Relative contributions by different
components to the redshift evolution of the average \oiiihb\ (red, top left), \niiha\ 
(beige, top right), \siiha\ (green, bottom left) and \oiha\ (blue, bottom right) 
ratios of the 20 simulated massive galaxies of Section~\ref{theory} and their 
main high-redshift progenitors. In each panel, black stars (joined by a solid line) show the 
relation obtained using a simplified base model of pure-SF nebular emission, with 
fixed SF ionization parameter, $\log(U_{\mathrm{sim,}\star}) = -3.5$, and global 
interstellar metallicity, $\log(Z_{\mathrm{gas,glob}})=\log(Z_\odot)$.
The other symbols/lines show the effect of adding to this model, step by step: AGN nebular
emission (triangles/dotted line); PAGB nebular emission (squares/dashed line); and
cosmic evolution of $Z_{\mathrm{gas,glob}}$ (upside-down triangles/triple-dot-dashed line) 
and $U_{\mathrm{sim,}\star}$ (circles/solid line and shaded $\pm1\sigma$ scatter around the
mean relation; identical to Fig.~\ref{Evol_lineratios}).}\label{Evol_lineratios_cumu}       
\end{figure*}
\begin{figure*}
\center{\large Progenitors at fixed stellar mass bin $10.5<\log(M_{\rm stellar}/\Msun)<11.0$}
\epsfig{file=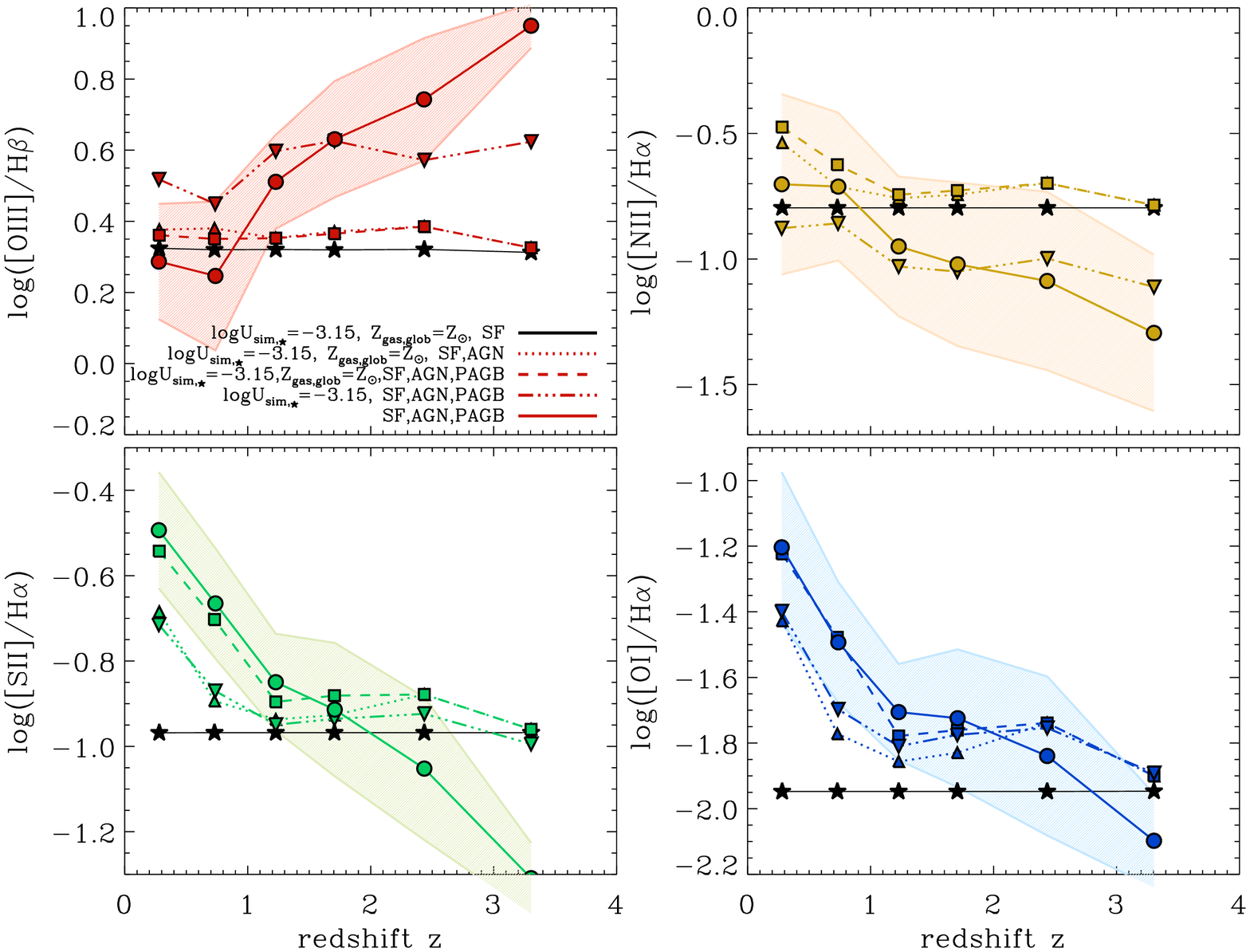, width=0.9\textwidth}
\caption{Same as Fig. \ref{Evol_lineratios_cumu}, but for the subsample of galaxies
and progenitors with stellar masses in the range $10.5<\log(M_{\rm
    stellar}/\Msun)<11.0$}\label{Evol_lineratios_cumu_massbin}       
\end{figure*}
\begin{figure*}
\epsfig{file=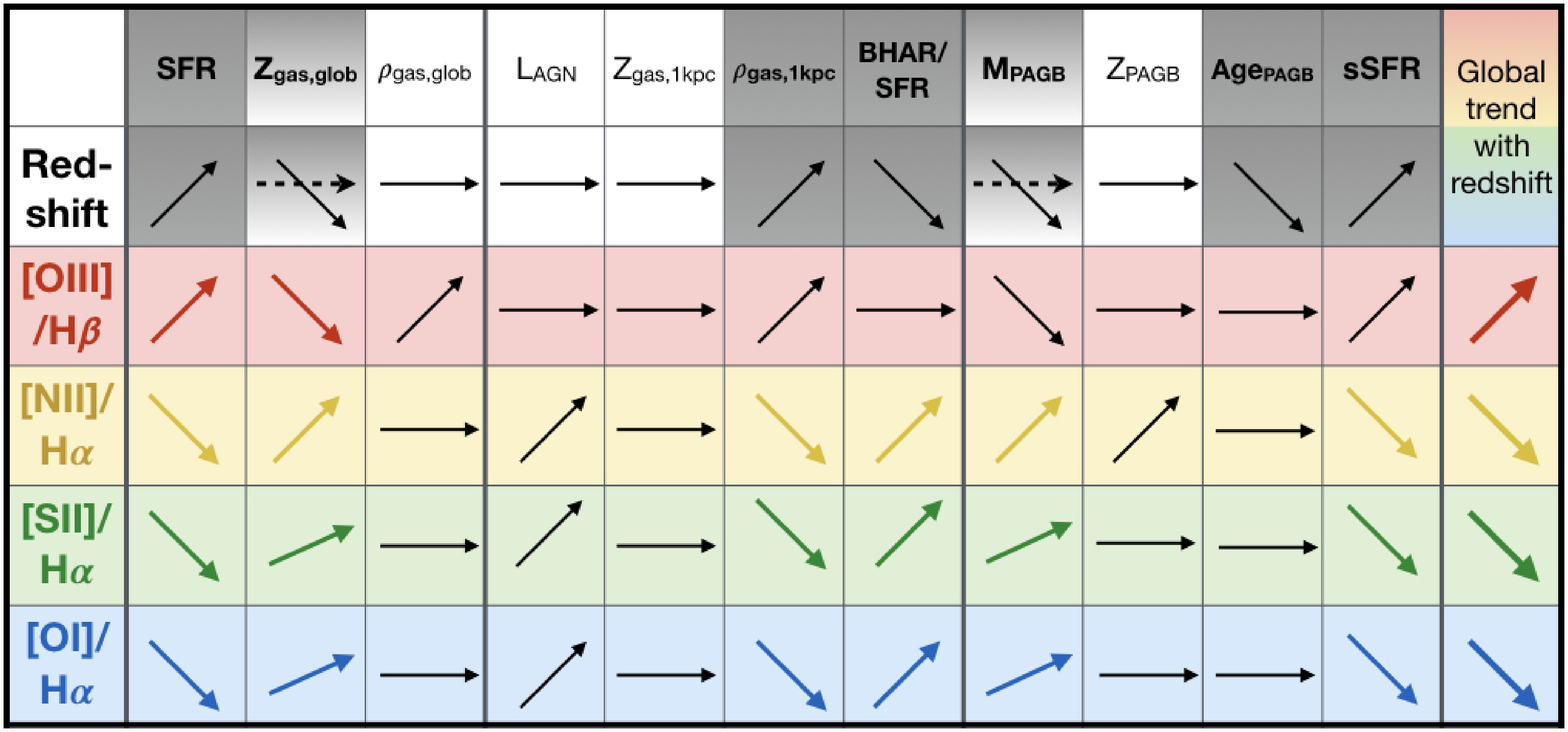, width=1.0\textwidth}\vspace{-4.5cm}
\caption{Schematic summary of Figs~\ref{Propevol},
  \ref{Lineratios_prop}, \ref{Evol_lineratios_cumu} and
  \ref{Evol_lineratios_cumu_massbin}: arrows visualise qualitatively
  how  redshift, \oiiihb , \niiha , \siiha\ and \oiha\ (different rows) change as different radiation and 
  ISM properties increase (different columns, referring to the same physical quantities 
  as shown in Fig. \ref{Propevol}), for the 20 simulated massive galaxies of Section~\ref{theory} 
  and their main high-redshift progenitors. Grey cells refer to properties found to vary with redshift,
  grey/white-gradient cells to properties whose dependence on redshift vanishes when
  selecting a fixed stellar-mass range, and white cells to properties independent of redshift.
Coloured arrows highlight those physical quantities expected to drive the cosmic evolution
of \oiiihb, \niiha, \siiha\ and \oiha, according to our analysis of Figs~\ref{Evol_lineratios_cumu}
and \ref{Evol_lineratios_cumu_massbin}.}\label{SchematicSummary}        
\end{figure*}

\subsubsection{\siiha\ and \oiha\ ratios}\label{SII} 

\begin{itemize}

\item{\it Influence of SF-related parameters:}
The top row of Fig.~\ref{Lineratios_prop} shows that the dependence of
\siiha\ (green lines) and \oiha\ (blue lines) on SF-related parameters
is similar to that of \niiha, for the same reasons as outlined for
that ratio in Section~\ref{NII}.   We note that \siiha\ and  \oiha\
depend less strongly than \niiha\ on global interstellar  metallicity,
because of the influence of secondary production in the case of
nitrogen. We conclude that, as for \niiha, global interstellar
metallicity and (specific) SFR are likely to contribute to the rise of
\siiha\ and \oiha\ from high to low redshift in
Fig.~\ref{Evol_lineratios}.  
 
\item{\it Influence of AGN- and PAGB-related parameters:} 
\siiha\ and \oiha\ depend in a similar way to \niiha\ on AGN-
and PAGB-related parameters in Fig.~\ref{Lineratios_prop}. Thus,
central  gas density (AGN ionization parameter), BHAR/SFR ratio
and, to a lesser extent than for \niiha\ (because of the shallower
dependence on  $M_\mathrm{PAGB}$ in Fig.~\ref{Lineratios_prop}) mass
of post-AGB stellar populations can also contribute to the cosmic
evolution of \siiha\ and \oiha.  

\end{itemize}

\subsection{Relative influence of the different drivers of cosmic line-ratio evolution}\label{relimpact}

In Section~\ref{levolparam}, we have seen that global and central
interstellar metallicities,  (specific) SFR, BHAR/SFR ratio and
mass of post-AGB stellar population can all contribute to the cosmic
evolution of the optical line ratios in Fig.~\ref{Evol_lineratios}.
To identify the {\em relative} influence of these different parameters
on the evolution of nebular emission, we now proceed with a
`cumulative' approach: we start  by combining our zoom-in simulations
of galaxy formation with a simplified model of  pure-SF nebular
emission, for which we adopt fixed $U_\star$ and $Z_\star$
(Sections~\ref{emlines_sf} and \ref{sfmatch}). Then, step by step, we
examine how the predicted line-ratio evolution changes after
introducing AGN and PAGB nebular emission and accounting for cosmic
evolution of $U_\star$ and $Z_\star$. For simplicity, we explore in
this way only the overall effect of adding the AGN and PAGB
components,  without distinguishing between, e.g., the influence of
central gas density and BHAR/SFR  ratio. In practice, we start from an
SF nebular-emission model with fixed $\log U_{\rm sim,\star}  = -3.5$
and $Z_\star=Z_\mathrm{gas,glob} =Z_\odot$. We show in
Fig.~\ref{Evol_lineratios_cumu}  the implied (flat) redshift evolution
of the mean \oiiihb, \niiha, \siiha\ and \oiha\ ratios obtained with
this base model for our sample of simulated galaxies (black stars and
solid lines). 

\subsubsection{\oiiihb\ ratio}

The top-left panel of Fig.~\ref{Evol_lineratios_cumu} shows that
adding the contribution from AGN nebular emission to the base model
(red triangles and dotted line) slightly raises \oiiihb\
relative to the black line, but without introducing any redshift
evolution. Thus, central gas density, and in turn, AGN ionization
parameter, do not appear to affect significantly the cosmic evolution
of \oiiihb. The inclusion of PAGB nebular emission (red squares and
 dashed line) reduces slightly \oiiihb\ at $z<1$, but the effect is
still negligible at higher redshift. Instead, the effect of
incorporating in $Z_\star$ the  evolution of the global interstellar
metallicity ($Z_\mathrm{gas,glob}$) predicted by the simulations  (red
upside-down triangles and triple-dot-dashed line) is drastic: this
causes \oiiihb\ to  drop sharply from high to low redshift. At $z<1$,
this trend is accentuated by the inclusion, through $U_\star$, of the
drop in average SFR predicted by the  simulations (red circles and
solid line, which together with the shaded area, are the same as in
Fig.~\ref{Evol_lineratios}). 

Hence, the cosmic evolution of \oiiihb\ predicted by our simulations
results  from combined effects of different parameters governing SF
and PAGB (but not AGN) nebular emission. Global interstellar
metallicity emerges as the main driver of this evolution, with minor
contributions from SFR evolution and post-AGB  stellar populations at
redshift $z<1$. 
 
\subsubsection{\niiha\ ratio}

The top right panel of Fig.~\ref{Evol_lineratios_cumu} shows that
adding the contribution from AGN nebular emission to the base model
(beige triangles and dotted line) make \niiha\ increase significantly
at redshift $z<1$. Including PAGB nebular emission reinforces slightly
this trend (beige squares and dashed line). As in the case of \oiiihb,
accounting for the evolution of $Z_\mathrm{gas,glob}$ predicted by the
simulations is crucial and makes \niiha\ increase significantly from
high to low redshift (beige upside-down triangles and
triple-dot-dashed line). This rise  is intensified at $z<1$ by the
inclusion of SFR evolution (beige circles and solid line). 

Our simulations therefore predict that the cosmic evolution of \niiha\ 
is regulated by different parameters governing SF, AGN and PAGB
nebular emission. At $z>1$, the evolution is dominated by that of 
global interstellar metallicity, and at $z<1$, by the drops in central
gas density (driving the AGN models) and SFR, together with the
build-up of metal-rich  populations of post-AGB stars.

\subsubsection{\siiha\ and \oiha\ ratios}

The bottom panels of Fig.~\ref{Evol_lineratios_cumu} show that adding
the contribution from AGN nebular emission to the base model makes
both \siiha\ and \oiha\ (green/blue triangles and dotted lines)
increase sharply at redshifts below unity. Including PAGB nebular
emission reinforces  significantly these trends (green/blue squares
and dashed lines). In strong contrast with \oiiihb\ and \niiha,
accounting for the evolution of global interstellar metallicity
predicted by the simulations has only a modest effect on the evolution
of \siiha\ and \oiha, and only at redshift $z>2$ (green/blue
upside-down triangles and triple-dot-dashed lines). Instead, the
inclusion of  SFR evolution strengthens the rise in \siiha\ and \oiha\
over the whole  range from high to low redshift (green/blue circles
and solid lines). 

Hence, as for \oiiihb\ and \niiha, our simulations predict that the
cosmic  evolution of \siiha\ and \oiha\ arises from combined effects
of different  parameters governing SF, AGN and PAGB nebular
emission. In this case, global interstellar metallicity plays only a
minor role, and the evolution of \siiha\  and \oiha\ appears to be
driven primarily by that of central gas density and SFR and the
build-up of metal-rich populations of post-AGB stars.

\subsection{Influence of the intrinsic stellar-mass evolution of simulated galaxies}\label{massbias_origin}

It is important to check how the global evolution of stellar mass 
in our simulations (Section~\ref{massbias}) affects our conclusions 
regarding the physical origin of cosmic evolution of optical line
ratios in Section~\ref{relimpact}. We already demonstrated that the
predicted cosmic  evolution of \oiiihb, \niiha, \siiha\ and \oiha\
persists even when considering  galaxies (progenitors) in a fixed
stellar-mass range at all redshifts
(Fig.~\ref{Lineratios_mass}). However, SFR and global interstellar
metallicity, identified  above as primary drivers of the evolution of
nebular emission, correlate with stellar  mass in the simulations,
reflecting the galaxy mass-metallicity relation and SF main sequence
\citep[see the model predictions by][]{Hirschmann13, Hirschmann16}. In
fact, the thick dashed lines in Fig.~\ref{Propevol} show the average
redshift  evolution of the parameters controlling the SF, AGN and PAGB
nebular-emission models, when considering only galaxies in the fixed
stellar-mass range $10.5<\log (M_{\rm stellar}/\Msun)<11.0$,
independent of redshift. The evolutionary trends are similar to those
obtained when including the full  sample (thick solid lines), except
for the loss of redshift dependence of the average interstellar
metallicity (top left panel) and the average mass of post-AGB
stellar populations (second panel of bottom row). We also checked that
the results of Fig.~\ref{Lineratios_prop} do not change when
considering galaxies in a fixed mass range.

In Fig.~\ref{Evol_lineratios_cumu_massbin}, we show the analog of 
Fig.~\ref{Evol_lineratios_cumu} obtained when considering only
galaxies with  masses in the range $10.5<\log (M_{\rm
  stellar}/\Msun)<11.0$  at all redshifts. As anticipated in the
discussion of Fig.~\ref{Lineratios_mass} (Section~\ref{massbias}), the
predicted global evolution of optical-line ratios remains very strong
in this case (circles and solid lines in all panels). Moreover, the
influence of different physical quantities on this evolution is
qualitatively similar to that in Fig.~\ref{Evol_lineratios_cumu},
although the {\it relative} contributions by global interstellar
metallicity and SFR change significantly.  At fixed stellar mass,
global interstellar metallicity now has only a weak impact on the
cosmic evolution of \oiiihb\ and \niiha\ down to $z=0$ (red/beige
upside-down triangles and triple-dot-dashed lines), and SFR is the
primary driver of the evolution of any line ratio at $z>1.5$ (circles
and solid lines). This is because, at fixed stellar mass, SFR evolves
more strongly with redshift than global interstellar metallicity
(thick dashed lines in Fig.~\ref{Propevol}). The bottom two panels of
Fig.~\ref{Evol_lineratios_cumu_massbin} further show than nebular
emission from AGN and post-AGB stellar populations still contributes
significantly to the  increase in \siiha\ and \oiha\ at $z<1.5$. We
note that, in this case, the rising  influence of the PAGB
contribution at fixed stellar mass does not come from an  increase in
$M_\mathrm{PAGB}$ (Fig.~\ref{Propevol}), but simply from the  drop in
specific SFR.   

Remarkably, therefore, we find that identifying the physical origin of
cosmic evolution  of optical-line ratios depends sensitively on the
stellar-mass properties of the considered sample. When considering our
full set of simulated galaxies, for which mass evolves with redshift,
the associated evolution of global interstellar metallicity accounts
for most of the predicted evolution of line ratios. Instead, when
considering galaxies in a fixed stellar-mass range of $10.5<\log
(M_{\rm stellar}/\Msun)<11.0$,  global interstellar metallicity has a
weaker influence on the predicted evolution,  which is primarily
driven by SFR (through the ionization parameter), at all redshifts for
\oiiihb\ and \niiha\ and at $z>1.5$ for \siiha\ and \oiha. 

To summarise the above trends, we show graphically with arrows in
Fig.~\ref{SchematicSummary}  how redshift, \oiiihb, \niiha, \siiha\
and \oiha\ (different rows) change qualitatively as different
radiation and ISM properties increase (different columns), 
for the 20 simulated massive galaxies of Section~\ref{theory} 
and their main high-redshift progenitors. Grey cells refer to
properties varying with redshift, grey/white-gradient cells to
properties whose dependence on redshift vanishes when selecting a
fixed stellar-mass range, and white cells to properties independent of
redshift. Coloured arrows highlight those physical quantities expected
to drive the cosmic evolution of \oiiihb, \niiha, \siiha\ and \oiha,
according to our analysis of Figs~\ref{Evol_lineratios_cumu} and
\ref{Evol_lineratios_cumu_massbin}. 

\begin{figure}
\epsfig{file=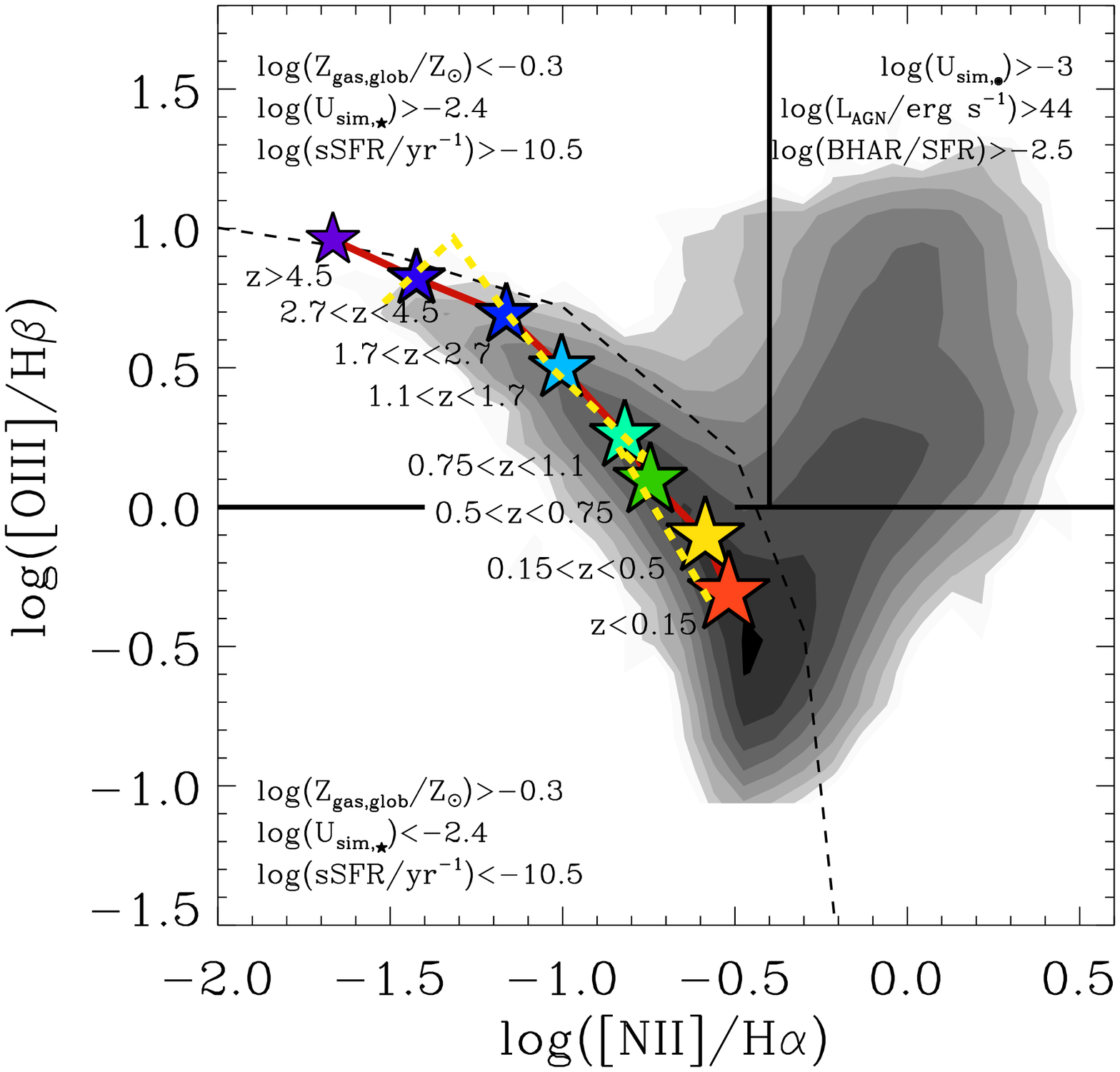, width=0.5\textwidth}
\caption{Redshift evolution in the \oiiihb\ versus \niiha\ diagram of
  the average location  of the {\it star-forming subset} [i.e., with
  $\log(\mathrm{BHAR/SFR})< -4$] of the 20  simulated massive galaxies
  of Section~\ref{theory} and their main high-redshift progenitors
  (stars colour-coded according to redshift, joined by the red
  line). The yellow dashed line shows the result obtained when
  including only galaxies in the   fixed stellar-mass range $10.5<\log
  (M_{\rm stellar}/\Msun)<11.0$. The reference SDSS   data (in grey)
  and black dashed line are the same as in Fig.~\ref{BPT_SFoffset}. SF
  galaxies in the top-left quadrant tend to have low global
  interstellar metallicities and high specific SFR (and ionization
  parameter), while those in the bottom half have on average high
  global interstellar metallicities and low specific SFR (ionization
  parameter).}\label{BPT_SF}         
\end{figure}

\subsection{Physical origin of the evolution of SF galaxies in the \oiiihb\ versus \niiha\ diagram} 

It is of interest to examine the evolution of the SF-galaxy population
predicted by our simulations in the \oiiihb\ versus \niiha\
diagram. This is shown in Fig.~\ref{BPT_SF}, where stars indicate the
average line ratios of all SF-dominated galaxies at different
redshifts (as indicated). The predicted evolution is primarily driven
by global interstellar metallicity and (specific) SFR (ionization
parameter).  SF galaxies in the top-left quadrant of Fig.~\ref{BPT_SF}
tend to have low global interstellar metallicities,
$\log(Z_\mathrm{gas,glob}/Z_\odot)<-0.2$, and high specific SFR,
$\log(\mathrm{sSFR/yr^{-1}})>10.5$, while those in the bottom half
have on average high global interstellar metallicities,
$\log(Z_\mathrm{gas,glob}/Z_\odot)>-0.2$, and  low specific SFR,
$\log(\mathrm{sSFR/yr^{-1}})<10.5$. The rise in \niiha\ with  redshift
is also influenced by an increase in BHAR/SFR ratio. Considering only
galaxies (progenitors) in a fixed stellar-mass range of $10.5<\log
(M_{\rm stellar}/\Msun)<11.0$ hardly affects the predicted evolution
in Fig.~\ref{BPT_SF}, as shown by the yellow dashed line. In this
case, global interstellar metallicity has a negligible influence on
the evolution,  which is mostly driven by the drop in (specific) SFR
from high to low redshift. It is worth pointing out that, if AGN
feedback is not included in the simulations,  the predicted star
formation histories are much flatter and do not show such significant
drop in (specific) SFR \citep[see][]{Choi16}. Hence, we find that AGN
feedback plays a key role  in the predicted \oiiihb\ evolution of
SF-dominated galaxies. We discuss this in more detail in
Appendix~\ref{AGNfeedback}. 

\section{Discussion}\label{discussion} 

In the last two sections, we demonstrated that the redshift evolution 
of optical emission-line ratios predicted by our simulations of
massive galaxies is widely consistent with available observations. We
investigated the physical origin of this evolution in terms of ISM and
ionizing-source parameters (as summarised by Fig.~\ref{SchematicSummary}). 
In this section, we discuss the potential influence
of other parameters, so far fixed at standard values in the
simulations (Section~\ref{model}), on these results. We also discuss
some caveats of our nebular-emission models and zoom-in
simulations. Finally, we put our results on the cosmic evolution of
emission-line ratios in the context of previous observational and
theoretical studies.    

\subsection{Influence of fixed model parameters}\label{fixparams} 

When coupling nebular-emission models with galaxy simulations in
Section~\ref{model}, we adopted standard values of parameters, such
as: the IMF upper mass cutoff,  $m_{\mathrm{up}}=100\,\Msun$; the
hydrogen density in \hii\ regions, $n_{\mathrm{H},
  \star}=100\,\mathrm{cm}^{-3}$; the slope of AGN ionizing radiation,
$\alpha =-1.7$; and the dust-to-metal mass ratio in ionized gas,
$\xi_\mathrm{d}=0.3$ (Table~\ref{Table_1}). We now investigate the
influence of these parameters of the predicted \oiiihb\ and \niiha\
ratios of simulated  galaxies (the influence on \siiha\ and \oiha\
being similar to that on \niiha). By analogy with  Fig.~\ref{BPT}, we
show in Fig.~\ref{freeparam} the locations of galaxies and their main
progenitors in the  \oiiihb\ versus \niiha\ diagram in three redshift
bins, $z=0$--0.5 (left column), 1--1.5 (middle column) and 2--3 (right
column). The top row shows the results obtained when using standard
values of all fixed parameters (identical to the first, third and
fifth panels in the left column of Fig.~\ref{BPT}). The other rows
show the results obtained when adopting $m_{\mathrm{up}}=300\,\Msun$
(second row),  $n_{\mathrm{H}, \star}=10^3\,\mathrm{cm}^{-3}$ (third
row), $\alpha =-1.2$ and $-2.0$ (fourth and fifth rows) and
$\xi_\mathrm{d}=0.1$ and 0.5 (sixth and  seventh rows). The average
SF-galaxy sequence in the top panels (yellow line) is reported in
other panels (yellow dashed line) for easy comparison with  the SF
sequence obtained using other parameters (red lines).  

\subsubsection{Upper mass cutoff of the IMF}

\begin{figure*}
\vspace{-0.5cm}
\epsfig{file=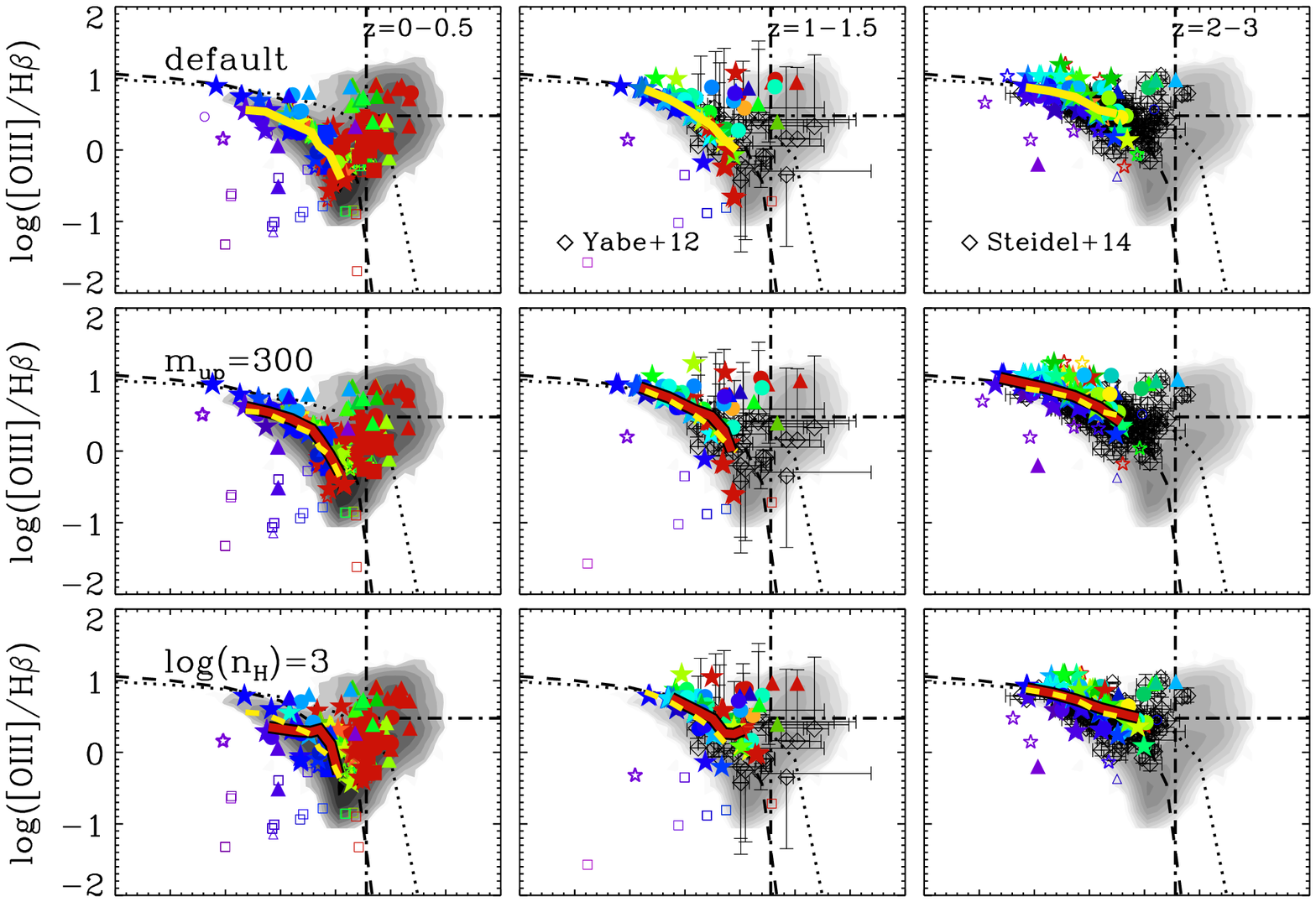,
  width=0.8\textwidth}\vspace{-1.3cm}\\
\epsfig{file=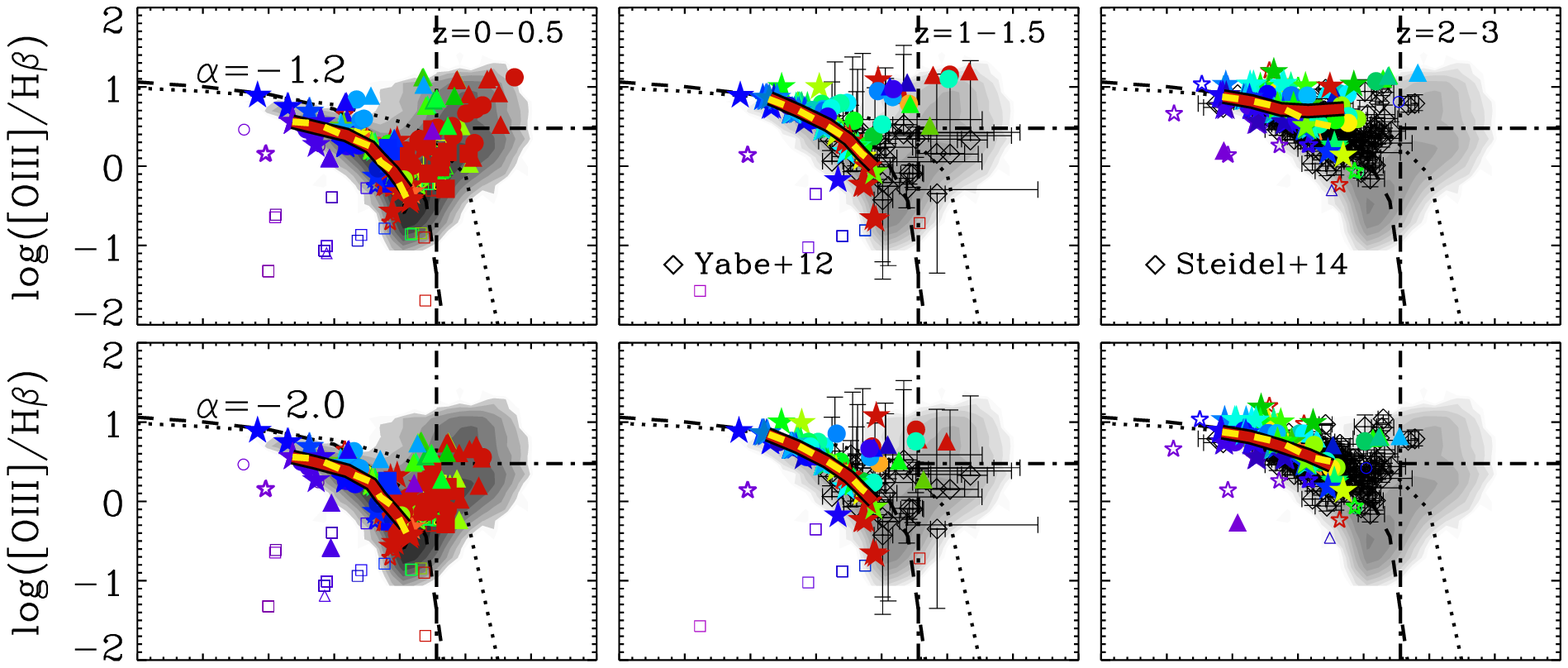,
  width=0.8\textwidth}\vspace{-1.3cm}\\
\epsfig{file=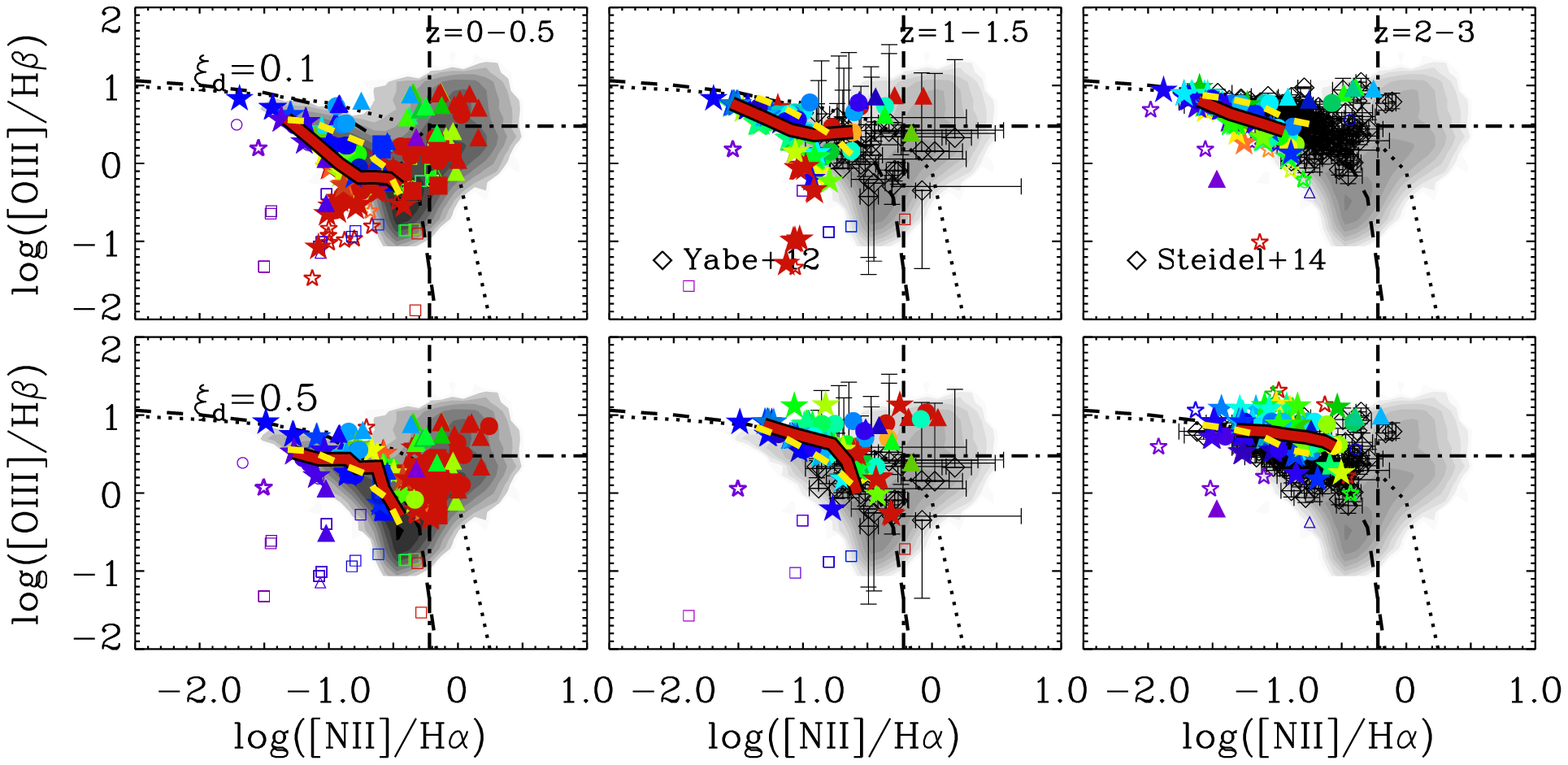,
  width=0.8\textwidth}
\vspace{-0.4cm}
\caption{\oiiihb\ versus \niiha\ diagram for the 20
   simulated massive galaxies of Section~\ref{theory} and their main high-redshift
   progenitors, in different redshift intervals (different columns), and for different 
   values of four parameters otherwise fixed by default (at $m_{\mathrm{up}} 
   =  100\,\Msun$, $n_{H, \star} = 10^2\,{\mathrm{cm}}^{-3}$, $\alpha = -1.7$ 
   and $\xi_\mathrm{d} = 0.3$, top row) in the simulations: adopting an upper
   IMF cutoff of $m_{\mathrm{up}} =  300\,\Msun$ (second row); an ionized-gas
   density of $n_{\mathrm{H}, \star}=10^3\,\mathrm{cm}^{-3}$ (third row), a slope
   of AGN ionizing radiation $\alpha =-1.2$ and $-2.0$ (fourth and fifth rows) and 
   a dust-to-metal mass ratio $\xi_\mathrm{d}=0.1$ and 0.5 (sixth and seventh rows).
    All symbols, lines and observational 
    data points have the same meaning as in Fig.~\ref{BPT}.
   The average SF-galaxy sequence in the top panels (yellow
line) is reported in other panels (yellow dashed line) for easy comparison with 
the SF sequence obtained using other parameters (red lines).}\label{freeparam}       
\end{figure*}

Increasing the upper mass cutoff of the IMF from $m_{\mathrm{up}}=100$  
to 300\,$\Msun$ leads to a harder ionzing spectrum of the stellar
populations, since stars with initial masses greater than 100\,$\Msun$
evolve at higher effective temperatures than lower-mass stars. As
shown by \citet{Gutkin16}, this causes only a slight increase in
\oiiihb\ and \niiha, illustrated by  the marginal difference between
the red and yellow lines in the second row of
Fig.~\ref{freeparam}. Thus, we do not expect that changes in
$m_{\mathrm{up}}$ over time would strongly affect the evolutionary
trends in these optical line ratios.   

\subsubsection{Hydrogen density in \hii\ regions}

As outlined in \citet[][see their section~3.3]{Gutkin16}, a rise in
the density of gas clumps in \hii\ regions from $n_{\mathrm{H},
  \star}=10^2$ to $10^3\,\mathrm{cm}^{-3}$ increases the probability
of collisional de-excitation over radiative cooling, expecially for
infrared fine-structure transitions, resulting in a slight increase of
cooling through optical transitions, and hence \oiiihb\ and
\niiha. The effect can be significant at large interstellar
metallicities, but vanishes at low metallicities \citep[fig.~5
of][]{Gutkin16}. This is illustrated by the small difference between
the red and yellow lines in the third row of Fig.~\ref{freeparam},
which decreases from low to  high redshift together with mean global
interstellar metallicity (Fig.~\ref{Propevol}).  We find a similarly
weak effect when increasing $n_{\mathrm{H}, \bullet}$ and
$n_{\mathrm{H}, \diamond}$ in the AGN and PAGB models. Hence,  even if
the ionized-gas density in high-redshift galaxies were typically
larger  than in local ones, as is plausible according to some
observational  studies \citealp[e.g.,][]{Steidel14,Sanders16,
  Kashino17}), our conclusions  about the main drivers of the cosmic
evolution of optical-line ratios would  not be altered.   

\subsubsection{Hardness of AGN ionizing radiation}

Stellar ionizing radiation in our models is controlled by the physical 
properties  of simulated galaxies. For SF models, the ionizing
radiation depends on the properties of young stars, whose metallicity
is assumed to be the same as that of star-forming gas (as justified in
Section~\ref{emlines_sf}). The impact of  stellar metallicity on
optical-line ratios is in any case negligible compared to that of
interstellar metallicity \citep[e.g.][]{Gutkin16}. For PAGB models,
the age  and metallicity of old, post-AGB stellar populations
predicted by our simulations define unambiguously the corresponding
ionizing radiation. In contrast, we set the hardness of AGN ionizing
radiation via a fixed power-law index $\alpha=-1.7$
(Section~\ref{agnmatch}). A larger (lower) $\alpha$ would imply a
harder (weaker)  spectrum, resulting in enhanced (reduced) \oiiihb\
and \niiha\  \citep{Feltre16}. This  is confirmed by
Fig.~\ref{freeparam}, which shows at all redshifts a marked decline in
these ratios for AGN and composite galaxies as $\alpha$ drops from
$-1.2$ (fourth row)  to $-2.0$ (fifth row). Fixing $\alpha$ at a value
different from $-1.7$ in the AGN models  would therefore affect the
emission-line properties of AGN and composite galaxies, but  not their
cosmic evolution.  Current observations provide hardly any constraints 
on the dependence of $\alpha$ on redshift and/or BH accretion rate and suggest 
simply a range of plausible values between $-1.2$ and $-2.0$ 
\citep[e.g.,][]{Feltre16}. Even if $\alpha$ dropped systematically from
high to low redshift, this would induce a corresponding drop in
\oiiihb, but also in \niiha, which would be unlikely to  contribute to
the observed evolutionary trends in Fig.~\ref{Evol_lineratios}.

\subsubsection{Dust-to-Metal mass ratio (and relation to N/O)}\label{noratio}

Increasing the dust-to-metal mass ratio depletes metal coolants from  
the gas phase. The electronic temperature rises, as does cooling
through collisionally excited optical transitions. This implies a
strong rise in \niiha\ and a less strong rise in \oiiihb,  since
oxygen is a refractory element strongly depleted from the gas phase.
This is illustrated by the fifth and sixth rows of
Fig.~\ref{freeparam}, showing predictions for $\xi_\mathrm{d}=0.1$ and
0.5,  and where the average SF galaxy sequence (red line) lies,
respectively, below  and above that for $\xi_\mathrm{d}=0.3$ (yellow
line).  A  systematic drop in $\xi_\mathrm{d}$ from high to low
redshift would therefore induce a corresponding drop in
\oiiihb. However, this would be accompanied by an even stronger drop
in  \niiha, which argues against a significant influence of the
dust-to-metal mass ratio on the observed evolutionary trends in
Fig.~\ref{Evol_lineratios}. Moreover, this would go against recent support
for a lower dust-to-metal mass ratio in metal-poor compared to metal-rich
galaxies \citep{RemyRuyer14}, also predicted by some theoretical studies
\citep{Inoue03, Asano13}. 

It is worth mentioning that, O being a refractory element and N not, 
our models account naturally for the rise in {\em gas-phase} N/O 
abundance ratio, (N/O)$_\mathrm{gas}$, implied by a rise in
$\xi_\mathrm{d}$  \citep[see fig.~1 of][]{Gutkin16}.\footnote{ 
A rise in (N/O)$_\mathrm{gas}$ in our models
could also  be induced by a change in C/O ratio \citep[fig.~1 of][]{Gutkin16}.
However, this is unlikely for the simulations of massive galaxies 
presented here, which show hardly any evolution of C/O over cosmic
time (Fig.~\ref{Propevol}). We note that other scenarios 
not considered in this work, such as chemically differential
  galactic winds \citep{Vincenzo16} or changes in the nitrogen yield,
  may also influence (N/O)$_\mathrm{gas}$.}
An elevated (N/O)$_\mathrm{gas}$ in star-forming
galaxies at high redshift has been suggested by some observational
studies \citep[e.g.][see Section~\ref{obstudies}
below]{Shapley15}. Our results indicate that, while this could
contribute for a modest part to the observed drop in \oiiihb\ from
high to low redshift (by less than 0.2~dex, according to the
difference between the red and yellow lines in Fig.~\ref{freeparam}),
most of the trend in \oiiihb, and all of that in \niiha, are likely to
be dominated by the drop in (specific) SFR and rise in  global
interstellar metallicity (which make \oiiihb\ decline by
$\sim$0.8~dex; see Fig.~\ref{Evol_lineratios_cumu_massbin}).  

Further support for a negligible role of the dust-to-metal mass
ratio (and hence N/O in our models) in the cosmic
evolution of \oiiihb\ comes  from recent observations
\citep[e.g.,][]{RemyRuyer14,DeCia16,Wiseman17}  and predictions from a
semi-analytic model accounting for dust formation  
and destruction \citep[][]{Popping16}. Both types of studies suggest that, 
at given gas metallicity, the dust-to-metal mass ratio in massive galaxies
hardly changes with redshift out to $z=6$, the models further predicting that this
is also the case at fixed stellar mass \citep[see figs~5 and 6 of][]{Popping16}. In 
these models, galaxies more massive than $\log(M_{\rm stellar}/\Msun)\sim9.5$ 
are predicted to have dust-to-metal mass ratios around 0.2, reasonably close to 
the typical observed values of 0.3--0.4 in present-day galaxies (with a scatter 
sampling the full range from 0.1 to 1.0). This reinforces our choice of 
adopting a fixed $\xi_\mathrm{d} = 0.3$ at all redshifts in our
simulations (Table~\ref{Table_1}).   

\subsection{Caveats}\label{caveats} 

\subsubsection{Nebular-emission models: shocks, leakage of ionizing photons and dusty tori}

\begin{itemize}
\item{\em Shock models:}
The SF, AGN and PAGB nebular-emission models used in this paper 
include several important features allowing accurate comparisons with 
observations, such as improved prescriptions for stellar interiors and
atmospheres, the self-consistent treatment of depletion of metals onto 
dust grains and the inclusion of secondary nitrogen production. This
results in robust predictions of emission-line fluxes for given
combinations of adjustable model parameters. In contrast to other
theoretical studies  \citep{Orsi14,Shimizu16}, we also account
self-consistently for nebular emission from AGN and post-AGB stars, in
addition to that from young stars. Yet, we do not include nebular
emission from radiative shocks potentially produced by, e.g.,
starburst- and AGN-driven outflows and cloud-cloud  collisions in
galaxy interactions \citep[e.g.,][]{Sharp10, Rich10, Rich11,Soto12,
  Weistrop12}.  
  
Since many observations indicate that galactic winds exist in a
significant fraction of galaxies out to high redshifts
\citep[e.g.,][]{Kornei12, Steidel10, Genzel11, Newman12}, shock
excitation may be an important contributor to optical-line ratios in
high-redshift galaxies. Theoretically, shock models were developed by
\citet{Dopita03}, \citet[fast shocks]{Allen08} and \citet[slow
shocks]{Rich10, Rich11}. Based on these models, \citet[their
fig.~6]{Kewley13} show that metallicity and shock velocity are the
primary parameters defining the location of pure-shock models  in the
\oiiihb\ versus \niiha\ diagram. In fact, shock models behave very
similarly to AGN models, with \niiha\ strongly increasing with
increasing metallicity at roughly constant \oiiihb. For shock models
of any velocity and metallicity, the results of \citet{Kewley13}
indicate that \oiiihb\ is  always predicted to be larger than unity
($\log\oiiihb>0$), which could,  when assuming that shocks primarily
arise in high-redshift galaxies,  contribute to the cosmic drop in
\oiiihb\ from high to low redshift.  

However, reality is more complex: shocks are generally not
  expected to dominate line emission from a galaxy and are usually
  associated with star-formation and AGN activity \citep[see,
  e.g.,][]{Kewley13}. Exploring self-consistently the contribution
from  shocks to the cosmic evolution of emission-line ratios would
therefore require that we identify shocks in our simulations and
couple their  properties (velocity, metallicity) with, for example,
the fast- and slow-shock models of \citet{Allen08} and \citet{Rich10,
  Rich11}. We postpone such an analysis to a future study.

\item{\em Leakage of ionizing photons:}
All SF, AGN and PAGB models adopted in this paper are
  ionization bounded, i.e., they do not include any
 leakage of ionizing photons into the circumgalactic medium. The escape 
 of ionizing radiation from high-redshift galaxies is a heavily debated issue, 
 with so far, at redshifts around 3, only few galaxies identified as 
genuine ionizing-photon leakers \citep{deBarros16, Shapley16,Bian17}. 
The escape of substantial ionizing radiation from density-bounded \hii\ 
regions is expected to decrease the emission contribution of
low-ionization species, affecting emission-line ratios of high-to-low 
ionization species, such as \oiii/\oii\ \citep[e.g.,][and references therein]{Jaskot13}.
Recent calculations by \citet[][their fig.~6]{Izotov17} further suggest that \oiiihb\ 
is hardly affected by the escape of ionizing photons from density-bounded
\hii\ regions. We might expect \niiha, which involves a lower-ionization line, to 
be more strongly affected. A detailed analysis of this issue is
  postponed to future work.   

\item{\em Dust obscuration on torus level:}
The AGN nebular-emission models used in this paper neglect the 
potential influence of a dust torus on emission-line luminosities. In
principle, accounting for the presence of a dust torus would reduce 
the amount of ionising photons emitted by the central AGN capable 
of reaching the narrow-line region, hence lowering all emission-line 
luminosities by a same factor (corresponding to the fraction of solid
angle occupied by the torus). For an AGN-dominated galaxy, the
predicted integrated line ratios would not change, but in galaxies with 
substantial SF and PAGB components, the relative 
contribution to these ratios by the AGN would be altered. Thus, some
galaxies identified as composite systems in the simulations could 
instead move to the SF and LINER regions in the
\oiiihb\ versus \niiha\ diagram (Figs~\ref{BPT} and \ref{freeparam}).
We can speculate that such changes might become more significant toward
high redshift, where torus obscuration could become more relevant
\citep[e.g.][]{Hasinger08}. While worthwhile, a more detailed investigation
of this effect is beyond the scope of the present study.
\end{itemize}

\subsubsection{Cosmological zoom-in simulations: statistical completeness}

A potential source of inaccuracy of all current cosmological
simulations is related to the necessary assumption of often rather
simplified sub-resolution models for baryonic
processes, which cannot accurately capture the properties of a
  multi-phase ISM \citep[see, e.g.,][]{Naab16}. Specifically, different
models for stellar \citep[e.g.,][]{Guedes11, Stinson13, Hirschmann13,
  Hirschmann16, Hopkins13} and AGN feedback
\citep[e.g.,][]{Dubois13, Steinborn15, Steinborn16, Choi16,
  Weinberger17} have been shown to be capable of
significantly affecting various galaxy properties at all times
. In this context, it is important to
stress that the  simulations presented in this paper rely on
physically motivated models for  stellar and AGN feedback, constrained
in part by small-scale simulations of the  ISM \citep[see
Section~\ref{simulations} above and][Hirschmann et al., in
preparation]{Choi16}. This provides fairly realistic massive  galaxies
in terms of chemical enrichment, SFR and BHAR histories, stellar
populations, etc. Specifically, we checked that the predictions of 
our simulations are widely consistent with observed relations
between SFR, gas-phase metallicity and galaxy stellar mass at redshifts
out to $z=2$ \citep[using observational constraints
from][]{Maiolino08, Maier15, Andrews12, Daddi07, Elbaz07}. This
success puts our predictions for emission-line   ratios -- controlled
by these physical quantities -- on firm grounds. 

A drawback of our suite of cosmological zoom-in simulations is
their sparse statistics,  since we analyse only 20 massive galaxies,
whose most massive progenitors we follow back in time. As a
result, our sample may be missing a contribution by less massive galaxies
with emission-line fluxes potentially above the adopted flux-detection limit.
On the one hand, this implies an intrinsic stellar-mass evolution of  simulated
galaxies, whose effect we addressed in Sections~\ref{massbias} and
\ref{massbias_origin}. On the other hand, the fraction of
AGN/composite galaxies in our limited set of zoom-in simulations may 
not be realistic and comparable to that of observationally sampled
galaxy populations. The fraction of AGN,  and hence, their relevance
to the cosmic evolution of emission-line ratios, may change when
considering a statistically complete population of galaxies and
AGN. This would affect more \niiha, \siiha\ and \oiha\ than \oiiihb,
which is insensitive to nebular emission  from AGN
(Figs~\ref{Evol_lineratios_cumu} and
\ref{Evol_lineratios_cumu_massbin}). Increasing our set of zoom-in
simulations or appealing to statistically complete samples of galaxy
simulated using semi-analytic models and large cosmological boxes
could help us improve this limitation, but lies beyond the scope of
the present study.  

\subsection{Comparison with previous explanations for the cosmic evolution of \oiiihb}\label{comparison}  

In Section~\ref{origin}, our combination of zoom-in galaxy simulations
with versatile  nebular-emission models allowed us to show that a
decline in (specific) SFR (ionization parameter) and a rise in global
interstellar metallicity are the main potential drivers of the
observed drop in \oiiihb\ of star-forming galaxies from high to low
redshifts. At fixed stellar mass, the  higher \oiiihb\ of distant
galaxies relative to local ones is primarily attributable to an
elevated (specific) SFR. In contrast, a larger IMF upper-mass cutoff,
higher ionized-gas density, harder ionizing radiation and larger
depletion of metals on to dust grains (and  associated elevated
gas-phase N/O ratio) are not expected to play any significant role in
the cosmic evolution of optical-line ratios. A number of different,
and sometimes contradictory explanations have been proposed in
previous, mostly observational  studies. We now replace our model
predictions in the context of these studies. 

\subsubsection{Observational studies}\label{obstudies}

Different observational investigations have favoured different
physical origins  of the cosmic evolution of \oiiihb, no unique
conclusion having been  drawn so far. Possible explanations include
 the prevalence at high redshift of increased SFR
\citep[e.g.,][]{Kashino17}, higher hydrogen/electron densities
\citep[e.g.,][]{Brinchmann08, Lehnert09}, increased  contribution by an
AGN to nebular emission \citep[e.g.,][]{Wright10}, elevated
(N/O)$_\mathrm{gas}$  ratio \citep[e.g.,][]{Shapley15}  and harder
stellar ionizing radiation \citep[e.g.,][]{Steidel14, Strom17} compared to
low redshift. We now examine the conclusions from these studies in
light of our theoretical predictions. 

\begin{itemize}
\item{\it Gas-phase metallicity:}
based on a sample of 701 star-forming galaxies at $z=1.4$--1.7 from  
the Fiber Multi-Object Spectrograph (FMOS)-COSMOS survey,
\citet{Kashino17} find that the empirically determined gas-phase
metallicities of galaxies with masses $\log(M_{\rm
  stellar}/\Msun)\ga11$ are similar to those of local SDSS
counterparts, while less massive galaxies exhibit a rise in gas-phase
metallicity from high to low redshift.  \citet{Kashino17} show
that gas-phase metallicity {\em can} affect \oiiihb\ and 
\niiha, but they further note that a rise in metallicity at
fixed ionization parameter cannot account for the observed cosmic
evolution of \oiiihb\ and \niiha. They conclude that a change in
ionization parameter is likely to be the primary  cause of this
evolution (their fig.~12 and section~4.1). These results are
consistent with our conclusion in Section~\ref{massbias_origin} that,
at fixed stellar mass in the range $10.5<\log (M_{\rm
  stellar}/\Msun)<11.0$, the cosmic evolution  of optical-line ratios
is primarily driven by  (specific) SFR -- i.e., ionization  parameter
-- rather than global interstellar metallicity.  

\item{\it Ionization parameter and star formation rate:}
\citet{Brinchmann08}, \citet[from the analysis of 251 star-forming
galaxies at $z \sim 2.3$ with the Keck/MOSFIRE
spectrometer]{Steidel14}, \citet[from the analysis of 118 star-foming
galaxies at $z\sim1.5$ with the Subaru Fiber Multi Object
Spectrograph]{Hayashi15} and  \citet[see above]{Kashino17} attribute
at least part of the enhanced \oiiihb\  ratio of high-redshift
galaxies relative to low-redshift ones to a higher  ionization
parameter \citep[see also,][]{Cullen16}. So far, no consensus has been
reached on whether  this 
quantity plays a primary \citep{Kashino17} or only minor
\citep{Steidel14} role. \citet{Hayashi15} and  \citet{Kashino17}
discuss different possible origins of an elevated ionization parameter
(higher star-formation efficiency, top-heavy IMF, harder ionizing
spectra, etc.)  without drawing any final conclusion. Our results in
Section~\ref{origin},  based on the self-consistent modelling of
galaxy formation and nebular  emission, confirm the importance of an
evolving ionization parameter in the cosmic evolution of \oiiihb,
especially at fixed galaxy stellar mass. In these models, (specific)
SFR is the main driver of the change in ionizing parameter over cosmic
time. 

\item{\it Hydrogen/electron density in \hii\ regions:}
densities larger than in typical local \hii\ regions have
been repeatedly measured in high-redshift star-forming galaxies and
put forward as a possible cause for enhanced \oiiihb\
\citep[e.g.][]{Brinchmann08,  Lehnert09,Steidel14,Sanders16}. Yet,
most observational studies tend to  agree that changes in ionized-gas
density alone cannot account for the cosmic evolution of \oiiihb\
\citep[e.g.,][]{Rigby11,Hayashi15,Kashino17,Strom17}. Specifically,
based on the analysis of 380 star-forming galaxies at $z=2$--3 with
the  Keck/MOSFIRE spectrometer, \citet{Strom17} argue that
measurements  of the density-sensitive [O{\sc
  II}]$\lambda\lambda3727,3729$ doublet for  galaxies at small and
large offsets from the local SF sequence in the  \oiiihb\ versus
\niiha\ diagram are nearly identical. This observational  conclusion
is consistent with our theoretical finding that changes in
$n_{\mathrm{H}, \star}$ hardly affect the cosmic evolution of these
optical-line ratios (third row in Fig.~\ref{freeparam}).

\item{\it Additional contribution from (weak) AGN:}
most recent observational studies rule out enhanced contribution  
by a central accreting BH as the main driver for the cosmic evolution
of  \oiiihb, since the high-redshift star-forming galaxies showing
this evolution do not exhibit any spectral signature of an AGN
\citep[such as strong  high-ionization
lines;][]{Steidel14,Strom17,Kashino17}. This is consistent  with our
theoretical prediction that the cosmic evolution of \oiiihb\ in
star-forming galaxies (and composite and AGN galaxies) is not driven
by nebular emission from an AGN, which can cause at most a slight
offset  at any redshifts (Figs~\ref{Evol_lineratios_cumu} and 
\ref{Evol_lineratios_cumu_massbin}).

\item{\it Enhanced (N/O)$_\mathrm{gas}$ ratio at fixed (O/H)$_\mathrm{gas}$ ratio:}
some studies favour an enhanced (N/O)$_\mathrm{gas}$ ratio at fixed  
(O/H)$_\mathrm{gas}$ as the primary driver of the cosmic evolution of 
\oiiihb\ and \niiha\ \citep{Masters14,Shapley15,Yabe15, Cowie16,
  Masters16, Sanders16}.\footnote{These studies find hardly any
  evidence at high redshift for harder ionizing radiation and enhanced
  ionization parameter  at fixed metallicity.} In fact, these studies
attribute the offset of the  `SF branch' in the line-ratio diagram at
high redshift to a rise in \niiha\ at fixed \oiiihb,  rather than to a
rise in \oiiihb\ at fixed \niiha. This interpretation is consistent
with the absence of any offset between high- and low-redshift galaxies
in the \oiiihb\ versus \siiha\ diagram noted by \citet[from the
analysis of 133 star-forming galaxies at $z \sim 2.3$ with the
Keck/MOSFIRE  spectrometer]{Shapley15}, who also find that only
low-mass galaxies,  with $\log(M_{\rm stellar}/\Msun)<10$, exhibit
higher \niiha\ at high redshift  than at low redshift. The fact that
(N/O)$_\mathrm{gas}$ could be enhanced at fixed (O/H)$_\mathrm{gas}$
at high redshift has been challenged by  \citet{Strom17}, who find
that the relation between these two abundance ratios is redshift
invariant. Moreover, \citet{Kashino17} {\em do} find an offset in
\oiiihb\ at fixed \siiha\ in their high-redshift sample relative to
local galaxies.  Our theoretical predictions are more in line with
these recent studies, since  we also find that, at fixed \siiha,
massive galaxies (with $M_{\rm stellar} >3 \times 10^{10} \Msun$) have
on average larger \oiiihb\ at high redshift than at low redshift
(Fig.~\ref{BPT}). Even if enhanced (N/O)$_\mathrm{gas}$ at fixed
(O/H)$_\mathrm{gas}$ at high redshift (corresponding to higher
dust-to-metal mass ratio in our models) can contribute to the cosmic
evolution of  \oiiihb\ and \niiha, our simulations strongly disfavour
this as the dominant process for the offset of the SF branch in this
line-ratio diagram at high redshift (Section~\ref{noratio} and bottom
row of Fig. 9),  at least for massive galaxies. For less massive
galaxies (with $M_{\rm stellar} <3 \times 10^{10} \Msun$), we cannot
draw any robust conclusion on this point.

\item{\it Harder stellar ionizing radiation:}
\citet{Steidel14,Steidel16} and \citet{Strom17} favour harder 
stellar ionizing radiation (presumably from metal-poor, massive
binaries)  as the main driver of the higher typical \oiiihb\ ratio of
high-redshift star-forming galaxies relative to local ones. Since the
\citet{Gutkin16} SF models used in our analysis do not include any
prescription for metal-poor massive binary stars, we cannot draw any
robust conclusion regarding the importance of this particular stellar
component. Interestingly, \citet{Gutkin16} reproduce remarkably 
well the observed ultraviolet and optical emission-line properties of
the composite  \citet{Steidel16} spectrum  while ignoring any
extra component of metal-poor,  massive binary
stars.\footnote{\citet{Gutkin16} did not consider the observed
  far-ultraviolet stellar-emission continuum spectrum, which
  \citet{Steidel16} used to constrain their models, in addition to
  emission line properties.} They 
also show that stellar metallicity, which controls the  hardness of
stellar ionizing radiation, has only a minor influence on \oiiihb\
relative to gas-phase metallicity. The fact that at fixed stellar
mass, global interstellar metallicity (which is coupled to stellar
metallicity in our approach) has only a negligible impact on \oiiihb\ 
(Section~\ref{massbias_origin}) suggests that, in our analysis, harder 
stellar ionizing radiation is unlikely to play a major role in the
cosmic evolution  of \oiiihb. 

\end{itemize}

In summary, our theoretical explanation for the cosmic evolution of
\oiiihb\ is closest to that recently proposed by \citet{Kashino17},
who favour the ionization  parameter as the primary driver of this
evolution (a minor part of which could  come from harder ionizing
radiation at high redshift), with potential additional  contributions
by interstellar metallicity and ionized-gas density. 

\subsubsection{Theoretical studies}

So far, only \citet{Kewley13} investigated theoretically possible
reasons for an  evolving \oiiihb\ ratio using galaxy properties
extracted from cosmological  hydrodynamic simulations. However, these
authors focused exclusively on chemical enrichment histories extracted
from simulations, which they  injected into nebular-emission models,
exploring independently the space  of other ISM parameters (hydrogen
density, ionization parameter) and an  AGN component (central gas
metallicity, ionization parameter). The results  of \citet{Kewley13}
indicate that the location of the star-forming sequence in the
\oiiihb\ versus \niiha\ diagram at any redshift depends mainly on ISM
conditions. Specifically, `extreme' ISM conditions at high redshift,
driven by (a combination of) higher ionization parameter, harder
ionizing radiation and higher electron density, can shift the sequence
toward larger \oiiihb\ than at  low redshift. The location of
composite and AGN galaxies also strongly depends on the ISM conditions
adopted for the narrow-line region, in particular the inner gas
metallicity. The exploratory approach adopted by \citet{Kewley13}
does not allow them to draw any conclusion on the {\it relative}
influence of  these different physical quantities on the cosmic
evolution of \oiiihb\ and \niiha.  

Our simulation results are consistent with the finding by
\citet{Kewley13} that ISM conditions can strongly affect the cosmic
evolution of \oiiihb\ at fixed stellar  mass. Our results further
suggest that (specific) SFR -- which controls the ionisation
parameter -- is the primary driver of this evolution, with potential
minor contributions by hydrogen density and the hardness of stellar
ionizing radiation.  

\section{Summary}\label{summary} 

In the previous sections, we have investigate theoretically the
physical origin of  the observed cosmic evolution of optical
emission-line ratios in galaxies, employing  for the first time a
self-consistent modelling approach.  
 
%

Specifically, we compute synthetic  \oiiihb, \niiha, \siiha\ and
\oiha\ ratios for galaxies in a cosmological framework, by coupling --
in post-processing --  newly developed spectral-evolution models,
based on photoionization calculations, with a set of 20
high-resolution cosmological zoom-in simulations of massive
galaxies. The latter are performed with the code SPHGal, a modified
version of Gadget3, including sophisticated prescriptions for star
formation, chemical enrichment \citep{Aumer13}, stellar feedback
\citep{Nunez17}, black-hole growth and AGN feedback \citep{Choi16}. 

We include nebular emission from young stars \citep{Gutkin16}, AGN  
\citep{Feltre16} and post-AGB stars (Section~\ref{emlines_pagb}). We
adopt direct predictions from our simulations for the redshift
evolution of  global and central interstellar metallicity, C/O
abundance ratio, star formation  rate, black-hole accretion rate,
global and central average gas densities, and the age and metallicity
of post-AGB stellar populations. Based on these, we select SF, AGN and
post-AGB nebular-emission models for each galaxy  and its most massive
progenitor at any redshift. By default, we adopt  fixed dust-to-metal
mass ratio, ionized-gas hydrogen/electron density and  power-law index
of AGN ionizing radiation.  

We can summarize our main results as follows:\vspace{-0.2cm}
%
\begin{itemize}
\item The synthetic  \oiiihb, \niiha, \siiha\ and \oiha\ emission-line ratios 
predicted by our simulations are in excellent agreement with observations 
of both star-forming and active SDSS galaxies in the local universe.
\item Toward higher redshifts, at fixed galaxy stellar mass, \oiiihb\ is 
predicted to increase and \niiha, \siiha\ and \oiha\ to decrease. These
evolutionary trends are consistent with observations by \citet{Yabe12}
and \citet{Steidel14}.
\item The physical origin of the cosmic evolution \oiiihb, \niiha, \siiha\
and \oiha\ is a complex mix of different evolving ISM and ionizing-radiation 
properties governing the nebular emission from young stars, AGN and 
post-AGB stars. 
\item When considering the entire sample of simulated galaxies and 
their main progenitors, interstellar metallicity appears to be a main driver
of the cosmic evolution of optical-line ratios, along with (specific) SFR,
which controls the ionization parameter. This dominant role of metallicity
arises primarily from the intrinsic stellar-mass evolution of simulated galaxies,
combined with the correlation between mass and metallicity.
Instead, at fixed stellar mass, interstellar metallicity evolves only weakly with 
redshift and has a negligible influence on the evolution of optical-line
ratios.
\item At fixed stellar mass, the drop in \oiiihb\ from high to low redshift 
in our simulations is driven primarily by that in (specific) SFR, via the ionization
parameter. Nebular emission from the growing population of post-AGB
stars can also play a minor role in the cosmic evolution of \oiiihb, but not 
that from accreting black holes.
\item	AGN feedback appears to play a key role in the predicted cosmic
evolution of \oiiihb, as test simulations not including AGN feedback 
exhibit much flatter star formation histories and hardly any redshift 
evolution of \oiiihb.
\item At fixed stellar mass, the rise in \niiha\ from high to low redshift
follows primarily from the decline in SFR (and ionization parameter),
which reduces the probability of multiply ionizing nitrogen at the expense 
of N$^+$. At redshift $z<1$, a drop in average central gas density and 
rise in BHAR/SFR ratio make the contribution by AGN emission 
contribute more significantly to the cosmic evolution of \niiha. Nebular
emission from post-AGB stellar populations hardly affects this evolution.
\item At fixed stellar mass, the rise in \siiha\ and \oiha\ from high to 
intermediate redshift ($z\sim1.5$) also follow from the drop in SFR (and
ionization parameter). The continued rise at lower redshift is driven
in roughly equal parts by nebular emission from AGN and post-AGB
stellar populations.
\item Applying observational flux limits to our sample of simulated 
galaxies and their main progenitors indicates that evolution effects 
are likely to dominate over flux-selection effects in determining
the cosmic evolution of optical-line ratios (although this result might 
change for a cosmologically representative sample). The ability with our approach 
to draw such a conclusion is particularly noteworthy, given the difficulty 
in disentangling these competing effects in emission-line studies of distant 
galaxies \citep{Juneau14}.  
\item We have checked that the dust-to-metal mass ratio, ionized-gas 
hydrogen/electron density and power-law index of AGN ionizing radiation, 
which are fixed by default in the nebular-emission models, have only a
minor (or even negligible) influence on the cosmic evolution of \oiiihb,
\niiha, \siiha\ and \oiha. Based on our investigation, we can speculate 
that adopting a harder ionizing radiation, higher ionized-gas density or
higher dust-to-metal mass ratio [which implies higher (N/O)$_\mathrm{gas}$
in our self-consistent modelling of metals and their depletion
on to dust grains] in high-redshift galaxies relative to local ones
may strengthen the redshift dependence of \oiiihb, but not account for
the bulk of the evolution.
\end{itemize}

The theoretical results presented in this paper provide useful
insight into the physical origin of observed cosmic evolution of
optical-line ratios. Nevertheless, it is important to keep in mind
the sparse statistics of our sample of 20 simulated massive 
galaxies and their main progenitors, which is likely to affect, for 
example, the predicted fraction of AGN-dominated galaxies
at any given cosmic epoch, and thus, the AGN contribution to the 
redshift dependence of \niiha, \siiha\ and \oiha\ (but not
\oiiihb).  In addition, our sample does not include galaxies
with present-day low masses, for which conclusions may be
  different. The contribution by radiative shocks to the cosmic
evolution of  optical-line ratios, which we neglected in this study,
must also be  quantified in detail in future work. This paper is the
first in a series. In follow-up studies, we plan to investigate
ultraviolet-line  diagnostics to help characterise the nature of
ionizing radiation  in very distant galaxies observed through
near-infrared spectroscopy.  We also plan to explore the contribution
by different ionizing sources to nebular emission in different regions
of a galaxy,  producing spatially resolved emission-line maps to
improve the interpretation of modern integral-field spectroscopic
observations in terms of galaxy physical parameters.

\section*{Acknowledgements}

 We thank the referee, A. Inoue, for carefully reading and providing
helpful comments on our manuscript. We also 
thank St\'ephanie Juneau and Emma Curtis-Lake for helpful
  advice  and Dan Stark, Bodo Ziegler, Christian Maier and the NEOGAL
  team  for fruitful discussions. MH, SC and AF acknowledge financial
  support from the European Research Council (ERC) via an Advanced
  Grant under grant agreement no.\,321323--NEOGAL.  AF acknowledges
  support from the ERC via an Advanced Grant under grant  agreement
  no.\,339659--MUSICOS. TN acknowledges support from the DFG priority
  program 1573 `Physics of the interstellar medium' from the DFG
  Cluster of Excellence `Origin and structure of the Universe'. RSS
  is grateful for the generous support of  the Downsbrough family, and
  acknowledges support from the Simons Foundation  through a Simons
  Investigator grant.

\bibliographystyle{mn2e}
\bibliography{Literaturdatenbank}


\begin{appendix}
\section{The role of AGN feedback in the cosmic evolution of \oiiihb\ }\label{AGNfeedback}

\begin{figure*}
\centering
\centering{\large Without AGN feedback}
\epsfig{file=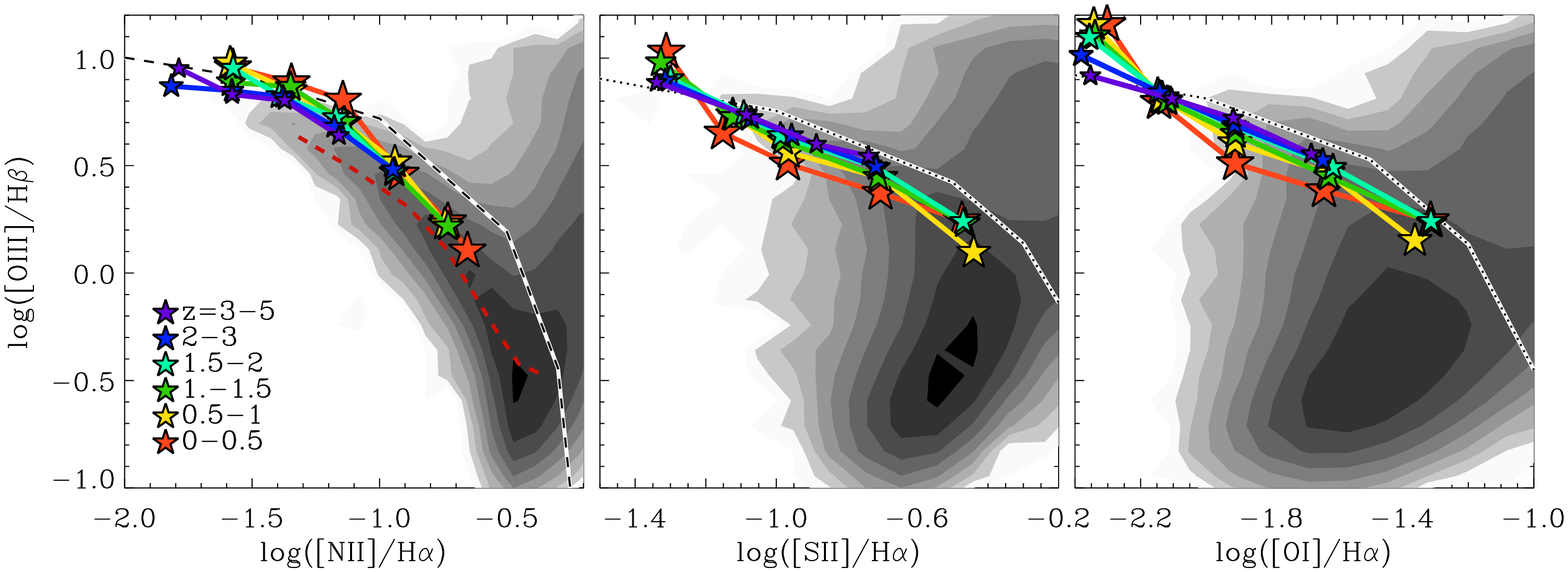,
  width=1.0\textwidth} 
 \caption{Same as Fig. \ref{BPT_SFoffset}, but for
 the simulation set of galaxies {\em without} AGN feedback, as described in Appendix~\ref{AGNfeedback}.}
 \label{BPT_SFoffset_NoAGN}      
\end{figure*}

\begin{figure*}
\centering{\large Physical quantities for SF models without AGN feedback}
\epsfig{file=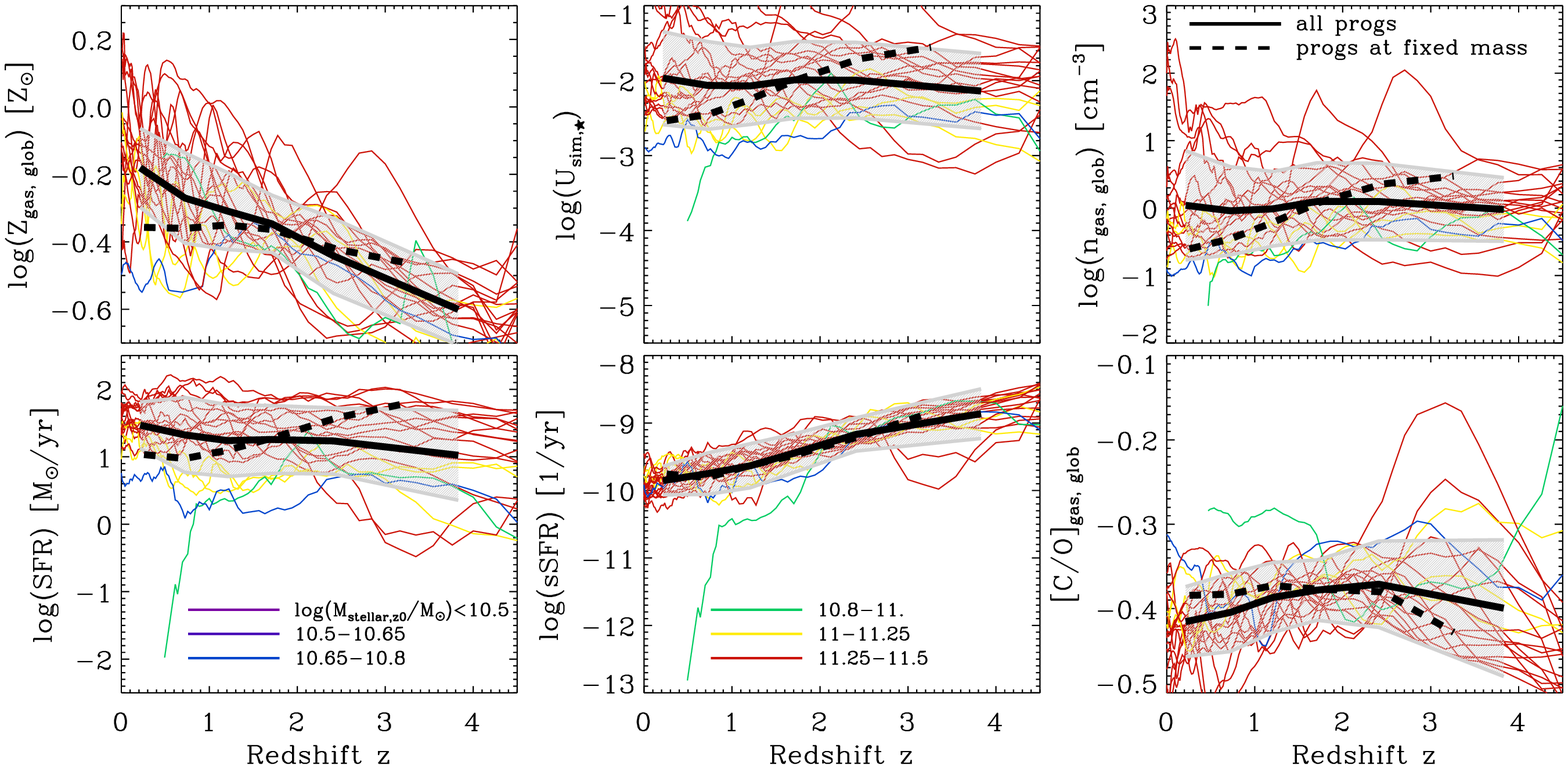,
  width=0.95\textwidth}
\caption{Same as the first two rows of Fig. \ref{Propevol}, but for
 the simulation set of galaxies {\em without} AGN feedback, as described in Appendix~\ref{AGNfeedback}.}
 \label{Propevol_NoAGN}        
\end{figure*}

Throughout this study, we have investigated predictions from
  cosmological zoom-in simulations including a prescription for AGN feedback.
Previous work \citep[e.g.,][]{Choi16} has shown that
  AGN feedback (and in particular our specific implementation of this process)
  can strongly affect galaxy properties, such as SFR and central and
  global gas densities, which control nebular emission from stars and AGN.
  Thus, we may expect {\em a priori} AGN
  feedback to be imprinted in the nebular emission from  
  galaxies. Our aim in this appendix is to test this hypothesis,
  focusing on the evolutionary trends of optical-line ratios.

 To achieve this, we performed a new suite of zoom-in simulations of  
  20 massive halos based on the same initial conditions as
  described in section \ref{setup}, but switching off BH growth and
  AGN feedback in our simulation code SPHGal. By design,
this set of re-simulated galaxies does not include any 
AGN contribution to nebular emission. As \niiha, \siiha\ and \oiha\ 
are known to be directly affected by the presence of nebular emission
from an AGN, unlike
\oiiihb\  (Fig.~\ref{Evol_lineratios_cumu}), it is most meaningful here
to examine the impact of AGN feedback on the cosmic evolution of
  \oiiihb\ in
SF-dominated galaxies.  

  Fig. \ref{BPT_SFoffset_NoAGN} shows the analog of
  Fig. \ref{BPT_SFoffset}, i.e., the evolution of the average \oiiihb\ of SF galaxies
in bins of \niiha\ (left panel), \siiha\ (middle panel) and \oiha\ (right panel), 
for the new simulations {\em without AGN feedback}. Compared to
  Fig. \ref{BPT_SFoffset} (based on simulations including AGN
  feedback), the drop in \oiiihb\ from high to low redshift is significantly
  reduced at given \siiha\ and \oiha. At fixed \niiha, we even
  find a slightly reversed trend of higher \oiiihb\ in local
  than distant galaxies.  

  To understand the origin for the hardly evolving SF branch in these
  optical diagnostic diagrams, we plot in Fig.~\ref{Propevol_NoAGN} the
  analog of the first two rows of Fig.~\ref{Propevol}, i.e., the redshift
  evolution of the different physical quantities used to select SF
  nebular-emission models, for the 20 massive galaxies and their
  main high-redshift progenitors simulated {\em without} AGN
  feedback. In contrast to the predictions including AGN
  feedback, where the drop in (specific) SFR has been identified as the
  main driver for the cosmic evolution of \oiiihb, Fig.~\ref{Propevol_NoAGN}
  shows much flatter histories of star formation, and hence, 
  ionizing-photon production ($U_{\mathrm{sim,}\star}$), whether 
  galaxies are selected by mass or not (thick black solid and dashed
  lines). In fact, at $z=0$, all massive galaxies in Fig.~\ref{Propevol_NoAGN} 
  are still highly star-forming, with $\log(\mathrm{sSFR/yr^{-1}})>-10$.  
  This is because of the lack of mechanical 
  and radiative AGN feedback, which, when present, can very efficiently 
  heat and expel cold, star-forming gas from massive galaxies, thereby 
  reducing late in-situ star formation  \citep[see e.g.][]{Choi15, Choi16}. As 
  a result, without AGN feedback, \oiiihb\ in local massive galaxies is 
  predicted to be typically as high as in distant galaxies.

From this analysis, we conclude that the drop in \oiiihb\  from high to low redshift
in our sample of simulated massive galaxies is ultimately caused by 
AGN feedback being the main responsible factor for the strong decrease in
(specific) SFR, and thus, in the SF ionization parameter.

\end{appendix}

\end{document}